\documentclass[twocolumn,aps,amssymb,footinbib]{revtex4-1}
\usepackage{notoccite}
\usepackage{amssymb}
\usepackage{graphicx}
\usepackage{amsmath}
\usepackage{mathrsfs}
\usepackage{times}
\usepackage{color}
\usepackage{subfigure}
\usepackage{feynmp}
\usepackage{bm}
\usepackage{bbold}
\usepackage{bm}
\usepackage{wasysym}
\usepackage{mathtools}
\usepackage{eufrak}
\usepackage[utf8]{inputenc}

\usepackage{calc}
\usepackage{array}

\newcommand{\rr}{\mathbf{r}}

\newcommand{\pp}{\mathbf{P}}

\newcommand{\kk}{\kappa}

\newcommand{\ve}{\varepsilon}

\begin{document}
\begin{abstract}
Planar cell polarity (PCP), the coherent in-plane polarization of a tissue on multicellular length scales, provides directional information that guides a multitude of developmental processes at cellular and tissue levels. While it is manifest that cells utilize both intracellular and intercellular mechanisms, how the two produce the collective polarization remains an active area of investigation. Exploring a generalized reaction-diffusion model, we study the role of intracellular interactions in the large-scale spatial coherence of cell polarities, and scrutinize the stages through which polarity emerges from subcellular to tissue-wide scales, as well as their dependence on intra-cellular and inter-cellular interactions. We demonstrate that nonlocal cytoplasmic interactions are necessary and sufficient for the robust long-range polarization of tissues---even in the absence of global cues---and are also essential to the faithful detection of weak directional signals. Non-locality of intracellular interactions makes a geometric readout a possibility, namely signatures of geometrical information in tissue polarity are expected to become manifest. Therefore, we investigate the deleterious effects of geometric disorder, and determine conditions on the magnitude and spatial extension of cytoplasmic interactions that guarantee the stability of polarization. These conditions get progressively more stringent upon increasing the geometric disorder, namely narrower ranges of parameters guarantee long-range polarization. Another situation where the role of geometrical information might be evident is elongated tissues. Strikingly, our model recapitulates an observed influence of tissue elongation on the orientation of polarity. Eventually, we introduce three major classes of {\textit{in silico}} mutants: lack of membrane proteins, lack of cytoplasmic proteins, and locally enhanced geometrical irregularities. In order to examine the predictability of our model, we adopt core-PCP as a model pathway, and interpret the model parameters and variables accordingly. Through comparing and matching the {\textit{in silico}} and {\textit{in vivo}} phenotypes, we interpret the effective functionalities of various core-PCP components. The success of our model in reproducing the \textit{in vivo} phenotypes, helps us shed light, in particular, on the roles of the cytoplasmic proteins Prickle and Dishevelled in cell-cell communication, and make falsifiable predictions regarding the cooperation of cytoplasmic and membrane proteins in long-range polarization, faithful readout of directional cues, as well as geometrical information. 
\end{abstract}

%unify the previous, seemingly contradictory, observations associated with the appearance of steady-state swirls and topological defects of polarization field in the absence of global cues, and 

\title{The Role of Intracellular Interactions in the Collective Polarization of Tissues\\ and its Interplay with Cellular Geometry}

%% Use letters for affiliations, numbers to show equal authorship (if applicable) and to indicate the corresponding author
%\author{Shahriar Shadkhoo}
%\author{Madhav Mani} 
%\affil{Kavli Institute for Theoretical Physics, University of California, Santa Barbara, CA 93106, USA}
%\affil{Department of Engineering Sciences and Applied Mathematics, Northwestern University, Evanston, IL 60208, USA}

\author{Shahriar Shadkhoo}
\author{Madhav Mani}
\affiliation{Kavli Institute for Theoretical Physics, University of California, Santa Barbara, CA 93106, USA}
\affiliation{Department of Engineering Sciences and Applied Mathematics, Northwestern University, Evanston, IL 60208, USA}

%\makeatletter
%\patchcmd{\@maketitle}{\LARGE \@title}{\fontsize{13}{10}\selectfont\@title}{}{}
%\makeatother
%\renewcommand\Authfont{\fontsize{11}{5}\selectfont}
%\renewcommand\Affilfont{\itshape\small}
%\sectionfont{\fontsize{11}{10}\selectfont}
%\subsectionfont{\fontsize{10}{10}\selectfont}

\maketitle

%\significancestatement{In spite of the advances in understanding the underlying circuitry of PCP, a unified view of how the interplay of cytoplasmic and intercellular mechanisms give rise to the observed patterns of polarization, is lacking. Additionally, the coupling between the cellular polarity and geometry is not well understood. This study proposes a model that clarifies the principal functions of cytoplasmic and membrane proteins in establishing long-range polarization in disordered tissues. The role of tissue elongation as a geometrical global cue is also investigated in detail. Furthermore, through a comparison of observed mutant phenotypes, with our \textit{in silico} mutants, we characterize the roles of cytoplasmic interactions in the core-PCP pathway.}

%\linespread{1.5}
%\linenumbers

\begin{small}

\section{Introduction}
The patterning of an organism requires the coupling of cellular states across multicellular scales. As such, the coordination of cellular processes on these longer length scales are crucial to the emergent phenotype of an organism and requires the faithful transduction of directional information across tissues. Planar cell polarity (PCP) is understood to be one of the core mechanisms responsible for such tissue-wide signaling \cite{axelrod2001unipolar, strutt2001asymmetric, gong2004planar, zallen2007planar, seifert2007frizzled, eaton2011cell, goodrich2011principles, gray2011planar, devenport2016tissue}. At the cellular level, polarity is defined as the asymmetric localization of membrane associated proteins on the apicolateral cell junctions, which is prompted and reinforced by cytoplasmic interactions and feedback loops \cite{strutt2001asymmetric, zallen2007planar, wang2007tissue, seifert2007frizzled, goodrich2011principles, eaton2011cell, peng2012asymmetric, devenport2016tissue}. Long-range polarization arises as a consequence of juxtacrine signaling through which adjacent cells align their polarities. How these two scales are connected, and the respective roles of molecular components in establishing long-range planar polarity is the focus of our study \cite{zallen2007planar, klein2005planar, classen2005hexagonal, aigouy2010cell}. These intra- and intercellular interactions are largely carried out via two PCP pathways: ``core-PCP'' and ``Fat/Dachsous'' \cite{zallen2007planar, goodrich2011principles, gray2011planar, devenport2016tissue}, each of which involves several interacting (trans-)membrane and cytoplasmic proteins. Throughout this paper, we adopt core-PCP as the reference pathway, according to which we interpret the results. However, the construction of the model is based on phenomenology, general arguments and physical assumptions, and its structure is independent of the molecular details of specific PCP pathways. \\

\noindent {\textbf{Molecular ingredients.}} Generically, PCP pathways consist of membrane-bound and cytoplasmic proteins. Core-PCP pathway consists of six known proteins: the membrane proteins Frizzled (Fz) and Van Gogh (Vang), which form complexes by binding to the transmembrane protein Flamingo (Fmi) on opposite sides of cell-cell junctions, and form the asymmetric heterodimers Fz:Fmi and Fmi:Vang. In addition to the above proteins, there exist three cytoplasmic proteins, believed to mediate intracellular interactions: Disheveled (Dsh) and Diego (Dgo) bind to Fz, and Prickle (Pk) binds to Vang \cite{strutt2001asymmetric,tree2002prickle, zallen2007planar, goodrich2011principles, peng2012asymmetric, devenport2014cell}. Although the presence of Dsh, Dgo, and Pk is found to be unnecessary for intercellular interactions, they facilitate the segregation of Fz and Vang to opposite sides of a cell, hence their absence impairs long-range polarization \cite{strutt2007differential, warrington2017dual, fisher2017information}. Furthermore, Fz:Dsh and Vang:Pk are believed to mutually suppress the activities of each other \cite{tree2002prickle,jenny2003prickle,klein2005planar,amonlirdviman2005mathematical,warrington2017dual}. One of the main goals of this paper is to address the significance of such cytoplasmic proteins in stabilizing cellular polarization, as well as their interplay with cell-cell signaling in establishing large-scale polarization. \\

\noindent {\textbf{Global cues.}} Although long-range polarization emerges spontaneously through cell-cell interactions, external cues are believed to be necessary for fixing the direction of polarization \cite{zallen2007planar, seifert2007frizzled, goodrich2011principles, aw2017planar, fisher2017information}. The graded distribution of regulatory factors across a tissue, i.e. morphogens \cite{wolpert1969positional, goodrich2011principles, bayly2011pointing}, mechanical signals \cite{aigouy2010cell, bosveld2012mechanical, eaton2011cell, sagner2012establishment, ma2008cell, julicher2017emergence, seifert2007frizzled, devenport2016tissue}, and geometrical cues \cite{lopez2004directional, blankenship2006multicellular, shi2014celsr1, chien2015mechanical, aw2016transient}, are speculated to provide such global orientational signals. Elongation in particular, has been observed to induce polarization. At a subcellular level, the polarization of microtubules and vesicle trafficking are also proposed to be acting as a bias to determine PCP orientation \cite{bertet2004myosin, shimada2006polarized, harumoto2010atypical, aigouy2010cell, matis2014microtubules, chien2015mechanical}. In the mammalian cochlea and skin, polarization is perpendicular to the elongation axis \cite{aw2017planar}. In mice, elongation along the medial-lateral axis has been suggested to orient the polarization along the anterior-posterior axis \cite{aw2016transient}. The present model aims to bring a mechanistic understanding to the potential role of cell geometry in the polarization of a patch of cells. \\

\noindent {\textbf{Geometry and timescales.}} Several studies (e.g. \cite{amonlirdviman2005mathematical, burak2009order, schamberg2010modelling, abley2013intracellular}) have proposed underlying physical mechanisms of PCP in ordered and isotropic (i.e. non-elongated) systems. Establishment of long-range polarization during the course of development, however, can precede the formation of an ordered lattice, e.g. margin-oriented polarity in the prepupal {\textit{Drosophila}} wing \cite{classen2005hexagonal, aigouy2010cell, eaton2011cell}. In particular, it is suggested that geometrical irregularities cause disruption in the polarization field \cite{ma2008cell}. Therefore, it is important to understand how PCP manages to propagate through disordered as well as elongated tissues. FRAP measurements of PCP proteins suggest turnover timescales to be much shorter than the timescales of cell rearrangements, justifying the study of PCP kinetics on a static tissue \cite{eaton2011cell}.\\

\noindent {\textbf{Modeling Planar Cell Polarity.}} Quantitative modeling of PCP and the underlying mechanisms has been of great interest to computational biologists and biophysicists. Several classes of model have been proposed each focusing on certain aspects, from subcellular molecular circuitry in charge of single-cell polarity, to intercellular communications that give rise to propagation of polarization over large distances. The coupling between the two modules has been a key question. While individual molecular components and their roles vary among different PCP pathways, networks of these components seem to share principal functionalities. In addition to the mechanisms of interaction among different components of a PCP pathway, detection of global cues is of great importance, as the direction of polarity is eventually set by such cues. As mentioned above global cues of various kinds have been observed in experiments. While the coupling of PCP proteins to the cues of chemical origins (morphogens) is more conceivable, the readout mechanisms of other cues such as geometrical and mechanical are not easy to decipher. Therefore, an important question is how each type of these cues influence the polarity. Some models have proposed mechanism through which cells are individually polarized by gradient cues {\cite{le2006establishment,amonlirdviman2005mathematical}}. Others have considered scenarios where rotational symmetry breaks spontaneously and collective polarization emerges even in the absence of global cues \cite{burak2009order,mani2013collective,abley2013intracellular,meinhardt2007computational}. The tissue polarity is then rotated in the right direction, through coupling to the global cues. The latter mechanism enjoys high sensitivity and faithful detection of global cues, and is robust against random misreading of the orientational information by individual cells. Since the origin of global cues is not always easily determined, knockdown of a certain gene does not necessarily unravel the underlying mechanism. 

%Most importantly the coupling between the two seemingly independent modules, has been a key question. While the molecular components vary among different PCP pathways, their effective roles have been tried to account for, by simplified interactions that capture the physically relevant features of individual, or a group of, components forming networks of feedforward/feedbacks interactions with certain functionalities. 

A popular class of mathematical models begin with intracellular interactions, which along with cell-cell couplings---that can involve the same components---give rise to long-range alignment of tissue polarity \cite{burak2009order,abley2013intracellular,mani2013collective, amonlirdviman2005mathematical}. Reaction-diffusion (RD) equations of different variations constitute the basis of the majority of these models. The characteristics of effective interactions between components of the PCP pathway are encoded in the RD equations. Generically, two types of protein complexes are considered that interact and localize asymmetrically on cell membranes. The interactions are assumed to activate/inhibit the like/unlike complexes, and can be of cytosolic and/or cross-membrane type. Several studied have focused on the necessary conditions, for the cytoplasmic interactions to establish long-range polarization. Well-established facts suggest that non-locality is among the essential features of these interactions \cite{burak2009order,abley2013intracellular,meinhardt2007computational}. These nonlocal interactions are mediated through diffusive cytosolic proteins or complexes thereof. It is suggested in Ref. \cite{burak2009order}, that nonlocal inhibition between opposite complexes are necessary and sufficient to establish long-range polarity in ordered tissues. This is fundamentally akin to the well-known local-activation--global-inhibition mechanism, which results in the accumulation of similar components on one side and the repulsion of the other components to the opposite side \cite{meinhardt2007computational}. In another study, Abley et.al. \cite{abley2013intracellular} considered various possibilities for local and nonlocal interactions between like and unlike components,  as well as different types of cell-cell couplings (i.e. direct and indirect). They concluded that intracellular interactions are crucial to the segregation of unlike complexes to the opposite compartments of cells. However, they claim that in spite of cytoplasmic partitioning, global cues are needed in order for the correlation length to exceed a few cell diameters; in the absence of cues, swirls (vortex-like) patterns of polarity appear in the steady state. \\

\noindent {\textbf{Physical considerations.}} Given the quantitative approach of this study, we find it crucial to clarify the term ``long-range'', used frequently throughout the paper. The Mermin-Wagner theorem states that ``true long-range'' ordering is prohibited in 2D systems with continuous (e.g. rotational) symmetries, except at zero stochastic noise. The long-range order is referred to as the algebraic decay of correlation functions with distance. Below we will see that the magnitude of noise in our system drops as $1/\sqrt{N_{\text{mol.}}}$, with $N_{\text{mol.}}$ the number of molecules participating in binding/unbinding reactions. Thus in the limit $N_{\text{mol.}}\to \infty$, long-range order is achieved. For finite $N_{\text{mol.}}$, a state of quasi-long range order can potentially exist. \\

\noindent {\textbf{Outline and Results.}} The objectives of this paper are threefold. Starting with a generalized reaction-diffusion equation for intracellular dynamics of proteins, we address the role of intracellular interaction in establishing tissue-wide alignment of polarization in tissues with disordered and/or elongated geometries. First we demonstrate that nonlocal interactions of both kind, namely stabilizing/destabilizing, promote the cellular segregation of unlike complexes and are crucial to the global alignment of polarity. By varying the associated length scales to stabilizing/destabilizing cytoplasmic interactions, we investigate and highlight the role of stabilizing interactions, and show that non-locality of the latter enhances the correlations of the PCP field in the presence of geometrical disorder. The role of geometrical disorder is studied, and it is found that the minimum length-scale of cytoplasmic interactions that stabilizes the long-range polarity increases for larger geometrical disorder of the tissue. Furthermore, we demonstrate that in elongated tissues, nonlocal interactions stabilize the polarization axis perpendicular to that of elongation. Finally, to further signify the necessity of nonlocal interactions, and facilitate a conversation between theory and experiment, we study three classes of {\textit{in silico}} mutants and identify phenotypic similarities with experimental observations. 

%two major questions: (a) what are the necessary and sufficient conditions on the intracellular interactions for 
%A consequence of nonlocal interactions is the coupling of geometrical information to the polarization field.

Before introducing the formalism, we shall disambiguate some terminology: ``edge'' and ``junction'', are used interchangeably throughout the paper, depending on the context emphasizing on the geometrical/mathematical or biological aspects of the problem, respectively. Therefore, one can think of an edge as a cell-cell junction. Additionally, we use for the interactions between like-like and like-unlike complexes, either activation/inhibition or stabilizing/destabilizing interactions, respectively; again depending on the context. Finally, ``defect'' has been used in two different contexts: (1) defects in the polygonal network of cells, that appear as a result of geometrical quenched disorder; and (2) topological defects in the polarization field, an example of which is swirls. The two appear in separate contexts, and should not cause any ambiguity; also the latter is usually preceded by ``topological".

\section{Model and Formalism}{\label{genmodel}}
In line with the known molecular interactions in core-PCP, we introduce a set of reaction-diffusion (RD) equations that govern the binding-unbinding dynamics of transmembrane complexes. Each cell is assumed to contain a finite pool of proteins Fz and Vang, which in their active state localize on the opposite sides of cell-cell junctions, and bind to a cross-junctional Fmi-Fmi homodimer and form asymmetric complexes Fz:Fmi-Fmi:Vang. The linear densities of total (bound plus free) Fz and Vang, are denoted by $f_0$ and $g_0$, which in the absence of global cues, are assumed to be identical for all cells across the tissue. Given that the transcriptional timescales typically far exceed the kinetic timescales of protein-protein interactions, $f_0$ and $g_0$ are treated as time-independent \cite{maree2006polarization}.

\begin{figure}[t]
\center\includegraphics[width=1\linewidth]{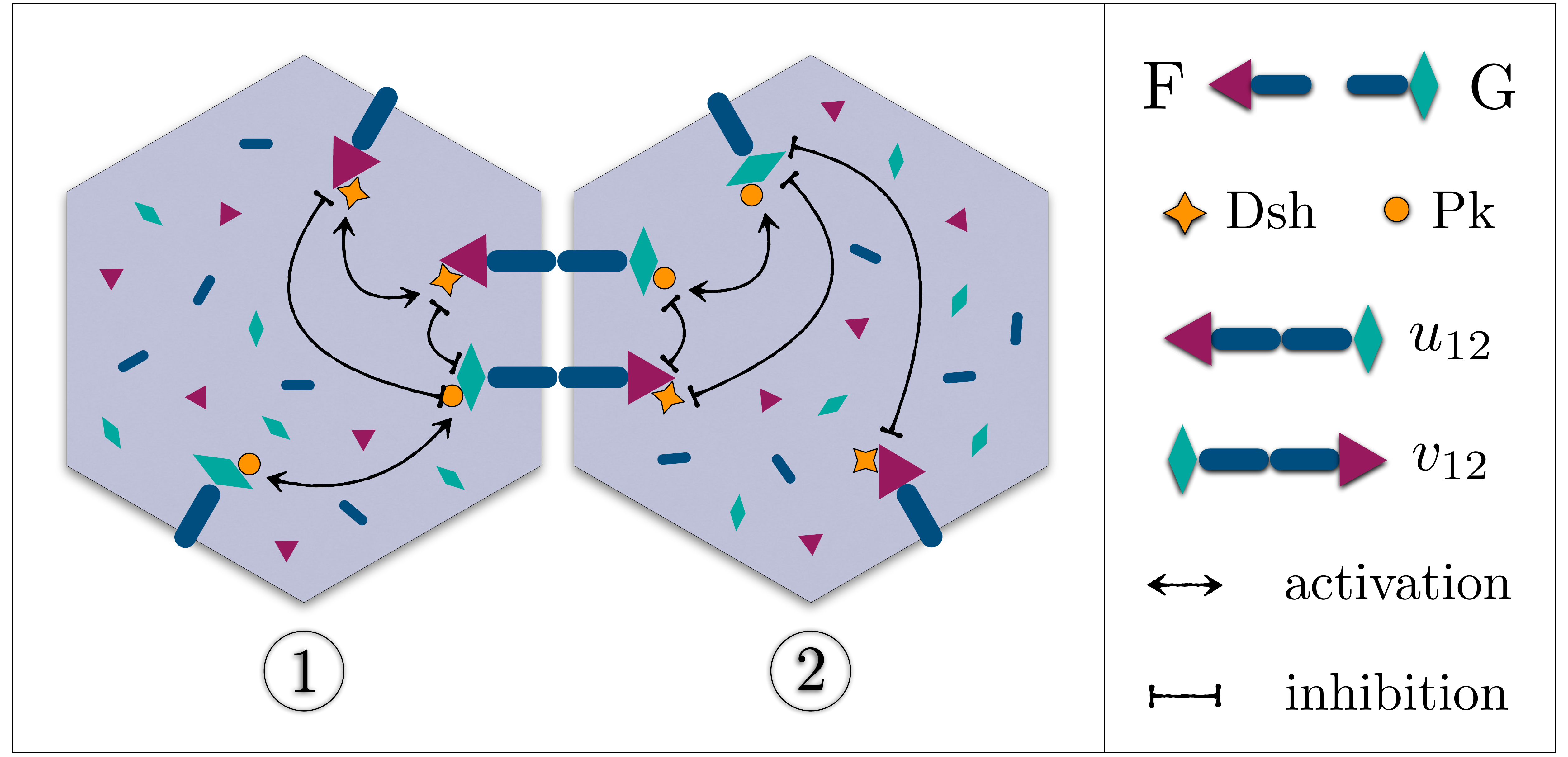}
\caption{A schematic of the relevant cytoplasmic interactions. Membrane proteins Fz (red triangles) and Vang (green diamonds) asymmetrically bind to the transmembrane proteins Fmi (dark blue rods), and form the heterodimers F-G across the junctions. At the junction shared by cells 1 and 2, two complexes of opposite directions are shown. The nonlocal interactions mediated by cytoplasmic proteins (Dsh and Pk), couple the bound proteins Fz and Vang on different junctions. All pairs of complexes in a given cell, interact with one another, with exponentially decaying magnitudes; like (unlike) complexes promote (inhibit) the membrane localization of one another. In order to keep the picture clear, we have not shown all the pairwise interactions, but only the generic ones.}
{\label{schematic}}
\end{figure}

It has been shown that Fz can in principle bind to Fmi-Fmi homodimers and make a Fz:Fmi-Fmi complex without a Vang molecule on the other side of the junction \cite{wu2008frizzled,struhl2012dissecting}. Therefore a thorough analysis requires separate RD equations for Fz and Vang. Ignoring this effect for simplicity, we define the unit of junctional polarity as a Fz:Fmi-Fmi:Vang complex for now; hence treating Fz and Vang on the same footing. We discuss the effect of this asymmetry in Sec. (\ref{Mut}). For notational convenience, we denote such complexes by F-G, where F $\equiv$ Fz:Fmi and G $\equiv$ Fmi:Vang. At any point $\rr$ on a junction shared by cells $i$ and $j$, the concentrations of bound [F$_i$-G$_j$] and [G$_i$-F$_j$], are denoted by $u_{ij}(\rr)$ and $v_{ij}(\rr)$, respectively. Consistent with this notation, we have $u_{ij}(\rr) = v_{ji}(\rr)$. The key assumption in this model is that within each cell, the formation of a dimer at a point $\rr$ is {\textit{nonlocally}} enhanced by like dimers, and its dissociation is enhanced {\textit{nonlocally}} by opposite dimers. In short, bound Fz nonlocally stabilizes Fz and destabilizes Vang; and vice versa. This represents an positive feedback between Fz and Vang in the adjacent cells: promoting (inhibiting) Fz in the same cell indirectly brings more (less) Vang to the other side of the junction. The nonlocal interactions are diffusively mediated through the cytoplasmic proteins Dsh, Dgo, and Pk, and/or their associated feedback loops. Figure (\ref{schematic}) is an illustration of the relevant cytoplasmic interactions.

The RD equations governing the binding/unbinding dynamics read:
\begin{align}{\label{mainRD}}
\frac{du_{ij}(\rr)}{dt} &=  \kappa f_i^{\text{ubd}}g_j^{\text{ubd}}\bigg(1+\alpha\sum_{\{k\}_i}\int_{i\cap k} d\rr'\mathcal K_{\text{uu}}(\rr-\rr')u_{ik}(\rr')\bigg)\nonumber\\
& - \gamma u_{ij}(\rr)\bigg(1+\beta\sum_{\{k\}_i}\int_{i\cap k} d\rr' \mathcal K_{\text{uv}}(\rr-\rr')v_{ik}(\rr')\bigg)\nonumber\\
& + \eta(\rr, t) + \mathcal M_{ij}(\rr)\,\exp({-t/\tau_0}).
\end{align}
The notations adopted here are as follows: $\sum_{\{k\}_i}$ is the summation over all neighbors $\{k\}$ of cell $i$; and $\int_{i\cap k}d\rr'$, represents integration over the junction shared by cells $i,k$. In the above equation, the first and second terms on the r.h.s. correspond to the formation and dissociation rates, respectively. $\kappa$ is the bare rate of formation. Based on the assumption that the diffusion of unbound proteins is rapid we posit that the pool of free proteins are uniformly accessible to the perimeter of a cell. Therefore, the formation rate of $u_{ij}(\rr)$ is proportional to the densities of unbound, i.e. cytoplasmic Fz in cell $i$, $f^{\text{ubd}}_i$, as well as that of Vang in cell $j$, $g^{\text{ubd}}_j$. In the second term, the dissociation rate is proportional to local concentration of the dimer itself, with the bare rate $\gamma$. The formation/dissociation processes are amplified by like/unlike dimers, respectively, through the nonlocal terms, $\alpha\sum_{\{k\}_i}\int_{i\cap k} d\rr'\mathcal K_{\text{uu}}(\rr-\rr')u_{ik}(\rr')$ and $\beta\sum_{\{k\}_i}\int_{i\cap k} d\rr' \mathcal K_{\text{uv}}(\rr-\rr')v_{ik}(\rr')$, that characterize cooperative formation and dissociation. The functional form of the kernels $\mathcal K_{\text{uu}}(\rr-\rr')$ and $\mathcal K_{\text{uv}}(\rr-\rr')$ and their coefficients $\alpha$ and $\beta$, are introduced below. Finally, the last term $\eta(\rr,t)$ is a stochastic Gaussian white noise: $\langle \eta(\rr,t)\rangle= 0$, and $\langle \eta(\rr,t)\eta(\rr',t')\rangle = \eta_0^2\delta(\rr-\rr')\delta(t-t')$, which arises from the molecular noise of chemical reactions and stochasticity in the upstream signaling pathways. The former, modeled as a Poisson process, is speculated to be the dominant source of noise \cite{burak2009order}, with its magnitude scaling as $\eta_0 \sim 1/\sqrt{N_{\text{mol.}}}$, where $N_{\text{mol.}}$ is the number of molecules per area of the lateral interfaces. More precisely, the number of participating molecules is $N_{\text{mol.}}\overline{u}/f_0$, where $\overline u$ is approximately the average value of $u$, and $f_0=1$ is the unit of concentration. Using the variance of the number of reactions per unit time, that is given by the r.h.s. of Eq. (\ref{mainRD}), the noise level, is estimated to be of order $\eta_0\simeq$ 0.01 -- 0.1, for $N_{\text{mol.}}\simeq$ 1 -- 5 $\times 10^3$, the approximate number of Frizzled molecules in the {\textit{Drosophila}} wing \cite{burak2009order}. The last term models global cues with local magnitude $\mathcal M_{ij}(\rr)$ and time scale $\tau_0$, over which the corresponding gene is expressed. For simplicity, we assume that the cues only couple to one of the complexes, say F. Global cues are discussed in Sec. (\ref{SLI-NLI}B), separately. Unless mentioned otherwise, the cues are assumed to be zero. 

The densities of unbound Fz and Vang are obtained by subtracting the densities of bound proteins from the total densities.
\begin{equation}{\label{totalAB}}
f_i^{\text{ubd}} = f_0 - \frac{1}{\mathcal{C}_i}\int_{\varhexagon_i} d\rr\,u_{i}(\rr),
\end{equation}
and a similar relation for $g_i^{\text{ubd}}$. Here $\mathcal C_i$ is the perimeter of cell $i$, and $u_i(\rr)$ represents a $u-$complex sitting at point $\rr$ on the perimeter of cell $i$, i.e. (F$_i$-G); we dropped the index $j$ of the neighbors, because we only care about the total bound F, not the specific cell with which this F-G complex is shared. The same notation is applied to $v_i(\rr)$:
The cell polarity with respect to the centroid of cell $i$ at $\mathbf R_i$, is defined as:
\begin{equation}
\pp_i = \frac{1}{2}\;\int_{\varhexagon_i}d\rr \,\frac{\rr-\mathbf R_i}{|\rr-\mathbf R_i|}\;\big(u_i(\rr) - v_i(\rr)\big).
\end{equation} 
All the quantities are expressed in units of $f_0 = \gamma = \ell_0 = 1$, where $\ell_0$ is the average length of the cell-cell junctions. In the following we will see that in certain situations, an alternative, but related, definition of polarization simplifies our analyses. We define two junctional variables: the cross-junctional polarity $p_{ij}(\rr) = u_{ij}(\rr) - u_{ji}(\rr) = u_{ij}(\rr) - v_{ij}(\rr)$, and the total concentration of localized complexes, $s_{ij}(\rr) = u_{ij}(\rr) + v_{ij}(\rr)$. Given $p_{ij},s_{ij}$, the cell polarity can be readily extracted; see SI. (2).\\

\noindent {\textbf{Nonlocal Interactions.}} The kernels $\mathcal K_{\text{uu}}(\rr-\rr')$ and $\mathcal K_{\text{uv}}(\rr-\rr')$, identify the functional form of the interactions between like and unlike complexes, respectively, and are taken to be exponentially decaying: $\mathcal K_{\text{uu}}(\rr) = N_{\text{uu}}^{-1} \exp(-|\rr|/\lambda_{\text{uu}})$ and $\mathcal K_{\text{uv}}(\rr) = N_{\text{uv}}^{-1} \exp(-|\rr|/\lambda_{\text{uv}})$, where $\lambda_{\text{uu}}, \lambda_{\text{uv}}$ are the characteristic length scales of $u$-$u$ and $u$-$v$ interactions, respectively. The prefactors $N_{\text{uu}}$  and $N_{\text{uv}}$ are normalization factors, to be determined shortly. Before calculating the normalization factors, we shall make a detour to explain an approximation we used that greatly reduces the computational cost simulations. The full integro-differential equation we introduced in Eq. (\ref{mainRD}) couples every pairs of points on the perimeter of a cell. Assuming uniform distributions of localized proteins on each junction, we simplify the equation to obtain effective equations for edges, where the integration over the kernels are replaced by matrix products, see Eqs. (\ref{approxRD}) and (\ref{matrix}). To avoid confusions with cell indices, we use Greek letter $\mu$ to label a single edge. First, we introduce junctional concentrations:
\begin{equation}
u_\mu = \int_\mu d\rr \,u_\mu(\rr),
\end{equation}
where the integral is taken over the concentration of complex $u(\rr')$, on all the points $\rr'$ along junction $\mu$. Same definition applies to $v_\mu$. With this approximation, one can recast Eq. (\ref{mainRD}) into:
\begin{align}{\label{approxRD}}
\frac{du_{\mu}}{dt} =& \kappa f_i^{\text{ubd}}g_j^{\text{ubd}}\bigg(1+\alpha\sum_{\nu}\widehat {\mathcal K}_{\text{uu}}^{\mu\nu}u_{\nu}\bigg) - \gamma u_{\mu}\bigg(1+\beta\sum_{\nu} \widehat {\mathcal K}_{\text{uv}}^{\mu\nu}v_{\nu}\bigg)\nonumber\\
& + \eta_\mu(t) + \mathcal M_{\mu}\,\exp({-t/\tau_0}).
\end{align}
In the above equation, $\eta_\mu$ and $\mathcal M_\mu$ are the noise and the global cue averaged over the length of edge $\mu$. The cooperative interactions are now reduced to simple matrix products of a matrix $\mathcal K_{\text{uu}}$ and a vector $u$, wherein the matrix elements $\widehat {\mathcal K}^{\mu\nu}_{\text{uu}}$ are purely geometrical constants obtained using the following relation:
\begin{equation}\label{matrix}
\widehat {\mathcal K}_{\text{uu}}^{\mu\nu} = \iint_{\mu,\nu}d\rr_{\nu}\,d\rr_{\mu}\;\mathcal K_{\text{uu}}(\rr_\nu - \rr_\mu).
\end{equation}
A similar relation holds for destabilizing interactions $\widehat {\mathcal K}_{\text{uv}}^{\mu\nu}$. The edge-edge coupling coefficients $\widehat {\mathcal K}_{\text{uu}}^{\mu\nu}$, are only a function of the cell geometries; once calculated for a given tissue, the full matrix can be used throughout the course of integrating the dynamics of protein concentrations; hence reducing computational runtime by orders of magnitude. 

With this background, we now calculate the normalization factors, introduced above. In order to discern the net effect of interaction ranges from the effective coefficients $\alpha, \beta$, we choose the normalization factors $N_{\text{uu}}$ and $N_{\text{uv}}$, such that the self-interaction of an edge of length $\ell_\mu$ equals $\alpha$, namely: $\alpha \widehat {\mathcal K}_{\text{uu}}^{\mu\mu} = \alpha$, or $\widehat {\mathcal K}_{\text{uu}}^{\mu\mu} = 1$. The same relation holds for $\beta$ and $\widehat {\mathcal K}_{\text{uv}}^{\mu\mu}$. This choice of normalization, by fixing the edge self-interactions, ensures that the observed behavior upon changing $\lambda$'s is purely due to nonlocal edge-edge coupling and not the effective coefficients of local interactions $\alpha,\beta$. Satisfying this condition for all edges simultaneously is not possible, except for ordered tissues. The normalization constant is thus calculated for a hypothetical edge with average length of all edges, $\ell_0 \equiv \overline\ell_\mu$. Using the definition of kernels, for the ``average'' edge $\widehat {\mathcal K}_{\text{uu}}^{\text{avg.}}$ from Eq. (\ref{matrix}), we obtain:
\begin{align}
N(\lambda) &= \iint d\rr d\rr' \exp(-|\rr' - \rr|/\lambda) \nonumber\\
&=  2\lambda^2\left(e^{-\ell_{0}\lambda^{-1}} + \ell_{0}\lambda^{-1} - 1\right).
\end{align}
Here, $\rr$ and $\rr'$ move along the length of a single edge with length $\ell_0$. We omitted subscripts ``$\text{uu}$'' and ``$\text{uv}$'' for simplicity. \\

\noindent {\textbf{Limit of Strictly Local Cytoplasmic Interactions (SLCI).}} In the limit of small $\lambda/\ell_\mu\to 0$, we get, $\widehat {\mathcal K}_{\mu\nu} = \delta_{\mu\nu}$, where $\delta_{\mu\nu}$ is the Kronecker delta. Thus in the SLCI limit, the equations read,
\begin{align}{\label{SLIRD}}
\frac{du_\mu}{dt} =&\,\kk f^{\text{ubd}}_ig^{\text{ubd}}_j(1+\alpha u_{\mu}) -\gamma u_{\mu}(1+\beta v_{\mu}) + \eta_\mu(t).
\end{align}

\noindent{\textbf{Interpretation of The Model Parameters.}} Besides the geometrical and stochastic disorder parameter, our model consists of four independent dimensionless parameters: $\kappa/\gamma , \alpha, \beta , \lambda/\ell_0$. Here we try to make connections between the effective roles of PCP components and these parameters. First we note that the cooperative interactions represent the {\textit{effective}} couplings of two {\textit{complexes}}, not the interactions between Fz's or Vang's proteins separately. In other words, $\alpha$ integrates the stabilizing interactions of Fz by Fz, and Vang by Vang; similarly for $\beta$, i.e. destabilizing effects between Fz and Vang. The functional form of the kernels can be interpreted as interactions mediated by diffusing cytoplasmic proteins with diffusion constant $D$ and the degradation rate $\tau_{\text{c}}^{-1}$, such that $\gamma\tau_{\text{c}}\ll 1$; thus $\lambda = \sqrt{D\tau_{\text{c}}}$ . The diffusion timescale of cytoplasmic proteins for $D\simeq 0.5$ ($\mu^2/s$), and $\ell_0\simeq 5$ ($\mu$), is of the order of $\simeq10$ (min), much shorter than polarization dynamics which occurs on timescales of a few hours. On the other hand, we know that Dsh and Pk, both promote localizing similar complexes, and suppress the opposite ones. In particular, Dsh {\textit{locally}} promotes the localization of Fz, whereas Pk destabilizes that in a {\textit{nonlocal}} fashion by inhibiting the membrane localization of Dsh and antagonizing Fz accumulation, hence indirectly stabilizing Vang localization \cite{tree2002prickle,warrington2017dual,amonlirdviman2005mathematical,fisher2017integrating,struhl2012dissecting}. Given that the diffusion lengths of these proteins are independent of their role (stabilizing/destabilizing), in the main text we assume $\lambda_{\text{uu}}=\lambda_{\text{uv}}=\lambda$. The cases of $\lambda_{\text{uu}}\ne \lambda_{\text{uv}}$ are elaborated on in SI. (4.4). Coefficients $\alpha, \beta$, which parametrize the strength and length scale of the cooperative interactions, depend on the concentrations of {\textit{bound}} cytoplasmic proteins Dsh and Pk, which in turn depend on the abundance of total cytoplasmic proteins, and their binding affinities with Fz and Vang; see SI. (1). In summary, both Dsh and Pk contribute to the magnitudes and length scales of the cooperative interactions $\alpha,\beta$, and $\lambda$. In order to shed light on the principal mission of Dsh and Pk in the long-range correlation of polarization, in Sec. (\ref{Mut}) we compare the \textit{in vivo} phenotypes of $\textit{dsh}^-$ and $\textit{pk}^-$ with the \textit{in silico} phenotypes of $\alpha,\beta$ and $\lambda$. Finally, the formation of the polar complex Fz:Fmi-Fmi:Vang, among other factors, is contingent on the presence of Fmi and formation of Fmi-Fmi dimers. Therefore, we expect $\kappa/\gamma$, the ratio of formation and dissociation rates of the complexes, to be an increasing function of the binding affinities of Fz:Fmi as well as Fmi:Vang.\\

\noindent {\textbf{Correlation Function.}} The correlation function of polarization is defined as a measure of alignment of polarity. In order to investigate the temporal behavior of the spatial extension of the alignment, we define the equal-time correlation functions as follows. Consider an arbitrary cell at $\rr_i$, and a vector $\rr$ connecting it to another cell at $\rr_j = \rr_i + \rr$. With no further assumption we can define a correlation function, that is dependent on the distance $r = |\rr|$ and the relative angle of $\rr$ and $\pp(\rr_i)$; we call it $\theta_{\rr,\textbf{p}}$. The latter appears due to the vectorial nature of polarization field; there is no {\textit{a priori}} reason for dipoles to be correlated equally in all directions. For a tissue of $N_c$ cells, the correlation function at time $t$ reads,
\begin{equation}{\label{corrfun}}
S(r,\theta_{\rr,\textbf{p}}\,;t) = N_c^{-1}\sum_{i} \pp(\rr_i\,;t)\cdot\pp(\rr_i + \rr\,;t).
\end{equation}
The above quantity calculates the average conditional probability that the dipole at point $\rr_j = \rr_i + \rr$, takes on the value and direction of $\pp(\rr_j)$, should the polarity at point $\rr_i$ be $\pp(\rr_i)$. For $\theta_{\rr,\textbf{p}} = 0, \pi$, and $\theta_{\rr,\textbf{p}} = \pm\pi$, we get parallel and perpendicular correlations, respectively. In spite of this angular dependence, averaging the above correlation function over $\theta_{\rr,\textbf{p}} \in [0 , 2\pi)$, returns a weighted averaged of correlation function as a function of $r=|\rr|$. This function is bounded by the the longitudinal and transverse correlations from above and below, respectively. Intuitive arguments are provided in this regard, in Sec. (\ref{SLI-NLI}). Thus, we define radial correlation function as follows.
\begin{equation}{\label{corrfun}}
S(\rr\,;t) = N_c^{-1}\sum_{i}\int_0^{2\pi}\frac{d\theta_{\rr,\textbf{p}}}{2\pi}\; \pp(\rr_i\,;t)\cdot\pp(\rr_i + \rr\,;t).
\end{equation}
Correlation length can be obtained from the above equation:
\begin{equation}{\label{corrlen}}
\xi(t) = \frac{\int_0^{R_c} dr\,r\,S(r;t)}{\int_0^{R_c} dr\, S(r;t)}.
\end{equation}
Here, $R_c = 40$ (cell diameter) is the farthest cell with which the correlation is calculated, for a given reference cell. A perfectly correlated polarization field, returns: $\xi = R_c/2$. 

A simpler measure for the global {\textit{orientational}} order is $\mathcal O(t) = \overline P(t)\big/\overline Q(t)$, where $\overline P(t) = |\langle \pp(t)\rangle|$, and $\overline{Q}(t) = \langle |\pp(t)|\rangle$, in which $\langle\bullet\rangle$ denotes spatial average. Thus $\mathcal O(t)$ saturates to unity for perfect alignment. However, we will see below that $\overline{Q}(t)$ can be misleading in interpreting the segregation of proteins, and in not a reliable  \\

%the integration variable $\theta$ is the angle between vector $\rr$ connecting cell $i$ to $j$, and the $x$-axis. The integral is calculated for a fixed $\rr_i$ as the reference cell, and fixed length $r$ of the relative position vector $\rr$. Changing $\theta$ makes a circle of radius $r$ around cell $i$, covering all cells $j$ at distance $r$ from cell $i$. Next, we average over all cells $i$ by changing the reference cell:

%The angular integration can only be interpreted if the angular invariance is preserved by the system. Indeed, both angular average, and the other average over the position of reference cells, require the system to extend to infinity. This is obviously neither true nor possible in reality or in the simulations. In order to make this approximation more reliable, we measure the correlations for a patch of 20-by-20 cells, located at the center of a lattice of size 100-by-100 cells, subject to periodic boundary conditions. Rotational invariance requires $S(r;t)$ to depend only on $r = |\rr|$. The correlation length is thus defined as

\noindent {\textbf{Geometrical Disorder.}} The edge lengths are $\ell_i = \ell_0(1 + \epsilon_i)$, where $-\epsilon_0\leq \epsilon_i \leq + \epsilon_0$ with uniform distribution, and $\langle \epsilon_i\epsilon_j\rangle = \epsilon_0^2\delta_{ij}/6$. In 2D, lattice defects, i.e. non-hexagonal cells appear above a certain level of quenched disorder corresponding to $\epsilon_0 \simeq 0.25$. In disordered cases, we use $\epsilon_0 \simeq 0.5$ and density of defects $n_d \simeq 0.6$, corresponding roughly to the statistics of larval and prepupal {\textit{Drosophila}} wing \cite{classen2005hexagonal}. In order to gain more intuition on the geometrical properties of the network of cells, as a function of $\epsilon_0$, we performed statistical analysis on synthetic tissues constructed from Voronoi tessellation of randomly positioned cell centroids. The results of these analyses are shown in SI. (3) and SI. Fig. (1).\\

\begin{figure}[h]
\center\includegraphics[width=1\linewidth]{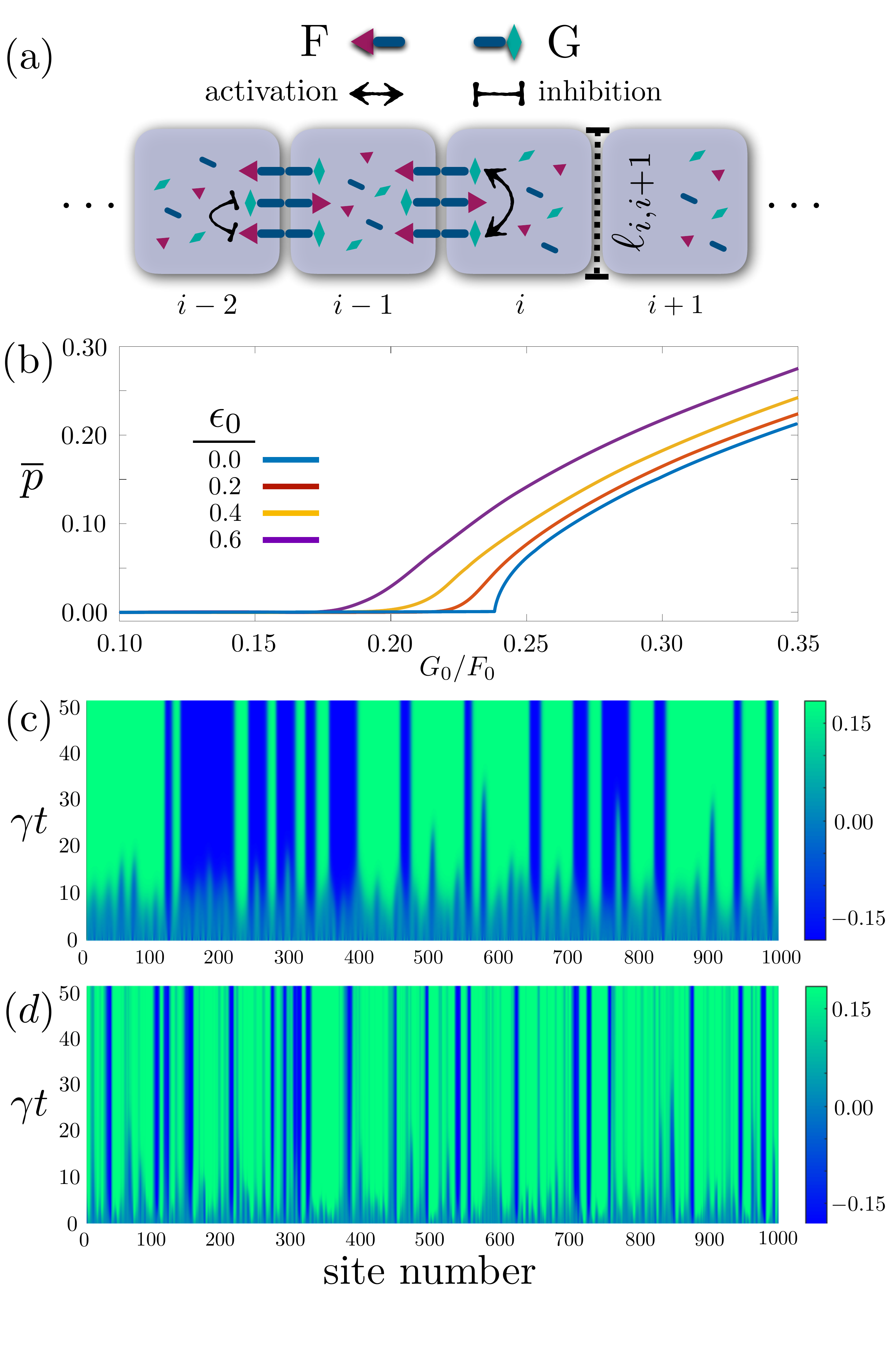}\\
\caption{(a) cartoon of a 1D array, with juxtaposing cells. Fundamental complexes F and G are shown on the top. Like/unlike complexes activate/inhibit each other on every interface (only two of them are shown here). The length of the junction between two cells are also shown as $\ell_{i,i+1}$. (b) shows numerical solutions of the average polarization against $G_0/F_0$, with $F_0, G_0$ the total number of proteins per cell, for different values of length disorder $\epsilon_0 = 0$ to $0.6$. In ordered arrays, the critical value is $g_0^* \simeq 0.23$. The plot is obtained by ensemble averaging over 1000 realizations of quenched disorder, in arrays of 1000 cells. (c), (d) show the heatmaps of polarization of different sites in units of $f_0\ell_0$, versus time (vertical axis) at $G_0/F_0 = 0.3$. (c) An ordered array with small bias, and (d) a highly disordered array $\epsilon_0 = 0.6$, with large initial bias.}
{\label{1D-MT}}
\end{figure}

\noindent{\textbf{One-dimensional Arrays of Cells.}} First, we briefly review the one-dimensional case as it is more amenable to analytical treatment, and captures some features of 2D systems. In particular when the sixfold symmetry of a hexagonal lattice is broken---either spontaneously or explicitly---down into a twofold ($+/-$) symmetry, the system behaves much like a one-dimensional array. Examples of this scenario in 2D include, partitioned cells in which the two complexes F and G are segregated to the opposite sides of the cells, and elongated cells where the sixfold rotational symmetry is broken. These cases are discussed in full details, in sections (\ref{SLI-NLI}) and (\ref{elongeom}), respectively. 

In one dimension, cells are juxtaposed on a line. A pair of adjacent cells $i$ and $i+1$, share a junction of length $\ell_{i,i+1}$, which hosts membrane-bound complexes F and G. A schematic of a 1D array is depicted in Fig. (\ref{1D-MT}a). At the mean-field (MF) level for an ordered system in 1D, edges separate cells $i$ and $i + 1$, i.e. for all edges $\forall \{i,i+1\} : \ell_{i,i+1} = \ell_0$, we get $u_{i,i+1} = u$ and $v_{i,i+1} = v$, hence $f^{\text{ubd}}_i = f^{\text{ubd}}$ and $g^{\text{ubd}}_i = g^{\text{ubd}}$. We switch variables from $(u,v)$ to $(p,s)$, where $p_{i,i+1} = u_{i,i+1} - v_{i,i+1}$, and $s_{i,i+1} = u_{i,i+1} + v_{i,i+1}$. The steady-state solution, defined as $\forall i \,:\, p_{i,i+1} = p$, and $\forall i \,:\, s_{i,i+1} = s$, exhibits a bifurcation from unpolarized to polarized state, as the control parameter $g_0/f_0$, is increased above a critical value \cite{mani2013collective}. The MF polarization reads:
\begin{subequations}
\begin{equation}{\label{pvss}}
s =(f_0 + g_0) - \sqrt{(f_0 - g_0)^2 + \frac{4\gamma}{\kappa\alpha}},
\end{equation}
\begin{equation}{\label{svss}}
p = \pm\bigg(s^2 - \frac{4}{\alpha\beta}\bigg)^{1/2}.
\end{equation}
\end{subequations}
From the second equation, the bifurcation takes place at $s^* = 2/\sqrt{\alpha\beta}$. In terms of actual control parameter $g_0$ we get,
\begin{equation}{\label{gcrit}}
\quad g_0^* = \frac{\gamma/\kappa\alpha}{1-\sqrt{1/\alpha\beta}}+\sqrt{\frac{1}{\alpha\beta}}.
\end{equation} 
This result indicates the divergence of the critical value $s^*$ (or $g_0^*$), for $\alpha\beta \to 0$, implying that the emergence of polarization requires cooperative interactions. Numerical solutions are presented in Fig. (\ref{1D-MT}b), for a system with constant $F_0, G_0$, the total number of Fz and Vang per cell. In ordered systems, one can choose $F_0 , G_0$ or equivalently $f_0 , g_0$ to be constant across the tissue. In disordered systems, the two conditions are not equivalent. Only in 1D and in order to emphasize a possible effect of disorder we assume $F_0 , G_0$ to be constant. As such, the concentrations available to the junctions of a cells, are no longer equal to those in the adjacent cells. In a disordered system with $\ell_{i,i+1} = \ell_0 + \epsilon_{i,i+1}$, the concentrations $f_{0,i} = 2F_0/(\ell_{i-1,i} + \ell_{i , i+1})$ and $g_{0,i} = 2G_0/(\ell_{i-1,i} + \ell_{i , i+1})$ are thus randomized. In SI. (4.1), we show that the relevant quantities for edge polarization are $g_{0,i+1}/f_{0,i}$ and $g_{0,i}/f_{0,i+1}$, both of which are nonuniform as a result of length disorder. The ratio $g_{0,i}/f_{0,i}$ which can be interpreted as ``local critical point'', is thus randomized as well. Therefore in disordered systems ($\epsilon_0\ne 0$), the collective singular behavior of the ordered systems, at the well-defined critical point, is smeared out, and the second-order transition is replaced by a smooth cross-over (Fig. (\ref{1D-MT}b)).
%Like a system with random critical point, the singularity is smeared out for . 

Numerical solutions, Fig. (\ref{1D-MT}c) and (\ref{1D-MT}d), suggest that in the limit of small stochastic noise and initial bias, the steady state is not guaranteed to be uniformly polarized. The initial imbalance of protein distributions is defined as $p_0=|u_0-v_0|$, with $u_0, v_0$, the spatial averages of initial dimers' concentrations. The bias is defined as $\delta p_0/p_0$, the normalized magnitude of spatial fluctuations of initial polarity. Thus, small and large bias limits correspond to $\delta p_0/p_0 \gtrsim 1$ and $\delta p_0/p_0 \lesssim 1$. While in ordered systems, a moderate initial bias suffices to achieve a uniform polarization, the patterns of polarity in highly disordered systems are robust and largely determined by the local geometry (disorder) of the array. Therefore we observe that already in 1D, the quenched disorder imposes undesirable solutions, impairing the faithful transduction of directional information through PCP signaling. As we will see in the following, the situation gets only more complicated in two dimensions, even for ordered tissues. One of the main goals of this paper is to find mechanisms to circumvent these issues associated with two-dimensional tissues. 

\section{Intracellular Interactions:\\ Local or Nonlocal?}{\label{SLI-NLI}}
The systems in one and two dimensions show inherently different behavior. In 1D, the proteins have only two junctions at which the can localize. This limited number of choices and the resultant predictability are absent in two dimensions. Due to the large number of possible steady states in 2D, the initial configuration, as well as stochastic and geometrical disorders, influence the final state. We show, in this section, that nonlocal cytoplasmic interactions (NLCI) destabilize a great portion of unpolarized fixed points, in favor of the polarized ones. Furthermore we find an optimal range of the NLCI length scale, $\lambda$, that assists with establishment of long-range alignment. In the following, we present the results for a set of parameters which lies within the polarized regime: in the units of $f_0 = \gamma = \ell_0 = 1$, we set $\alpha = \beta = 5, \kappa = 10$, and $g_0 = 1$. The qualitative changes upon varying the above parameters is discussed as well. 

The numerical solutions are presented below, but first let us attempt to gain some insight using analytical MF analysis for ordered tissues. We define the MF approximation in 2D as uniform distribution of $f_i^{\text{ubd}}g_j^{\text{ubd}}$ and $f_j^{\text{ubd}}g_i^{\text{ubd}}$ for all junctions $\{ij\}$. The validity of this approximation can be justified by noting the diffusive nature of $p, s$ dynamics, i.e. the amount of membrane-bound proteins (see Ref. \cite{mani2013collective}), which in turn is concomitant with the diffusive dynamics of the concentrations of free cytoplasmic proteins. We checked this assumption numerically. The value of $f_i^{\text{ubd}}g_j^{\text{ubd}}$, normalized by the mean value, is plotted for all edges for initial and final states in SI. Fig. (3). In steady state, the standard deviation of the distribution normalized by the mean, is very much independent of initial condition as well as the model parameters, and remains below $\simeq 0.05$, within the ranges explored in this paper, for both SLCI and NLCI regimes. Owing to the sixfold symmetry of equilateral cells, we expect the steady-state {\textit{magnitudes}} of $p, s$ to be the same on all edges. Thus, three edges will be carrying inward, and the other three outward dipoles; otherwise the MF criterion is violated. 

\begin{figure}[h]
\center\includegraphics[width=1\linewidth]{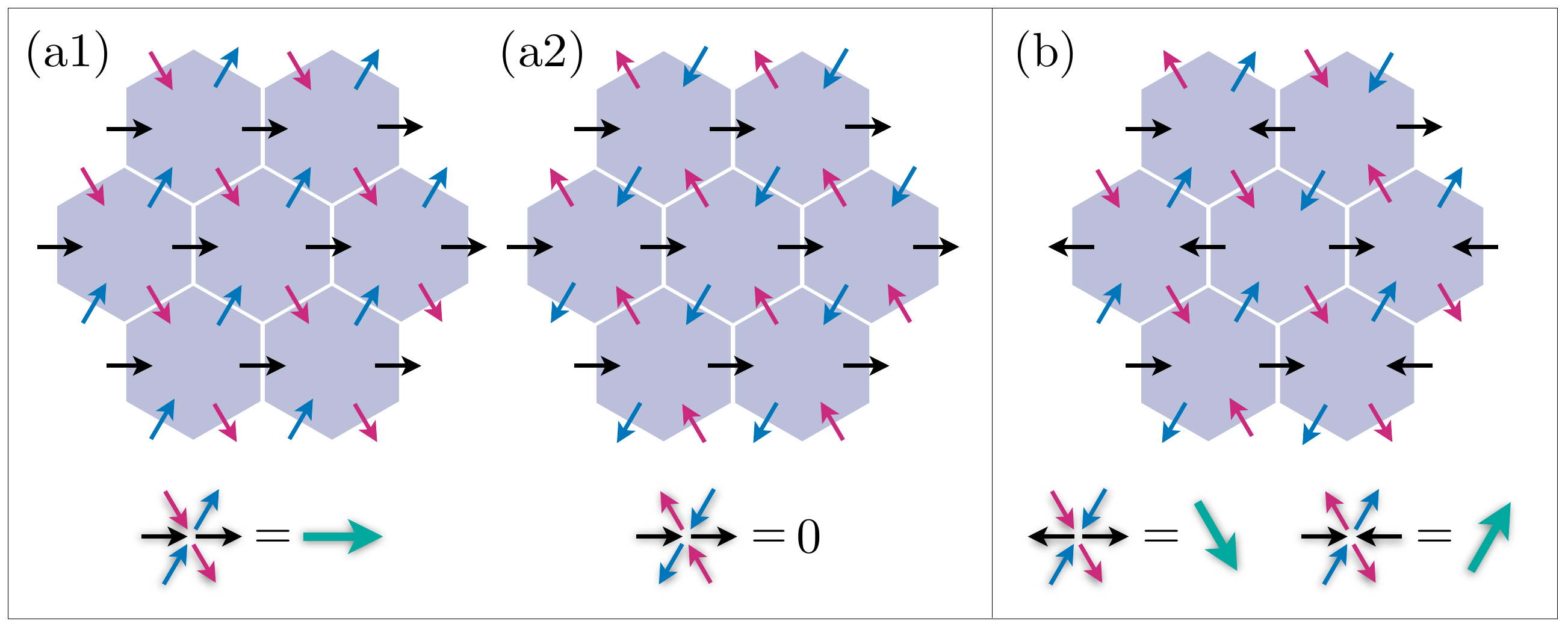}\\
\caption{\small{Cartoons of trivial (a1,a2), and nontrivial (b) mean-field solutions. In trivial solution, the translational invariance holds along each axis with nonzero (a1) and zero (a2) polarities. The latter is destabilized by any finite-range cytoplasmic interactions that induces segregation. In (b) the only constraint is uniform $f^{\text{ubd}}g^{\text{ubd}}$ across the tissue, hence three incoming and three outgoing dipoles, and unequal cell polarities. In (b) the dipoles of the central cell and its right neighbor are shown for example.}}
{\label{2Ds}}
\end{figure}

The simplest of the MF solutions, here referred to as the trivial solutions, preserve translational invariance along each of the main three axes of a hexagonal lattice separately. There exist two types of trivial MF solutions: polarized and unpolarized. The two types can be seen in Figs. (\ref{2Ds}a1) and (\ref{2Ds}a2): six configurations with nonzero net polarization, obtained by 60 (deg) rotations of the configuration (a1); and two unpolarized states, obtained by flipping all the dipoles in (a2). There exist other types of solutions meeting the MF criterion, in which translational invariance of cell polarity is not preserved, but $f_i^{\text{ubd}}g_j^{\text{ubd}}$ is uniform for all edges; see Fig. (\ref{2Ds}b). We call these nontrivial MF solutions. The cellular polarities are randomly oriented and long-range correlation is absent in the nontrivial MF solutions. Straightforwardly, as one can see in Fig. (\ref{2Ds}b), the number of possible configurations of this type hugely outnumbers the eight trivial solutions. Analytical investigation of such solutions are beyond the scope of this study. Important intuitive arguments are provided regarding this class of solutions, and their behavior is elaborated on in SI. (4.3). We shall emphasize that the above classification of MF solutions is independent of the locality or lack thereof of the cytoplasmic interactions, and is purely based upon MF criterion, and symmetry arguments. However, the class of solutions which is realized in a system, is strongly dependent on the cytoplasmic interactions. We will see below that systems with strictly local interactions are incapable of segregating the proteins to the opposite sides of cells, whereas nonlocal interactions mediate repulsive interaction between the unlike proteins that amounts to partitioning the cells. Therefore, the nontrivial MF solutions are realized in the SLCI regime, and the polarized trivial solutions are the result of NLCI regime. Unpolarized trivial solutions Fig. (\ref{2Ds}a2), although might appear by accident in SLCI systems, are destabilized by nonlocal interactions. 

% Here, we only discuss the trivial solutions in ordered systems, in part because we will eventually see, that NLCI suppresses a great deal of nontrivial solutions, as well as unpolarized trivial solutions.
Since the analysis of nontrivial solutions provides an intuitive argument as to why SLCI is insufficient to obtain long-range polarization, we highly encourage the reader to peruse SI. (4.3). The RD equations in 2D are precisely the same as those in 1D, except the pools of proteins Fz and Vang are shared between six junctions instead of two. Therefore, the MF concentrations in steady state, are identical to those of 1D case. The corresponding equations and solutions in 2D are derived in SI. (4.2). This can also be understood intuitively, by noting that in a polarized trivial MF state depicted in Fig. (\ref{2Ds}a1), each cell is partitioned into a positive and a negative side. Each partition can be thought of as an ``effective edge'' in the 1D case. Although the lengths of these ``effective edges" are different from those in 1D case, the concentrations are equal. Now, consider a hexagonal lattice; each edge carries the same $|p|,s$, equal to those found in 1D case. For the polarized trivial MF solution, the net cellular polarization equals $p_{\text{cell}}=p_{\text{edge}} (1 - 2 \cos(2\pi/3))=2\sqrt{s_{\text{edge}}^2-4/\alpha\beta}$ , where $p_{\text{edge}}, s_{\text{edge}}$ are junctional polarity and sum of localized proteins, as were defined in the case of a 1D array of cells. In Sec. (\ref{genmodel}), we saw that in ordered systems, minimum concentration of G above which polarized state is stable, i.e. $g_0^*$, increases with increasing $\alpha\beta$, as well as $\gamma/\kappa$ (see Eq. (\ref{gcrit})). Using the units and parameters introduced in Sec. (\ref{genmodel}), we get $g_0^* \simeq 0.23$, for the SLCI critical point in MF approximation and in ordered systems. For $g_0 > g_0^*$ the polarized steady state, namely the polarized trivial MF solution, can be realized in special cases where the initial distribution of the proteins is not far from the steady state, and a small global cue assists with redistribution of proteins. Therefore, the efficacy of SLCI is strongly dependent on the initial condition.

%Straightforwardly, as one can see in Fig. (\ref{2Ds}b), the number of possible configurations of this type hugely outnumbers the eight trivial solutions. As such, starting from a random initial condition with no global cue, the system will almost surely settle in a nontrivial fixed point, like the one depicted in Fig. (\ref{2DMT-SLI-NLI}b). A natural system in which such solutions appear are mutants induced by loss-of-function of cytoplasmic proteins. In such systems, the partial randomly oriented polarization is achieved through local activation/inhibition between the similar/opposite complexes (see Sec. (\ref{Mut}), for further discussion). 

\subsection{Strictly Local Cytoplasmic Interactions (SLCI)}{\label{2DSLI}}
The regime of SLCI is defined as $\lambda/\ell_0\to 0$, i.e. local cytoplasmic interactions. In Sec. (\ref{genmodel}) we simplified Eq. (\ref{mainRD}) in the SLCI limit. Previously we argued qualitatively, that due to the absence of ``repulsive'' interactions between unlike complexes in SLCI limit, the segregation is not accomplished properly. Moreover, using MF arguments we showed that there exist steady-state solutions where segregation is not enforced by cytoplasmic interactions. As such, starting from a random initial condition with no global cue, a system with SLCI will almost surely settle in a nontrivial fixed point, by only local redistributions of proteins on the same junction, which is possible through SLCI. An example of such configurations is depicted in Fig. (\ref{2DMT-SLI-NLI}b). A natural system in which such solutions appear are mutants, in which mutation is induced by loss-of-function of cytoplasmic proteins. In such systems, the partial randomly oriented polarization is achieved through local activation/inhibition between the similar/opposite complexes (see Sec. (\ref{Mut}), for further discussion). 

Simulations of systems with random initial states and weak global cues testify to the lack of long-range order of polarity in SLCI regime. A generic steady state of such systems, the rose-plot of the angular distribution of dipoles, the time evolution of $\overline{Q}(t)$, $\overline{P}(t)$, their ratio $\mathcal O(t)$, as well as the correlation lengths $\xi(t)$, are plotted in Figs. (\ref{2DMT-SLI-NLI}a1), (\ref{2DMT-SLI-NLI}b1), (\ref{2DMT-SLI-NLI}c1), and (\ref{2DMT-SLI-NLI}c2), respectively. To facilitate the comparison, Figs. (\ref{2DMT-SLI-NLI}c1) and (\ref{2DMT-SLI-NLI}c2) include corresponding quantities of other cases of study, that are discussed in the next two subsections. One important observation is the rapid dynamics of polarization in the case of SLCI, in particular $\overline{P}(t)$, acquiring its steady state value within a time scale of the order of $t \lesssim 5\, \gamma^{-1}$. This is exactly due to the huge basin of attraction of nontrivial fixed points in SLCI limit; any initial state is close to a fixed point, to which it is quickly attracted in the absence of cues. 

\begin{figure}[t]
\center\includegraphics[width=1\linewidth]{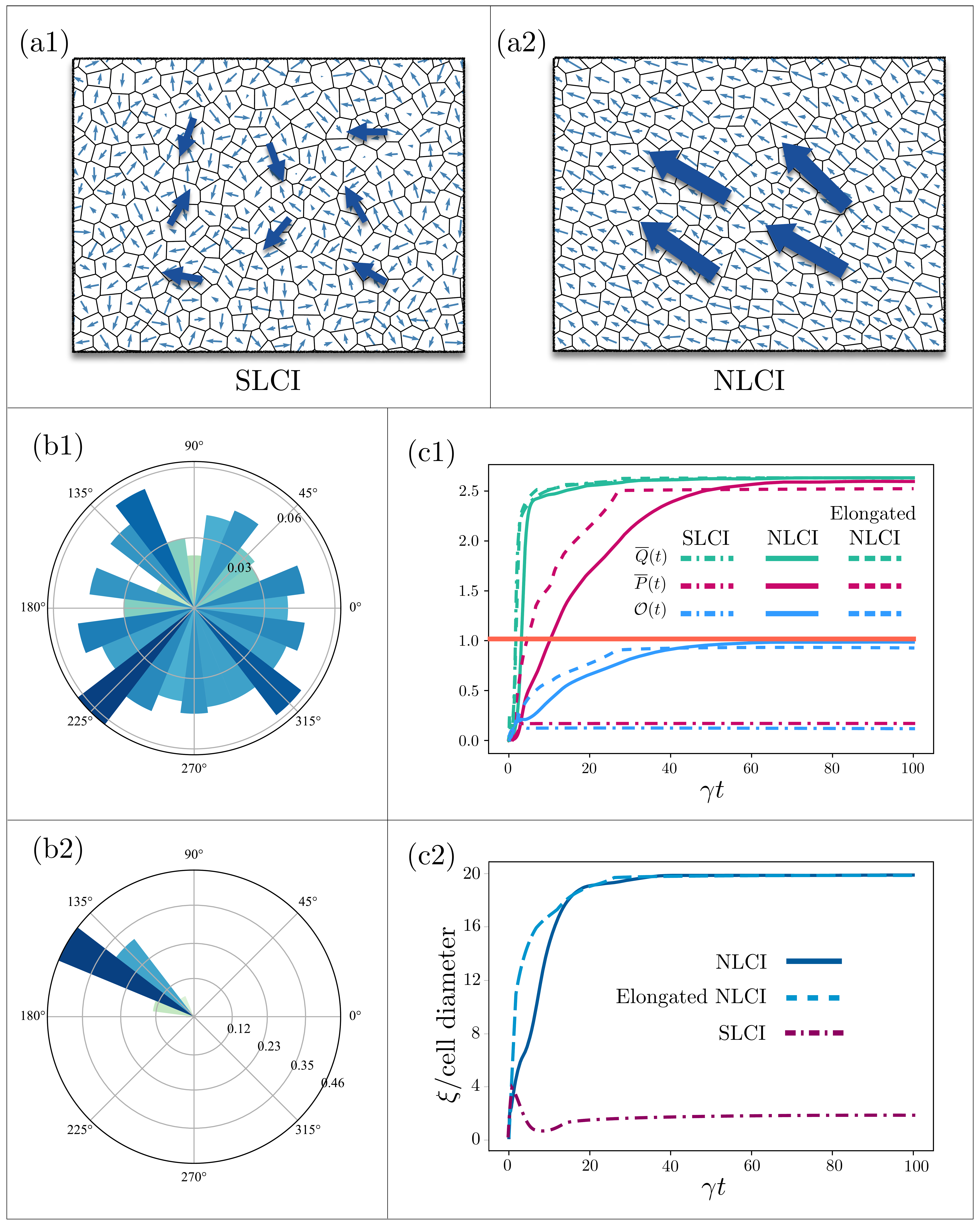}\\
\caption{Generic steady states of two identical systems (i.e. geometry and parameters) and random initial conditions, in (a1) SLCI limit ($\lambda/\ell_0 = 0.01$), and (a2) NLCI of range $\lambda/\ell_0 = 0.5$, length disorder $\epsilon_0 = 0.5$. The big arrows are to clarify the direction of dipoles. The rose-plots of the two cases are shown in (b1) and (b2), indicating the angular distribution of the dipoles. The time evolution of $\overline Q$, $\overline P$, and $\overline O$, for isotropic systems with SLCI and NLCI, as well as elongated systems with NLCI are shown in (c1). The curves corresponding to $\overline Q$ overlap to a large extent, which implies SLCI systems are capable of polarizing individual cells, but fail to align the dipoles on larger scales; also clear from $\overline P$ and $\mathcal O$ in (c1) and $\xi$ in (c2), for SLCI and NLCI. The correlation length in (c2) remains at around 4 cell diameters for SLCI, whereas that of NLCI systems grows until they reach $R_c/2$, i.e. 20 cells. For clarity, the small fluctuations of the curves due to the stochastic noise are removed. }
{\label{2DMT-SLI-NLI}}
\end{figure}

\subsection{Nonlocal Cytoplasmic Interactions (NLCI)}{\label{NLI}}
Nonlocal interactions, as introduced in Sec. (\ref{genmodel}), are mediated by cytoplasmic proteins. Intuitively, NLCI facilitates the segregation of unlike complexes to the opposite sides of cells, by nonlocally activating the like, and inhibiting the unlike complexes formation; one can think of this is effectively ``attracting'' the like, and ``repelling'' the unlike complexes. Segregation makes the system behave more like a one dimensional lattice, by splitting each cell into two compartments. Indeed, the trivial MF solution of type (a1) in Fig. (\ref{2Ds}), becomes increasingly more valid as the range of nonlocal interactions is increased up to an upper limit for $\lambda$ to be found shortly. Segregations implies that adjacent edges of a cell prefer to carry the same polarities, and alternating between inward and outward is not favorable. As such, it is conceivable that configurations like the two trivial MF solutions with zero-net polarization in Fig. (\ref{2Ds}, are destabilize by NLCI. The rest of this section discusses the results of our simulations. 

While in relatively ordered tissues with NLCI, and in the absence of orientational cues, the orientation of polarization is determined purely by chance, namely stochastic noise and initial conditions, highly disordered systems show robustness against such random factors, and the fixed points of polarization fields are determined collectively by the geometry of the lattice. In finite-size systems, the geometrical disorder provides a bias towards one orientation over others. This effect gets progressively more pronounced with increasing range of NLCI and/or level of the quenched disorder. Below, where we discuss topological defects in polarity patters, we provide more evidence for this behavior. External cues of sufficiently large magnitudes, however, reorient the polarity towards the favored direction. See below for further discussion. A typical configuration of the steady states is illustrated in Fig. (\ref{2DMT-SLI-NLI}a2). We find a range of interaction length scale, $0.15\lesssim\lambda/\ell_0\lesssim 0.7$, for which the NLCI guarantees the long-range alignment of polarization, with the standard deviation of the dipoles' directions less than 30 (degrees). The directional correlation shows a peak at around $\lambda/\ell_0 \simeq 0.4-0.5$. The aforementioned range of $\lambda$ also depends on the disorder. Geometrical disorder hinders the establishment of polarization, and increases the lower bound of the range. For example, with $\epsilon_0 = 0.6$, the range changes to $0.25\lesssim\lambda/\ell_0\lesssim0.7$. Not surprisingly, the angular correlation at a certain disorder is dependent also slightly lower in tissues with disorder on the geometrical disorder; see SI. Fig. (4). However, we believe that this is, in part, due to the simplification associated with uniform junctional distribution of proteins, in the absence of which the dipoles are have more rotational degree of freedom. Before addressing the functional range of $\lambda$, we note that for $\lambda/\ell_0\gtrsim 0.7$, all edges within a cell strongly couple to each other, hence preventing the segregation and making the system fragile to stochastic noise. Interestingly, this regime becomes relevant in one of the mutants discussed in Sec. (\ref{Mut}).

For the range of interest, the dynamics of $\overline Q$ and $\overline P$, shown in Fig. (\ref{2DMT-SLI-NLI}c1), imply that the spontaneous emergence (i.e. with no global cue) of collective polarization from an initially random distribution, consists of two distinct stages: (i) the segregation of PCP proteins within each cell, and saturation of the amplitude of polarity, accompanied by the formation of polarized local domains, which is followed by (ii) the subsequent coarsening and alignment of the domains across the tissue. The first and second stages are carried out mostly through intracellular and intercellular interactions, respectively. Below, we argue that the former, is indeed the key to long-range polarization. \\

\noindent{\textbf{Segregation mechanisms in SLCI vs. NLCI.}} Through a comparison of the dynamics of SLCI and NLCI in Figs. (\ref{2DMT-SLI-NLI}b1) and (\ref{2DMT-SLI-NLI}b2), the role of cytoplasmic nonlocal interactions in cell-cell interactions becomes evident. While the average polarization $\overline P(t)$ remains negligible in SLCI limit, it saturates to the average magnitude $\overline Q(t)$ in systems with NLCI, which reflects the angular correlation of cell-cell polarities. More elaborately, during the first stage of dynamics, the nonlocal cytoplasmic interactions prepare each and every cell for later intercellular communications. The coarsening and propagation of polarization is then carried out by cell-cell interactions, which increase with the magnitude of cellular dipoles $Q_i(t)$. Here, an important question arises, regarding the interpretation of $\overline Q(t)$: Can we think of $\overline Q(t)$ as a measure of cellular segregation of proteins? A na\"ive guess would be that since $Q_i(t)$ is oblivious of the direction of polarity, it only measures the magnitude of the dipoles per cell, which is a candidate for quantifying segregation. This, however, warrants a careful investigation, as one can see that $\overline Q(t)$ shows arguably similar behavior in both SLCI and NLCI cases. Does this rule out the lack of segregation in SLCI? One possibility is that while the average $\overline Q(t)$ increases rapidly like in the NLCI case, the segregation is not accomplished consistently in all cells, namely some cells are highly segregated while others are not. This is in part due to initial condition. Consider a tissue with randomly distributed proteins on the membranes. The magnitude of a cell's dipole increases due to the localization of some of the free proteins on the membranes. Above the polarization instability, namely for large enough $g_0$, edges with higher initial concentration of a certain protein absorb more free protein of the same kind due to the cooperative interactions. Therefore, the final polarity depends, among other factors, on the initial condition. This is a separate effect from nonlocal interactions, and is built in the nonlinearity of RD equations regardless of the length-scale of nonlocal interactions. In order to directly measure the segregation, we define partial polarities as follows:

\begin{figure}
\center\includegraphics[width=1\linewidth]{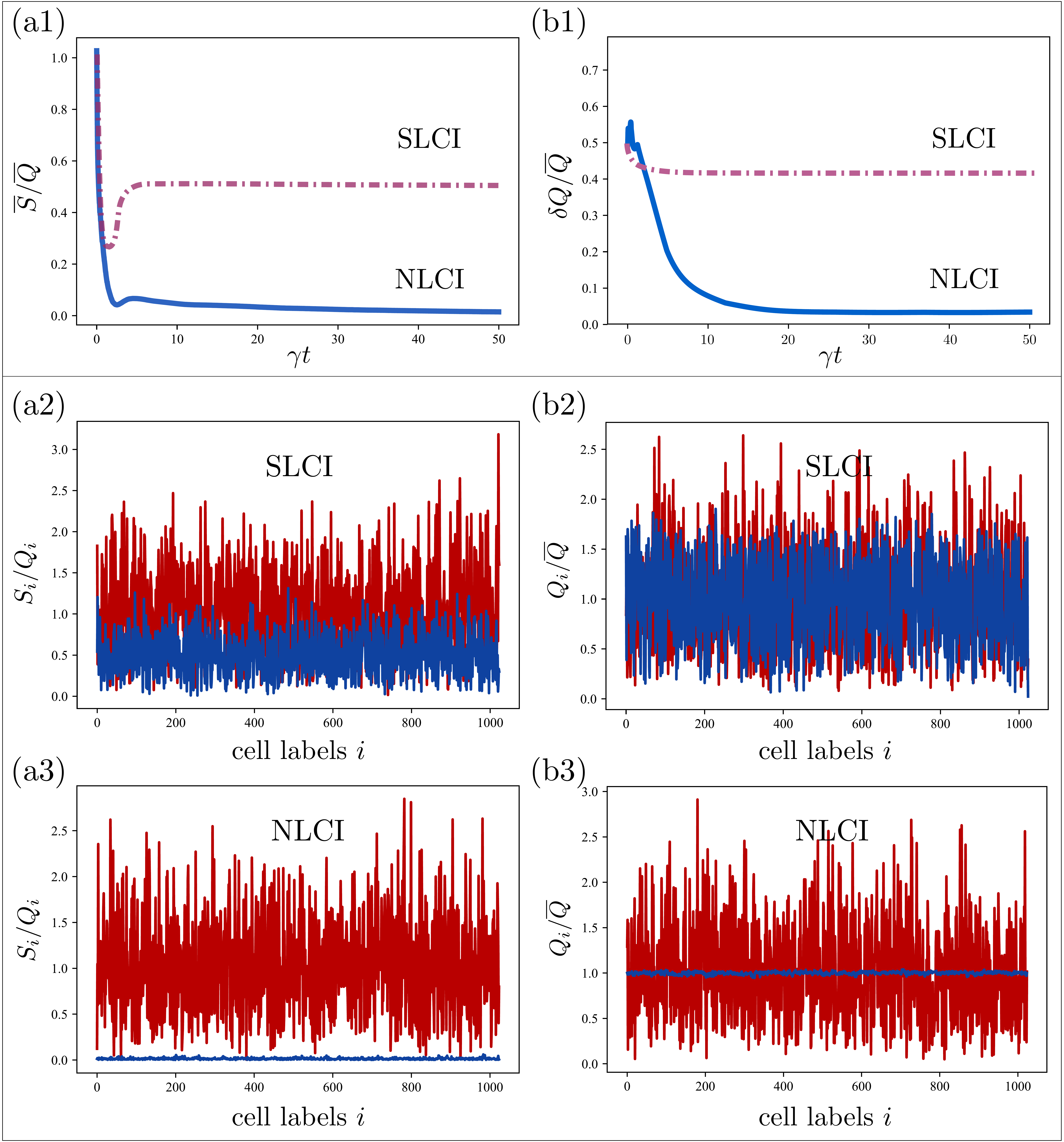}\\
\caption{(a1) The average value of the vector sum of the partial polarities as defined in Eq. (\ref{SumSeg}), divided by the average $Q$, as a function of time for SLCI and NLCI. Evidently, the ratio drops to zero for NLCI, implying full segregation. (b1) The normalized standard deviation of cell polarities defined in Eq. (\ref{delQ}). Zero standard deviation of NLCI case, implies that segregation is achieved in these systems, as opposed to SLCI.}
{\label{segreg}}
\end{figure}

\begin{equation}
\pp_i^{F} = \int_{\varhexagon_i}d\rr \,\frac{\rr-\mathbf R_i}{|\rr-\mathbf R_i|}\;u_i(\rr),
\end{equation}
and similarly $\pp_i^{G}$ is obtained by substituting $v_i(\rr)$ for $u_i(\rr)$. Perfect segregation then corresponds to $\pp_i^{F} = - \pp_i^{G}$ in the steady state, regardless of the initial distributions. The less the segregation strength, the less deviation from the initial condition over the time evolution: $\pp_i^{F} \sim \pp_i^{F}(t=0)$ and $\pp_i^{G} \sim \pp_i^{G}(t=0)$. Therefore, by comparing the final to initial partial polarities, one can easily signify the differences between NLCI and SLCI mechanisms, in terms of the cytoplasmic segregation. In order to quantify the segregation level, and compare that in the two mechanisms of SLCI and NLCI, we introduce the following measures: \\

\noindent (a) the spatial average of the magnitude of: $\mathbf{S}_i = \pp_i^{F} + \pp_i^G$. Note the vector sum, thus for perfect segregation we get: $\mathbf S_i = 0$. (Overlines mean spatial average over all cells at a given time).
\begin{equation}{\label{SumSeg}}
\frac{\overline S(t)}{\overline Q(t)} = \frac{\overline{|\mathbf{S}_i(t)|}}{\overline Q(t)}= \frac{\overline{|\pp_i^F(t) + \pp_i^G(t)|}}{\overline Q(t)}.
\end{equation}
(b) the standard deviation of $Q_i$ normalized by its mean, 
\begin{equation}{\label{delQ}}
\frac{\delta Q(t)}{\overline Q(t)} \equiv \frac{\bigg(\overline {|Q_i(t) - \overline Q(t)|^2}\bigg)^{1/2}}{\overline Q(t)},
\end{equation}
The former characterizes the asymmetry of protein distributions, and the latter measures the consistency in segregation, among the cells. We plot the above quantities as functions of time in Figs. (\ref{segreg}a1) and (\ref{segreg}b1), respectively. In (a1), while the ratio approaches zero in steady state for the system with NLCI, it remains finite $\simeq 0.5$ in the SLCI case, clearly showing the lack of segregation. (b1) We see that in SLCI, the normalized standard deviation drops slightly from $0.5$ at $t=0$ to $0.4$ in steady state, whereas in the NLCI case, it drops to nearly $0$. The latter implies that segregation is fully achieved in all cells, and the individual cell polarities are very much close to the average polarity. In SLCI, as suspected, only the average value of $\overline Q$ grows, whereas cells are not coherently polarized across the tissue. 

In order to show explicitly the cell-by-cell distributions of initial and final values of $S_i$ and $Q_i$, and compare SLCI and NLCI cases, we plotted these quantities in Fig. (\ref{segreg}). In (a2) and (a3), the initial (red) and final (blue) cell-by-cell distributions of $S_i/Q_i$ for SLCI and NLCI, respectively. Again, the near-zero final values of the ratio for NLCI shows perfect segregation. In (b2) and (b3), cellular polarity normalized by spatial average at the corresponding time-point, i.e. initial and final. The width of the distribution shrinks dramatically in NLCI, whereas it remains comparable to its initial value in the SLCI case. \\

%We take one step further to characterize the underlying mechanism for growth of $\overline Q(t)$ in SLCI, i.e. whether or not the initial condition is responsible for the growth of the magnitude of dipoles, without much reorientation. If this is underlying scenario, we expect the magnitude of $|\pp_i^{F} - \pp_i^{G}|$ to grow proportionally with $\pp_i^{F} + \pp_i^{G}$, since the polarity is only due to accumulation of more proteins and not redistributions. The early stage of dynamics is absent in SLCI limit, thus $Q_i(t)$ and the magnitude of the cell-cell interactions remain close to their initial random values. 

\noindent{\textbf{Unequal interaction ranges}} ($\lambda_{\text{uu}}\neq\lambda_{\text{uv}}$). Here, we discuss various cases of unequal $\lambda_{\text{uu}}$ and $\lambda_{\text{uv}}$. In order to thoroughly investigate and distinguish the role of the two interactions, we run simulations on identical lattices, with identical initial distributions, for the following cases: (i) $0.1 \lesssim \lambda_{\text{uu}} = \lambda_{\text{uv}} \lesssim 0.8$, (ii) $\lambda_{\text{uu}} = 0.01$ and $0.1\lesssim\lambda_{\text{uv}} \lesssim 0.8$, and (iii) $0.1\lesssim \lambda_{\text{uu}}\lesssim 0.8$ and $\lambda_{\text{uv}} = 0.01$. In summary, while the length-scale of nonlocal interactions vary from $\simeq 0.1$ to $\simeq 0.8$, local interactions are modeled by $\lambda = 0.01$. The magnitude of geometric disorder takes the values, $\epsilon_0 = 0 , 0.2 , 0.45 , 0.6$. Borrowing the abbreviation of the celebrated local-activation--non-local-inhibition mechanism, i.e. LA-NLI, corresponding to (ii), the three regimes can be labeled as, (i) NLA-NLI, (ii) LA-NLI, and (iii) NLA-LI. In SI. (4.4) these three regimes are discussed in more detail, and the results of several simulations are illustrated in SI. Fig. (4), supporting the following conclusions. The most important findings are as follows: (1) First and foremost, as expected, regime (iii) NLA-LI is not able to stabilize long-range polarity. A localized complex, practically does the opposite of what is required for segregation, by activating similar complexes nonlocally (i.e. on other edges), and not repelling unlike complexes to the opposite side, but only inhibiting them in a small vicinity of itself. Therefore, we mostly focus on the first two regimes, namely NLA-NLI and LA-NLI: (2) While for $\epsilon_0\lesssim 0.5$, the angular correlations of polarization fields are comparable in NLA-NLI and LA-NLI, for $\epsilon_0 > 0.5$, the angular correlation arising from NLA-NLI is arguably better than LA-NLI's, suggesting the importance of nonlocal activation in highly disordered tissues. (3) For small geometrical disorder, the range of $\lambda_{\text{uv}}$ for which the polarization is established, extends from the above to $\lambda_{\text{uv}}\simeq 0.85$ in LA-NLI case, whereas in NLA-NLI $\lambda_{\text{uu}} = \lambda_{\text{uv}} \simeq 0.7$. (4) For small $\epsilon_0 \simeq 0.4$, where both LA-NLI and NLA-NLI mechanisms work equally well, NLA-NLI manages to establish the polarity faster, by up to a factor of two. This discrepancy grows with geometrical disorder; of course for $\epsilon_0 \gtrsim 0.6$, LA-NLI fails to polarize the tissue, yet we observe the patterns reach the steady state more slowly compared to the NLA-NLI case. \\

\begin{figure}[h]
\center\includegraphics[width=1\linewidth]{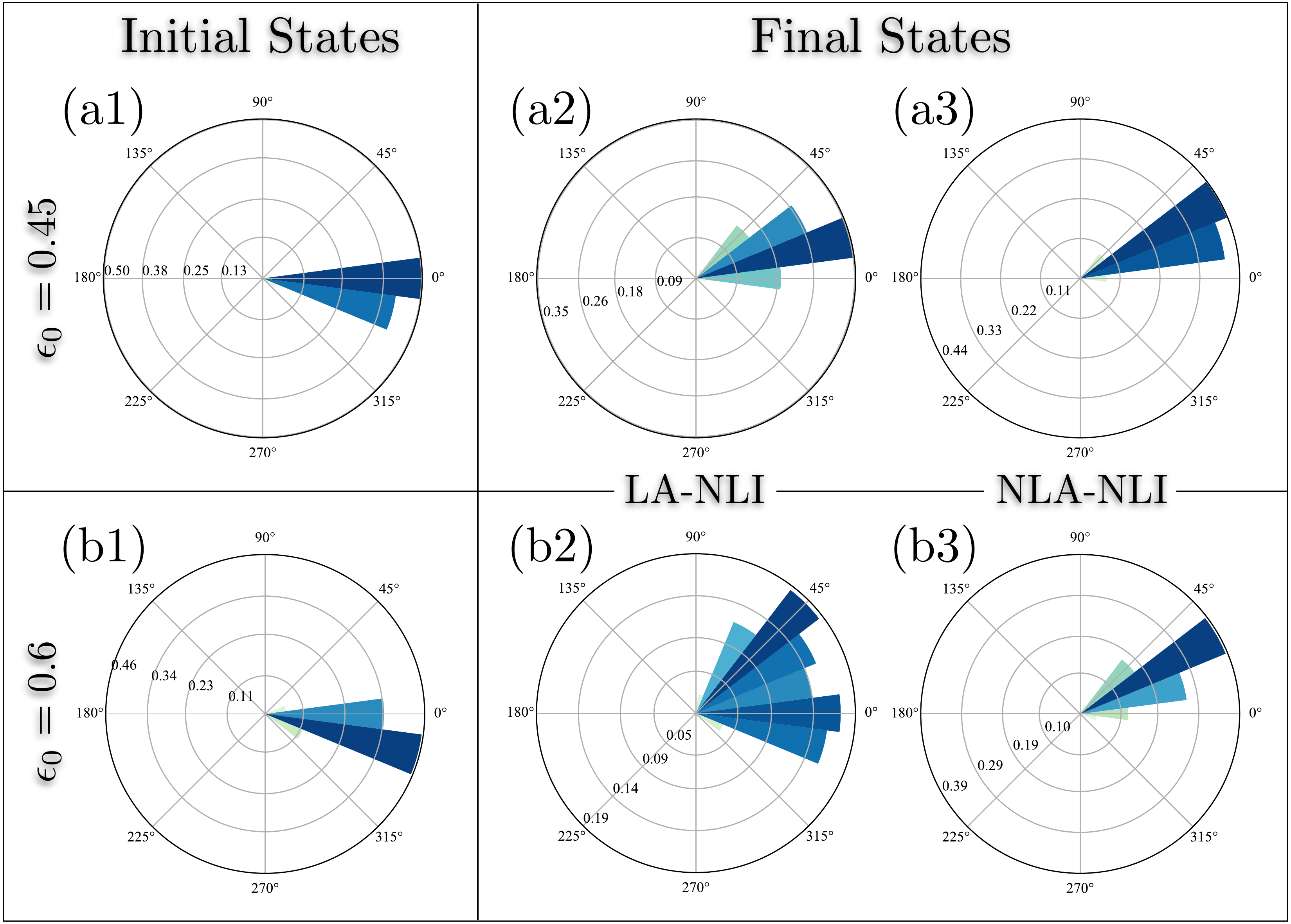}\\
\caption{The results of stability analysis for systems with LA-NLI and NLA-NLI with identical geometries and initial conditions. The magnitude of stochastic noise in both cases equals $\eta_0 = 0.1$. (a1) shows the polarized initial distribution of dipoles in a tissue with geometrical disorder $\epsilon_0 = 0.45$. (a2) and (a3) show the final distributions of the systems corresponding to LA-NLI and NLA-NLI. The same quantities are shown in the bottom panels for $\epsilon_0 = 0.6$. By comparing the final distributions of LA-NLI with those of NLA-NLI, we realize that LA-NLI is insufficient to establish polarity in highly disordered tissues.}
{\label{stabcheck}}
\end{figure}

\noindent{\textbf{Stability analysis.}} We performed numerical stability analysis on the above cases. Starting from a state of aligned dipoles, for identical geometries and stochastic noise, we find that unlike NLA-NLI systems, in highly disordered tissues, long-range correlation is destroyed in LA-NLI systems for large enough stochastic noise. Figures (\ref{stabcheck}) demonstrate the results for $\epsilon_0 = 0.45$ and $\epsilon_0 = 0.6$. In each case the initial condition is the same for both LA-NLI and NLA-NLI cases Figs. (\ref{stabcheck}a1) and (\ref{stabcheck}b1). For NLA-NLI and LA-NLI, we use $\lambda_{\text{uu}} = \lambda_{\text{uv}} = 0.5\ell_0$, and $\lambda_{\text{uu}}=0.1\ell_0$, $\lambda_{\text{uv}} = 0.5\ell_0$, respectively. The final distributions are shown in Figs. (\ref{stabcheck}a2), (\ref{stabcheck}a3), and (\ref{stabcheck}b2), (\ref{stabcheck}b3) for $\epsilon_0 = 0.45$ and $\epsilon_0 = 0.6$, respectively. By comparing the final distributions we realize that while for $\epsilon_0 = 0.45$, the final distributions of LA-NLI and NLA-NLI remain relatively narrow around the mean values, in the case of large geometrical disorder $\epsilon_0 = 0.6$, LA-NLI fails to establish correlated polarity. Note that even in the case of NLA-NLI, the final polarities are rotated compared to the initial condition, which is due to the disordered geometry determining the final direction of polarity; the stochastic noise is large enough to drive the polarization from its false fixed point, to the actual one dictated by geometry, yet the angular coherence of the polarization is preserved. In agreement with our simulations (not shown here), this rotation is absent in (nearly) ordered tissues. Therefore NLA-NLI seems to be necessary for long-range polarization to be stabilized in highly disordered geometries.\\ 

\noindent{\textbf{Directional cues.}} We consider two types of cues; bulk and boundary signals, each of which may be persistent or transient. Bulk cues couple to the F complex across the entire tissue, whereas boundary cues couple only at the boundaries. For bulk cues in, say $+x$ direction, we use the gradient cues of constant slope in each cell: $\mathcal M_{ij}(\rr) = \mathcal M_0\,(x_{ij} - X_i)$; where $\mathcal M_0$ is the slope of the gradient, $x_{ij}$ is the $x$-coordinate of the points on junction $(ij)$, and $X_i$ is that of the centroid of cell $i$. We simulated the response of the polarization field and observed that NLCI significantly enhances the sensitivity of the polarization field to such global cues. Before proceeding, we shall mention that there exist two time scales in this analysis: the response time scale of polarization field $\tau_{\text{res}}$, and the persistence time scale of the cue $\tau_{0}$. The results in a nutshell, are as follows. (1) Reorientation of dipoles over a certain $\tau_{\text{res}}$ and $\tau_{0}$, requires weaker cues in systems with NLCI than those with SLCI. For persistent cues ($\gamma\tau_0 \gtrsim 100$), NLCI responds to signals as small as $\mathcal M_0 \simeq 0.05$ over $\gamma\tau_{\text{res}} \simeq 2$, whereas SLCI requires at least $\mathcal M_0 \simeq 0.5$. The minimum $\mathcal M_0$ increases for smaller $\tau_0$'s. For example, in NLCI with $\mathcal M_0\simeq 0.05$ a nearly persistent signal is required, whereas for larger $\mathcal M_0 \gtrsim 1$, even a rapid transient signal $\gamma \tau_0\simeq 1$ is sufficient to rotate the dipoles over the same time scale. (2) In accord with (1), and due to the small correlation length in SLCI systems, the detection of a cue in these systems happens over exceedingly larger time scales, compared to NLCI with the same magnitude. (3) In the case of boundary cues (i.e. a column of polarized cells), in nearly ordered ($\epsilon_0\lesssim 0.2$) tissues with zero stochastic noise, SLCI suffice to detect the signal. Presence of geometrical disorder and/or stochastic noise, however, necessitates NLCI for the dipoles to align with the cue. Given that the onset of PCP alignment precede the geometrical ordering of the tissue \cite{classen2005hexagonal}, NLCI seems to be the key to the detection of directional cues. Finally, an interesting observation is that (4) NLCI systems appear to detect sufficiently large {\textit{initial}} boundary signals. Initial boundary signal is implemented by polarizing a column (or row) of cells, with significantly larger asymmetry compared to the bulk cells. This implies that a temporary boundary signal would in principle be able to rotate the dipoles, should the cytoplasmic interactions be nonlocal. \\

\noindent {\textbf{Longitudinal vs. transverse correlations.}} A few remarks are in order regarding correlations. As discussed above, correlation function and length as introduced in Eqs. (\ref{corrfun}) and (\ref{corrlen}), are dependent on the relative angle of the reference dipole and the connecting position vector, and at least in our case is stronger in the longitudinal compared to transverse directions. Intuitively, the polarity of a given cell points towards the edges with higher concentrations of localized proteins. These edges are shared with neighbors that are located rather longitudinally with respect to the axis of polarity. The polarities of these neighbors too, are influenced by the shared edges. Therefore longitudinal correlations are stronger than lateral correlations. This discrepancy leads to formation of (transient) vortex-like structures. The correlation lengths shown in Fig. (\ref{2DMT-SLI-NLI}) do not take into account this effect, and are angular averages of the correlation length. \\

%Starting with randomized or uniform initial distributions, the angular dependence is more pronounced in early stages.

\noindent {\textbf{Vortices and saddles.}} Several theoretical {\cite{amonlirdviman2005mathematical,burak2009order,abley2013intracellular}} and experimental studies \cite{cetera2017planar,chen2008asymmetric,ma2008cell} have observed swirls and saddles as different forms of the so-called ``topological defects''. Such defects appear either as steady or transient patterns. Steady defects can be an indication of mutations of various origins; geometry \cite{ma2008cell}, or genetics \cite{ma2003fidelity,struhl2012dissecting,fisher2017integrating,amonlirdviman2005mathematical,strutt2007differential,warrington2017dual,wu2008frizzled,struhl2012dissecting}. In {\textit{Drosophila}} wing, where Fz and Vang localize distally and proximally, respectively, the coupling to global cues is believed to be dependent on the existent of Fat \cite{ma2003fidelity, ma2008cell}. While small \textit{fat}$^-$ clones ($\lesssim$ 10 cell diameters), exhibit little deviation from wild-type polarization, larger clones show swirls, where the polarity is aligned over clusters of (roughly) 10 cells; also implying that Fz feedback loops are left intact in \textit{fat}$^-$ patches. Therefore, the propagation of polarization across neighboring cells is carried out through Fz feedback loop, and the global alignment is achieved through coupling to the cues. Our model predicts both transient and steady swirls, depending on the sector of the parameter space wherein the model parameters lie. Generally speaking, long-lived (steady) defects show up in parts of parameter space that are in between a polarized and an unpolarized sector. We observed two distinct types of steady defects: (a) As $\lambda$ is increased from SLCI to NLCI regime, there exists a narrow range $0.05\lesssim\lambda\lesssim 0.15$ (for $\kappa = 1$), over which vortices and saddles appear as long-lived structures; Fig. ({\ref{defects}a}). (b) Another situation that shows qualitatively similar behavior is for $g_0\lesssim g_0^*$, i.e. under-expression of one of the membrane-bound proteins, that is interpreted as a global mutation within the context of our model; Fig. ({\ref{defects}b}). Such patterns are indeed observed in Vang mutants \cite{cetera2017planar}. Local mutations of various kinds are fully discussed in Sec. (\ref{Mut}), and SI. Sec. (6). An interesting observation regarding the second type is that for a specific disordered tissue, upon cranking up $g_0$, from $\simeq 0.2$ to $\simeq 0.3$, the characteristics of the steady-state pattern of polarity remains very much the same, except the magnitude of polarity is increases. The swirls and branches gradually merge and align for $g_0\gtrsim 0.3$, and long-range polarity is stabilized. The range of $g_0$ over which similar patterns are stable depends on the level of disorder. The more disordered the tissue, the wider the range and the more stable the patterns. The above behavior is independent of the initial condition and stochastic noise, implying that  the polarity pattern is fixed by the the microscopic geometry of the tissue when in regimes where the correlation lengths are of the order of a few cell diameters. Finally, we would like to make a remark on the stability of patterns. In tissues with small disorder, ``steady'' defects might eventually disappear over very long timescales, and for sufficiently large stochastic noise.  
In general, as briefly mentioned above as well, for NLCI, the geometrical information is read by nonlocal interactions which locally biases the dipoles. As $g_0$ increases, the correlation length increases and the global direction is chosen collectively. Nonetheless, the direction is determined by the bias provided by geometry, in the presence of large geometrical disorder and/or small stochastic noise. 

\begin{figure}[h]
\center\includegraphics[width=1\linewidth]{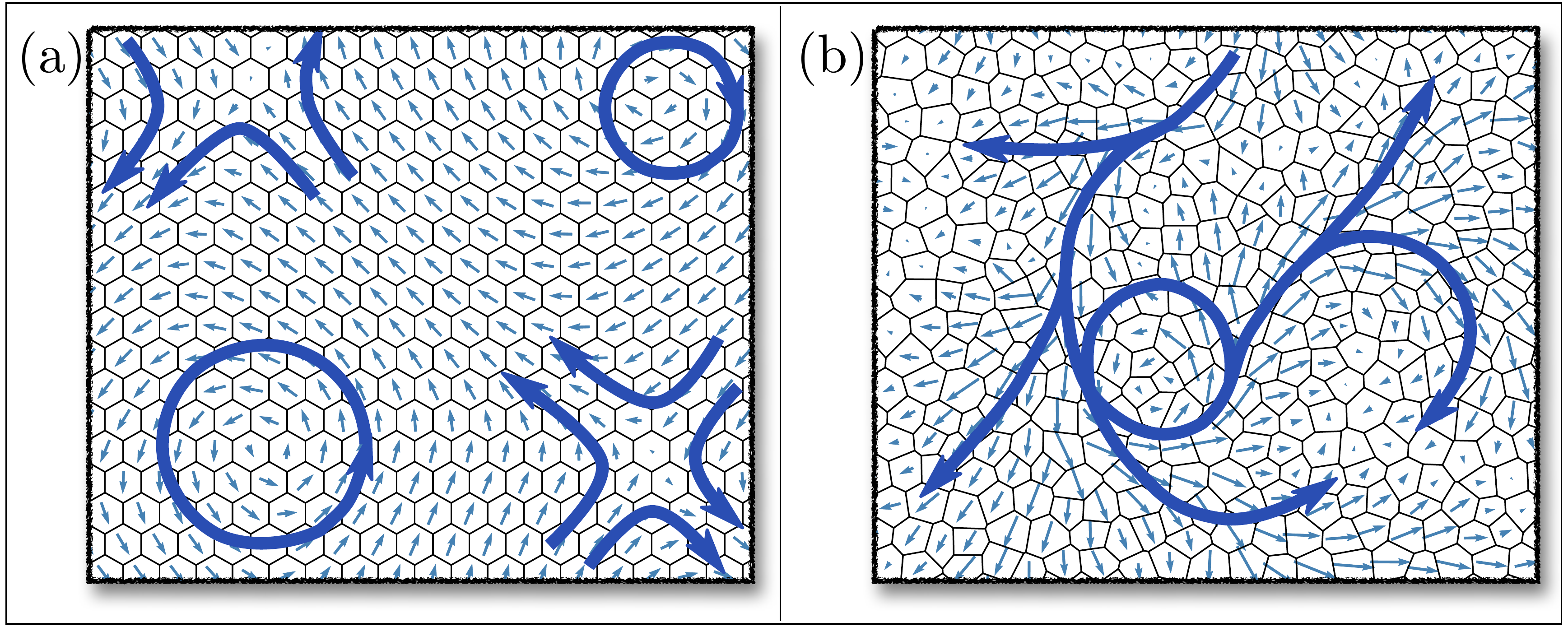}\\
\caption{Two examples of steady defects. (a) shows a system with $\lambda/\ell_0 = 0.1$. Other parameters are fixed at the values mentioned in the text, including $g_0 = 1$. (b) a system with $g_0 = 0.25$; again same parameters as above and $\lambda/\ell_0 = 0.5$. We chose one from ordered systems and the other from disordered. However in both cases, defects appear in ordered as well as disordered systems. }
{\label{defects}}
\end{figure}

%The correlation length in these region is of the order of a few cell diameters.

%For instance, while at $\kappa = 1$, NLCI with $\lambda = 0.15$ is sufficient for correlated polarization, upon increasing $\kappa$ to 10 swirls appear that seem to survive at least within the simulation timescales, which extend beyond the biologically relevant timescales. Another situation where swirls appear as steady patterns, is in the vicinity of the critical point. A weak global cue, however, would remove the swirls. {\textbf{FIGS.}}

%Finite-size tissues with disordered geometry are in principle biased in a preferred directions. The response of the global polarization to geometrical bias is larger near the critical point, and a small bias can effectively act a global cue; hence the stable polarized fixed point is reached, and swirls disappear quickly. Fixing the model parameter, as the geometrical disorder is decreased, the lifetime of the swirls increases slightly. 

%extends the lower bound of acceptable $\lambda$ (for which aligned polarization is achieved), from $\simeq 0.25$ to $\simeq 0.1$

\section{The Effect of Tissue Elongation on Polarization}{\label{elongeom}}
Elongation is suggested to be acting as a global cue in some systems, e.g. mammalian cochlea and mice medial-lateral skin \cite{aw2016transient}. Furthermore, it has been shown in the same study that the perpendicular polarization is not due to a na\"ive incorporation of length in the definition of polarization, but that the short junctions are indeed depleted of proteins. Here we show that NLCI, through increasing the strength of the cooperative self-interactions $\alpha_s,\beta_s$, enhance the stability of F-G complexes on longer junctions. Intuitively, unbound proteins receive, on average, stronger attractive and repulsive signals from complexes localized on longer junctions. This effect results in larger $\alpha_s,\beta_s$, thus enhanced junctional polarity. In SI. Fig. (5b), we plot the dependency of self-interactions on the length of junctions, for different $\lambda$'s.

The elongation of a cell, is characterized by a traceless and symmetric nematic tensor, with the diagonal and off-diagonal elements, $\pm\varepsilon_{i,1}$ and $\varepsilon_{i,2}$, respectively. The index of tissue elongation reads $\mathcal E = N_c^{-1}\sum_{i=1}^{N_c}(\ve_{i,1}^2 + \ve_{i,2}^2)^{1/2}$; see SI. (5). In ordered tissues, there are two possible choices for elongation axis. Elongation parallel to a pair of parallel edges reduces the sixfold symmetry to a twofold associated to long junctions, and a fourfold, Fig. (\ref{elonMT}a1); and vice versa if the elongation is perpendicular to a pair of edges; Figs. (\ref{elonMT}a2,\ref{elonMT}a3). While (a1) and (a2) are polarized perpendicularly to the axis of elongation, (a3) exhibits parallel polarization. The latter is destabilized by NLCI inhibiting localization of unlike complexes on adjacent junctions. In geometrically disordered tissues, elongation is a mixture of (a1) and (a2), both of which give rise to perpendicular polarization.

\begin{figure}[h]
\center\includegraphics[width=1\linewidth]{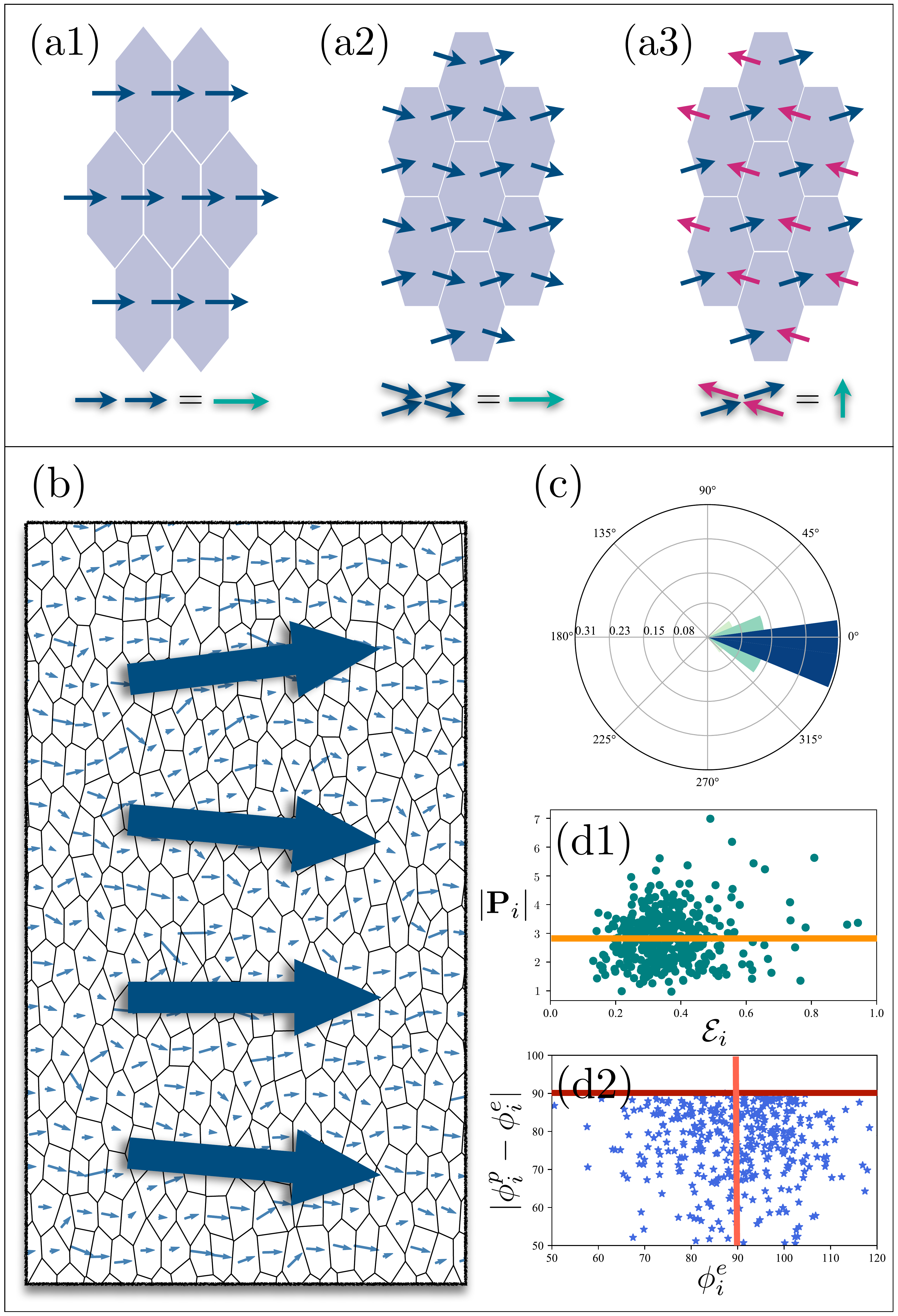}\\
\caption{In (a1) the axis of elongation passes through a vertex, and the edges parallel to elongation wins the polarization competition; twofold symmetry like in 1D. (a2) and (a3) correspond to elongation perpendicular to an edge, therefore the two pairs of elongated edges compete: (a2) represents a polarized state, whereas (a3) is polarized parallel to the elongation axis, but is precluded by nonlocal interactions. (b) shows the final state of polarization in an elongated tissue along the horizontal axis, with $\mathcal E = 0.4$. (c) the rose plot of (b). The cell-by-cell magnitudes of dipoles $|\mathbf{P}_i|$ vs. those of their elongations $\mathcal E_i$, are shown in scatter plot (d1). The orange line in shows the mean magnitude of cell polarities. In (d2), the relative angles between cellular polarities and elongations $|\phi_i^p - \phi_i^e|$, are plotted against the angles of cellular elongation measured from $x$-axis, i.e. $\phi_i^e$, with the average of 90 degrees, marked by vertical light red line. The relative angles $|\phi_i^p - \phi_i^e|$ are naturally smaller than 90 degrees (dark red line).}
{\label{elonMT}}
\end{figure}

Elongated systems involve three different length scales. With $L$ the length of long junctions, we have: (i) $\lambda\lesssim\ell_0< L$, (ii) $\ell_0\lesssim\lambda< L$ and (iii) $\ell_0< L<\lambda$. For (i) and (ii), NLCI leads to the separation of positively and negatively polarized edges. The third regime $\lambda\gtrsim \ell_0 , L$, however, is unstable, like the regime $\lambda\simeq \ell_0$ discussed in isotropic case. Time evolution of $\overline Q(t)$, $\overline P(t)$, $\mathcal O(t)$ and $\xi(t)$ are shown in Fig. (\ref{2DMT-SLI-NLI}c1) and (\ref{2DMT-SLI-NLI}c2). Figure (\ref{elonMT}b) illustrates the steady-state of the polarization field in an elongated tissue with $\mathcal E = 0.4$. In order to investigate the detection of this global cue, we ran simulations on identical tissues but with different values of elongation $\mathcal E = $ 0 -- 0.5; see SI. Fig. (5c). We find that the perpendicular polarization appears at $\mathcal E^* \simeq 0.1$, which is in a very good agreement with that in \cite{aw2016transient}. In order to see (a) whether the observed polarity is a collective effect or is due to single-cell geometry, and (b) that polarity is not a trivial geometrical effect, we make the scatter plots of magnitude and angle of the polarities vs. the magnitude and angle of their elongation for individual cells; see Figs. (\ref{elonMT}d1) and (\ref{elonMT}d2). The infinitesimal correlations between the two indicate that the polarization vector is not a local but a collective effect. 

\section{Local Mutations and The Associated Phenotypes}{\label{Mut}}
PCP mutants exhibit lack of orientational order, which is induced autonomously and/or non-autonomously by mutant clones \cite{adler1997tissue, adler2000domineering, goodrich2011principles, wang2007tissue, chang2013responses}. The phenotypes resulting from loss- and gain-of-function of various components are commonly used to specify the roles of the corresponding proteins. Here we introduce three distinct classes of mutations within the context of our model: Absence of (I) one or both of the membrane proteins, (II) cytoplasmic proteins, in clones embedded in wild type (WT) or mutant backgrounds; and (III) enhanced geometrical irregularity in a patch of cells. 

Since the polarized junctional complexes are assumed to be Fz:Fmi-Fmi:Vang, Fmi-associated mutations cannot be tested separately; in our model $\textit{fmi}^{-}$ translates into double mutants $\textit{fz}^{-}\textit{Vang}^{-}$. While some studies have reported distinct phenotypes in $\textit{Vang}^-\textit{fz}^-$ and $\textit{fmi}^-$ (e.g. \cite{wu2008frizzled,struhl2012dissecting}), others such as \cite{strutt2007differential, chen2008asymmetric} have seen great resemblance between the two, namely minimal non-autonomy. This is considered as a piece of evidence in favor of bidirectional signaling hypothesis \cite{fisher2017information}, and lends more support to our simplifying assumption of similar roles of Fz and Vang in the primary complexes. 

As discussed in Sec. (\ref{genmodel}), the roles of Dsh and Pk are captured effectively by the magnitude of cooperative interactions $\alpha,\beta$, as well as the length-scale of the cytoplasmic interactions $\lambda$. The dependencies of these parameters on Dsh and Pk are complicated and depend on the concentrations as well as the feedback loops between the two components. In order to uncover the major roles of Dsh and Pk in core PCP pathway, we compare the \textit{in silico} phenotypes induced by $\alpha , \beta \to 0$ and $\lambda \to 0$, to the \textit{in vivo} phenotypes $\textit{dsh}^-$ and $\textit{pk}^-$. Experimental observations reveal minimal non-autonomy of $\textit{dsh}^-$ and $\textit{pk}^-$ clones in WT backgrounds \cite{amonlirdviman2005mathematical}. However, the autonomous effects of the two are discernible: while $\textit{dsh}^-$ cells remain nearly unpolarized, $\textit{pk}^-$ clones seem to be almost perfectly polarized parallel to the WT background \cite{amonlirdviman2005mathematical}. Comparison with our results (see SI. Fig. (6)), suggests that the mutants generated by diminished cooperative interactions look very similar to $\textit{dsh}^-$ clones, with no cellular polarization, and almost zero non-autonomy. The $\lambda$-mutants, on the other hand, resemble, to some extent, the $\textit{pk}^-$ clones, though with imperfect angular coherence within the clone (a realization of non-trivial MF solutions depicted in Fig. (\ref{2Ds}b)); indeed they look more like $\textit{dsh}^-\textit{pk}^-$ double mutants. As such, we predict that while both Dsh and Pk contribute to the magnitude and the \textit{length scale} of nonlocal cytoplasmic interactions; Dsh is mainly in charge of the local interactions ($\alpha , \beta$), whereas nonlocal interactions are mediated by both Dsh and Pk. This hypothesis has the following important prediction. We know that the minimum concentration of Vang required for the emergent polarization, increases as $\alpha,\beta$ decrease; Sec. (\ref{genmodel}). Therefore we predict that under-expression of Vang can be---to some extent---compensated for, by over-expression of cytoplasmic proteins, mainly Dsh. This is an elegant manifestation of the collaboration between cytoplasmic and membrane proteins in establishing long-range polarization. Also note that our results suggest that $\lambda$ is not a simple diffusion length of proteins, but is indeed modified as the concentrations of cytoplasmic proteins change, implying nonlinear effects, perhaps due to their interactions that assist with cytosolic diffusion.

Results of simulations and the predicted phenotypes of our model for all mutant classes are shown in SI. (6), and their schematics are tabulated in Fig. (\ref{mutants}). The first and second rows belong to type-I mutants. The left and right panels of the third row corresponds to type II and III, respectively. Comparing the phenotypes of membrane proteins with cytoplasmic ones, we notice two distinctions: (1) Non-autonomous effects of the former are stronger than those of the latter. This is consistent with the fact that cell-cell communications occur through membrane proteins, whereas cytoplasmic proteins are the carriers of intracellular interactions. (2) Both cytoplasmic mutants attract the dipoles, implying the localization of Fz at the WT-mutant boundaries in mutants lacking cytoplasmic proteins.

\begin{figure}[h]
\includegraphics[width=1\linewidth]{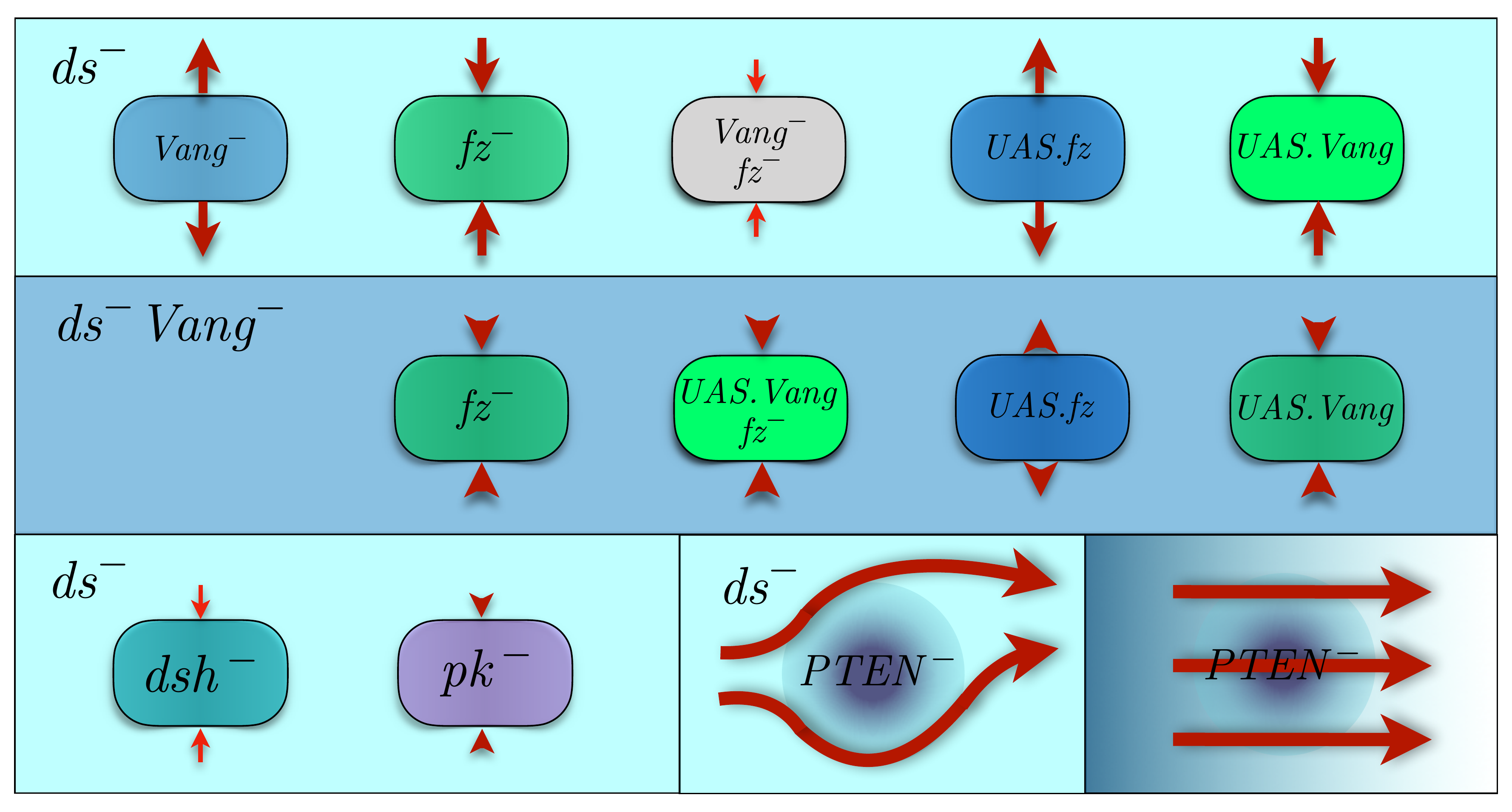}\\
\caption{First and second rows are illustrations of various clones embedded in $\textit{ds}^{-}$ and $\textit{ds}^{-}\textit{Vang}^{-}$ backgrounds, where $\textit{ds}^{-}$ implies the absence of global cue. Since Fz and Vang are treated similarly, the background mutants are only shown for $\textit{Vang}^{-}$. The corresponding phenotypes in $\textit{fz}^-$ background, are obtained by replacing Fz and Vang and flipping the arrows. The colors of the background and clones are chosen as follows: blue and green represent Fz and Vang, respectively. The presence of both Fz and Vang makes the cyan background in the first row. Lack of either one is represented by the complementary color. The left panel of the third row show mutants lacking a cytoplasmic proteins in $\textit{ds}^{-}$ in WT background. The read arrows indicate the direction of the polarity distortion with respect to the wild-type polarity; not the direction of the resulting polarity (see SI. Fig. (6) for further clarification). Thick (thin) \textit{arrows} in the above cases, show large (small) non-autonomous effects extended to multiple cell diameters, whereas the large (small) \textit{arrowheads} show large (small) those limited to 1-2 cells from the clones' boundaries. The right two panels of the third row represent geometrically disordered clones with and without a global cue. The arrows show the resulting polarization fields.}{\label{mutants}}
\end{figure}

In order to test our model's predictions in the case of clones with high geometrical irregularities, studied in Ref. \cite{ma2008cell}, we simulate type III mutants both in the absence and presence of a global cue. In experiments geometrical irregularities are induced by $\textit{PTEN}^{-}$, (marked by darker shades Fig. (\ref{mutants})). Ma, et.al. \cite{ma2008cell}, found that while the alignment of polarization field is preserved in single mutants $\textit{fat}^-$ or $\textit{PTEN}^-$, the angular correlation is disrupted significantly in double mutants $\textit{fat}^-\textit{PTEN}^{-}$, implying that geometrical disorder is an obstacle to the faithful propagation of polarization. In our simulations we used the statistics of cell area in $\textit{PTEN}^{-}$ patches from Ref. \cite{ma2008cell}, and introduced a patch with strong geometrical disorder and shrunken cellular area, that smoothly dissolves into a rather ordered background. In the absence of a global cue, the polarity field shows strong aberrations with swirl-like patterns centered at the mutant patch. Adding the global cue unwinds the swirls and alignment of WT polarization field reappears; see SI. (6). Disrupted polarity in cells with altered geometry can be understood in our model, by noting that NLCI sustains polarity, only within a certain range of $\lambda/\ell_0$. Upon decreasing the cell size, this ratio exceeds the upper bound of the NLCI functional range, and the polarization is destabilized. 

Comparison with experimental observations \cite{struhl2012dissecting,fisher2017integrating,amonlirdviman2005mathematical} (type I), and \cite{strutt2007differential,amonlirdviman2005mathematical,warrington2017dual} (type II), and \cite{ma2008cell} (type III) reveals qualitative similarities between the \textit{in vivo}, and \textit{in silico} phenotypes, suggesting that our model is capable of capturing the salient functions of different PCP components, and their coupling with global cues as well as cell geometry, all of which are concomitants of NLCI. 

\section{Summary and Discussion}{\label{discussion}}
In an attempt to understand the role of cytoplasmic interactions in PCP, and based upon the well-established facts deduced from the experimental studies, we devised a generalized reaction-diffusion model by incorporating nonlocal cytoplasmic interactions. Although we utilized the knowledge on core-PCP to construct the components of our model, no pathway-specific assumptions are made regarding the molecular details and interactions. Thus, we suspect as long as the dominant mechanism of cytoplasmic transport of proteins can be modeled by reaction-diffusion-like dynamics, our model should be able to capture and predict, at least the qualitative behavior of tissue cell polarity. We explored different scenarios of intra- and intercellular interactions, and particularly specified for a generic set of model parameters, the optimal range of cytoplasmic interactions length scales to achieve long-range polarization: $0.2\lesssim \lambda/\ell_0\lesssim 0.7$. Investigating the cases of unequal $\lambda_{\text{uu}}$ and $\lambda_{\text{uv}}$, reveals that in disordered systems with $\epsilon_0 \gtrsim 0.6$, the angular correlation drops significantly compared to that with identical $\lambda$'s. We further examined the response of polarization to external cues, and concluded that NLCI is essential to detecting directional signals in even moderately disordered tissues. 

A direct consequence of NLCI is the readout of cellular geometry. Of particular interest to our study is tissue elongation, as a symmetry-breaking cue. We showed that NLCI is responsible for collective stabilization of polarity perpendicular to the elongation axis. Agreement with the observed value of elongation at which the detectable perpendicular polarization appears, is suggestive of the NLCI as the dominant PCP mechanism in systems like mammalian cochlea and skin. We shall emphasize that this prediction is only valid under the following assumptions: (a) polarization is predominantly induced by reaction-diffusion processes, and (b) lattice dynamics are negligible on the native PCP kinetic timescales. Other mechanisms such as the polarization of microtubules, anisotropic stress, and relative timescales of cell division and PCP relaxation, are also known to influence the direction of polarity \cite{harumoto2010atypical,aigouy2010cell,eaton2011cell}. In order to examine the predictive power of our model, we studied three classes of mutants and found arguably similar phenotypes to the experimental observations, which helps with interpretation of the model parameters and prediction of the roles of Pk and Dsh in core-PCP pathway. \\

\noindent{\textbf{Comparison with other models.}} Although our model is not the first semi-phenomenological approach to the problem of PCP, we believe that the features included in this model, capture a broader range of recently observed phenomena. Some prior studies (e.g. \cite{mani2013collective, fisher2017information}) consider one-dimensional arrays of cells. Two-dimensional systems, however, call for a careful investigation of the cytoplasmic mechanism of segregation. Other successful models such as that studied in Ref. \cite{aigouy2010cell}, infer effective interactions from the observed response of the polarization to a combination of processes; cell elongation, cell rearrangements, and divisions. However the model does not aim at explaining the underlying cellular level mechanisms. Among the models derived from intracellular interactions, the ones put forward by Burak and Shraiman in Ref. \cite{burak2009order}, and Abley et.al. \cite{abley2013intracellular}, are closely related to ours. The former investigates the role of nonlocal inhibitory interactions and demonstrates that LA-NLI is sufficient to fully drive the intracellular segregation of Fz and Vang, in {\textit{ordered}} tissues, and for correlated polarity to emerge in the absence of external cues. However, as the authors point out, the role of nonlocal activation (i.e. NLA-NLI), as well as geometrical disorder remain to be investigated. We tried in this paper to address these questions. Apart from the modeling point of view and interesting aspects of the behavior of polarity, we found it crucial to consider the possibility of nonlocal activation from a phenomenological perspective. Given that the stabilizing and destabilizing interactions between the two complexes are carried by the same set of cytoplasmic proteins, it is a conceivable possibility for the two length scales to be comparable, i.e. $\lambda_{\text{uu}}\simeq\lambda_{\text{uv}}$. Therefore, for identical set of model parameters as well as initial conditions, we examined the behavior of tissue polarity for different values of the two length scales, as the geometrical disorder increases. Interestingly, we observe that the two seemingly unrelated factor that are missing in \cite{burak2009order}, become important in relation with one another. In highly disordered tissues, nonlocal stabilizing interactions are more efficient in the cytoplasmic segregation of PCP proteins; hence NLA-NLI is a reliable mechanism to stabilize the long-range polarity, at least in disordered tissues. Further evidence was also provided through stability analysis of initially polarized disordered tissues, for systems with NLA-NLI versus LA-NLI. 

The second study by Abley et.al. \cite{abley2013intracellular}, includes the two components missing in the former paper, and finds that local polarity alignment is achieved through intracellular partitioning, accompanied by direct cell-cell coupling. However, they find that long-range polarization requires global cues, in the absence of which longitudinal and lateral coordination together give rise to steady-state swirls over domains of size of a few cell diameter. Swirling patterns have been observed repeatedly, and as discussed in Sec. (\ref{SLI-NLI}), are a consequence of unequal longitudinal and lateral correlations in the early stages \cite{burak2009order}. Burak and Shraiman found that, in ordered lattices, large stochastic noise causes swirls in the absence of global cues, that disappear over long timescales, perhaps beyond relevant developmental dynamics. Although we observe steady swirls in parts of our parameter space, they appear only transiently, and long-range polarization is stabilized---even without global cues---in other sectors of parameter space. This is observed in both ordered and disordered systems. As far as the phenomenology of nonlocal activation and inhibition between complexes goes, the model proposed in \cite{abley2013intracellular} appears to be quite similar to ours. Therefore, as also mentioned in Sec. (\ref{SLI-NLI}), we suspect that although their simulations cover a relatively wide range of parameters, the discrepancy in the findings originate from the choice of parameter; e.g. distance from the critical point $g_0/f_0$, or relative formation and dissociation rates of the complexes $\kappa/\gamma$. In summary, we find stable fixed points with long-range spatial correlations in the absence of global cues, though they might require global cues as a drive to settle in the fixed point within the developmental timescales. 

%For instance, we observe that if $\kappa/\gamma$ is increased from 1 to 10, the lower bound of $\lambda$ for which polarization alignment is achieved, increases to $\lambda \simeq 0.25$, whereas steady swirls appear for $0.1\lesssim\lambda\lesssim 0.25$. 

%It is noteworthy that as discussed in Sec. (\ref{SLI-NLI}), geometrical disorder per se, provides an infinitesimal bias; in the absence of global cues and for small stochastic noise, geometrical disorder can determine the final orientation of polarization. Same argument holds true for initial condition. These effects are more pronounced in smaller systems. However, for perfectly uniform initial distribution of proteins, and ordered lattices we observe long-range polarization. 

%Therefore, disappearance of topological defects in disordered tissues is not unexpected, though in \cite{abley2013intracellular}, both ordered and disordered systems are observed to require global cues. 
%Therefore, to the best of our knowledge, models based on reaction-diffusion dynamics, equipped with nonlocal intracellular interactions are capable of producing long-range polarization in the absence of directional cues, should the right set of parameters are chosen.

With regards to the mutant phenotypes, former studies such as \cite{ma2008cell,amonlirdviman2005mathematical,zhu2013damped,abley2013intracellular,le2006establishment}, have proposed mathematical models that successfully reproduce the {\textit{in vivo}} phenotypes. Adopting an effective phenomenological approach to determine the minimal set of criteria essential to the large-scale PCP alignment, we tried to keep the number of model parameters as few as possible, and focus on the major mechanisms at work. The successful recapitulation of the perpendicular axes of polarity and elongation, as well as the \textit{in vivo} phenotypes, speaks to the predictive power of our model in identifying the roles of PCP components. We believe that in spite of adopting the core-PCP as the reference system, the prediction of our model are applicable to pathways other than core-PCP, provided that the polarity is predominantly governed by cytoplasmic RD-type equations; for instance Fat/Dachsous PCP. \\

\noindent{\textbf{Limitations of the model.}}
First and foremost, we shall emphasize again that some parameters ($\alpha,\beta,\kappa,\lambda$) represent effective quantities arising from combinations of molecular networks coupled through feedback/feedforward loops. Therefore, our model is only to account for, and investigate, the primary roles of the molecular components. There exist other quantitative models including pathway-specific details with constants inferred from experiments; e.g. \cite{amonlirdviman2005mathematical,abley2013intracellular,le2006establishment}.

The second point is the assumption of uniform distribution of proteins on each junction. We know from experiments that proteins form clusters at certain loci on the junctions called puncta. However, our approximation makes the simulations faster by orders of magnitude. This assumption, however, has a downside too. Should we allow the proteins to distribute nonuniformly on the junctions, the polarity degrees of freedom vary more smoothly around the cell, and the polarization pattern shows smaller fluctuations compared to what we obtained. Therefore, we suspect that the PCP correlations could be improved, had we solved the full RD equations on the perimeters of all the cells.\\

\noindent{\textbf{Predictions and outlook.}} 
A list of our model's predictions is as follows. (1) The minimum concentration of Vang required for the collective polarization to appear depends inversely on the concentration of Dsh. Therefore, polarity alignment can be retained in tissues with under-expression of Vang, by over-expressing Dsh. This prediction is crucial to understanding the collaboration of membrane and cytoplasmic proteins in transmitting directional information between abutting cells; and can be readily tested by tuning the expression levels of Dsh and Vang. (2) Since NLCI is suggested to be carried by Pk (and in part by Dsh). Thus, the role of elongation as a global cue is dependent on Pk, and the knockdown of \textit{pk} and \textit{dsh} must invalidate the guaranteed orthogonality of polarization and the elongation axis  (3) NLCI enhances the sensitivity to the gradient cues, namely smaller magnitudes and/or shorter time scales of the gradients are required for the dipoles to reorient, in NLCI compared to SLCI systems. This discrepancy becomes more pronounced in tissues with large geometrical disorder. Furthermore, transient global cues are sufficient to reorient the polarity in disordered tissues with NLCI, provided the decaying timescale of the cue ($\tau_0$) is larger than the intrinsic ordering timescale of polarity, i.e. that in the absence of global cue. The latter is estimated to be of the order of $\sim 10$ (hrs) in \textit{Drosophila} wing; see e.g. \cite{tree2002prickle,axelrod2001unipolar}. Transient cues can be provided through temporary expression of a gene using techniques that allow for spatio-temporal control of gene expression, such as heat-shock and light-switchable promoters \cite{adler1997tissue,shimizu2002light}. A natural conclusion of (2) and (3) is that Pk (and Dsh) are essential to the detection of both geometrical and molecular cues. Thus, their absence impairs the readout of all such global cues. (4) The length scale of cytoplasmic interactions are \textit{dependent} on the cytoplasmic protein concentrations, due to possible nonlinear effects on diffusion constants. Interpreting $\lambda$ to be an increasing function of Pk (and possibly Dsh), and given that the polarization is destabilized for $\lambda\gtrsim 0.8$, our model predicts that excess Pk destroys the PCP alignment. Interestingly, this effect was observed in a study by Cho, et.al. in Ref. \cite{cho2015clustering}. Since the NLCI is contingent on the presence of Pk (and Dsh), predictions (2), (3) and (4) can be tested---in the absence of the corresponding global cues---by under- or over-expressions of Pk, as the representative of NLCI. (5) The direction of the polarity in disordered tissues is chosen by the geometry, is independent of initial distribution, and shows robustness against stochastic noise and small external cues. This can be tested experimentally, by comparing the response of already polarized tissues, to global cues in transverse direction, for ordered and disordered geometries, e.g. before and after the ordering in \textit{Drosophila} wing. 

Finally, in spite of several insightful findings regarding the mutual interplay of PCP and tissue mechanics \cite{classen2005hexagonal,ma2008cell,aigouy2010cell}, the relevant molecular and physical mechanisms are yet to be explored. Planar polarity and cell packing are known to mutually influence one another. On the other hand cell packing is highly susceptible to mechanical tension. A natural question would be, how does PCP couple to tension at the molecular level? Furthermore, given the role of microtubules in parallel polarization through biasing the transport of membrane proteins, another important question is, under what conditions does this mechanism dominate the diffusive transport, whereby polarity seems to develop perpendicularly to the axis of elongation? Our study lays the groundwork for further investigations, by uncovering one of the scenarios through which PCP couples to cell geometry.\\

\noindent\textbf{Methods.} Dynamical simulations are carried out using Runge-Kutta method of 4th order, with the time steps of $10^{-3}\,\gamma^{-1}$, on lattices of size $40 \times 40$ cells. For each cell, starting from a randomized distribution of F and G proteins, the concentration of proteins are evolved according to the RD equations. For each value of geometrical disorder, and fixed model parameters, simulations are run for 500 initial conditions. A white and Gaussian stochastic noise is also added to the RD equations. All points on the perimeter of a given cell interact with each other through the kernels introduced in Sec. (\ref{genmodel}). Using the assumption of uniform proteins distribution along all junctions, it suffices to compute the geometrical coefficients ($\mathcal K_{\mu\nu}$) of junction-junction interactions by integrating the kernels along the two junctions, and for all pairs of junctions within a cell. Therefore, the integrals reduce to matrix products. The matrices are calculated for each configuration of disordered lattices. Boundary conditions are chosen to be periodic along both axes. The bulk cues are modeled as constant gradients in each cell. A ``drive" term is added to the RD equation, which for each point on the perimeter of a given cell, is proportional to its distance from the centroid of the cell. The boundary cues are incorporated by polarizing a column of the boundary cells through either a permanent or a transient bias, similar to the bulk cues. The strengths of the biases decay exponentially in time, in both bulk and boundary cues.

\noindent\textbf{Acknowledgement.} S.S. was supported by the Gordon and Betty Moore Foundation through Grant GBMF2919.01. M.M. would like to thank the Simons Foundation MMLS program for support.

\end{small}

\begin{tiny}
\bibliography{pcpref}{}

\begin{thebibliography}{10}

\bibitem{axelrod2001unipolar}
Jeffrey~D Axelrod.
\newblock Unipolar membrane association of dishevelled mediates frizzled planar
  cell polarity signaling.
\newblock {\em Genes \& development}, 15(10):1182--1187, 2001.

\bibitem{strutt2001asymmetric}
David~I Strutt.
\newblock Asymmetric localization of frizzled and the establishment of cell
  polarity in the drosophila wing.
\newblock {\em Molecular cell}, 7(2):367--375, 2001.

\bibitem{gong2004planar}
Ying Gong, Chunhui Mo, and Scott~E Fraser.
\newblock Planar cell polarity signalling controls cell division orientation
  during zebrafish gastrulation.
\newblock {\em Nature}, 430(7000):689--693, 2004.

\bibitem{zallen2007planar}
Jennifer~A Zallen.
\newblock Planar polarity and tissue morphogenesis.
\newblock {\em Cell}, 129(6):1051--1063, 2007.

\bibitem{seifert2007frizzled}
Jessica~RK Seifert and Marek Mlodzik.
\newblock Frizzled/pcp signalling: a conserved mechanism regulating cell
  polarity and directed motility.
\newblock {\em Nature Reviews Genetics}, 8(2):126--138, 2007.

\bibitem{eaton2011cell}
Suzanne Eaton and Frank J{\"u}licher.
\newblock Cell flow and tissue polarity patterns.
\newblock {\em Current opinion in genetics \& development}, 21(6):747--752,
  2011.

\bibitem{goodrich2011principles}
Lisa~V Goodrich and David Strutt.
\newblock Principles of planar polarity in animal development.
\newblock {\em Development}, 138(10):1877--1892, 2011.

\bibitem{gray2011planar}
Ryan~S Gray, Isabelle Roszko, and Lilianna Solnica-Krezel.
\newblock Planar cell polarity: coordinating morphogenetic cell behaviors with
  embryonic polarity.
\newblock {\em Developmental cell}, 21(1):120--133, 2011.

\bibitem{devenport2016tissue}
Danelle Devenport.
\newblock Tissue morphodynamics: translating planar polarity cues into
  polarized cell behaviors.
\newblock In {\em Seminars in cell \& developmental biology}, volume~55, pages
  99--110. Elsevier, 2016.

\bibitem{wang2007tissue}
Yanshu Wang and Jeremy Nathans.
\newblock Tissue/planar cell polarity in vertebrates: new insights and new
  questions.
\newblock {\em Development}, 134(4):647--658, 2007.

\bibitem{peng2012asymmetric}
Ying Peng and Jeffrey~D Axelrod.
\newblock Asymmetric protein localization in planar cell polarity: mechanisms,
  puzzles and challenges.
\newblock {\em Current topics in developmental biology}, 101:33, 2012.

\bibitem{klein2005planar}
Thomas~J Klein and Marek Mlodzik.
\newblock Planar cell polarization: an emerging model points in the right
  direction.
\newblock {\em Annu. Rev. Cell Dev. Biol.}, 21:155--176, 2005.

\bibitem{classen2005hexagonal}
Anne-Kathrin Classen, Kurt~I Anderson, Eric Marois, and Suzanne Eaton.
\newblock Hexagonal packing of drosophila wing epithelial cells by the planar
  cell polarity pathway.
\newblock {\em Developmental cell}, 9(6):805--817, 2005.

\bibitem{aigouy2010cell}
Beno{\^\i}t Aigouy, Reza Farhadifar, Douglas~B Staple, Andreas Sagner,
  Jens-Christian R{\"o}per, Frank J{\"u}licher, and Suzanne Eaton.
\newblock Cell flow reorients the axis of planar polarity in the wing
  epithelium of drosophila.
\newblock {\em Cell}, 142(5):773--786, 2010.

\bibitem{tree2002prickle}
David~RP Tree, Joshua~M Shulman, Rapha{\"e}l Rousset, Matthew~P Scott, David
  Gubb, and Jeffrey~D Axelrod.
\newblock Prickle mediates feedback amplification to generate asymmetric planar
  cell polarity signaling.
\newblock {\em Cell}, 109(3):371--381, 2002.

\bibitem{devenport2014cell}
Danelle Devenport.
\newblock The cell biology of planar cell polarity.
\newblock {\em J Cell Biol}, 207(2):171--179, 2014.

\bibitem{strutt2007differential}
David Strutt and Helen Strutt.
\newblock Differential activities of the core planar polarity proteins during
  drosophila wing patterning.
\newblock {\em Developmental biology}, 302(1):181--194, 2007.

\bibitem{warrington2017dual}
Samantha~J Warrington, Helen Strutt, Katherine~H Fisher, and David Strutt.
\newblock A dual function for prickle in regulating frizzled stability during
  feedback-dependent amplification of planar polarity.
\newblock {\em Current Biology}, 27(18):2784--2797, 2017.

\bibitem{fisher2017information}
Katherine~H Fisher, Alexander~George Fletcher, and David~I Strutt.
\newblock Information flow in planar polarity.
\newblock {\em bioRxiv}, page 236836, 2017.

\bibitem{jenny2003prickle}
Andreas Jenny, Rachel~S Darken, Paul~A Wilson, and Marek Mlodzik.
\newblock Prickle and strabismus form a functional complex to generate a
  correct axis during planar cell polarity signaling.
\newblock {\em The EMBO journal}, 22(17):4409--4420, 2003.

\bibitem{amonlirdviman2005mathematical}
Keith Amonlirdviman, Narmada~A Khare, David~RP Tree, Wei-Shen Chen, Jeffrey~D
  Axelrod, and Claire~J Tomlin.
\newblock Mathematical modeling of planar cell polarity to understand
  domineering nonautonomy.
\newblock {\em Science}, 307(5708):423--426, 2005.

\bibitem{aw2017planar}
Wen~Yih Aw and Danelle Devenport.
\newblock Planar cell polarity: global inputs establishing cellular asymmetry.
\newblock {\em Current opinion in cell biology}, 44:110--116, 2017.

\bibitem{wolpert1969positional}
Lewis Wolpert.
\newblock Positional information and the spatial pattern of cellular
  differentiation.
\newblock {\em Journal of theoretical biology}, 25(1):1--47, 1969.

\bibitem{bayly2011pointing}
Roy Bayly and Jeffrey~D Axelrod.
\newblock Pointing in the right direction: new developments in the field of
  planar cell polarity.
\newblock {\em Nature Reviews Genetics}, 12(6):385--391, 2011.

\bibitem{bosveld2012mechanical}
Floris Bosveld, Isabelle Bonnet, Boris Guirao, Sham Tlili, Zhimin Wang, Ambre
  Petitalot, Rapha{\"e}l Marchand, Pierre-Luc Bardet, Philippe Marcq,
  Fran{\c{c}}ois Graner, et~al.
\newblock Mechanical control of morphogenesis by fat/dachsous/four-jointed
  planar cell polarity pathway.
\newblock {\em Science}, 336(6082):724--727, 2012.

\bibitem{sagner2012establishment}
Andreas Sagner, Matthias Merkel, Benoit Aigouy, Julia Gaebel, Marko
  Brankatschk, Frank J{\"u}licher, and Suzanne Eaton.
\newblock Establishment of global patterns of planar polarity during growth of
  the drosophila wing epithelium.
\newblock {\em Current Biology}, 22(14):1296--1301, 2012.

\bibitem{ma2008cell}
Dali Ma, Keith Amonlirdviman, Robin~L Raffard, Alessandro Abate, Claire~J
  Tomlin, and Jeffrey~D Axelrod.
\newblock Cell packing influences planar cell polarity signaling.
\newblock {\em Proceedings of the National Academy of Sciences},
  105(48):18800--18805, 2008.

\bibitem{julicher2017emergence}
Frank J{\"u}licher and Suzanne Eaton.
\newblock Emergence of tissue shape changes from collective cell behaviours.
\newblock In {\em Seminars in Cell \& Developmental Biology}. Elsevier, 2017.

\bibitem{lopez2004directional}
Hern{\'a}n L{\'o}pez-Schier, Catherine~J Starr, James~A Kappler, Richard
  Kollmar, and AJ~Hudspeth.
\newblock Directional cell migration establishes the axes of planar polarity in
  the posterior lateral-line organ of the zebrafish.
\newblock {\em Developmental cell}, 7(3):401--412, 2004.

\bibitem{blankenship2006multicellular}
J~Todd Blankenship, Stephanie~T Backovic, Justina~SP Sanny, Ori Weitz, and
  Jennifer~A Zallen.
\newblock Multicellular rosette formation links planar cell polarity to tissue
  morphogenesis.
\newblock {\em Developmental cell}, 11(4):459--470, 2006.

\bibitem{shi2014celsr1}
Dongbo Shi, Kouji Komatsu, Mayumi Hirao, Yayoi Toyooka, Hiroshi Koyama, Fadel
  Tissir, Andr{\'e}~M Goffinet, Tadashi Uemura, and Toshihiko Fujimori.
\newblock Celsr1 is required for the generation of polarity at multiple levels
  of the mouse oviduct.
\newblock {\em Development}, 141(23):4558--4568, 2014.

\bibitem{chien2015mechanical}
Yuan-Hung Chien, Ray Keller, Chris Kintner, and David~R Shook.
\newblock Mechanical strain determines the axis of planar polarity in ciliated
  epithelia.
\newblock {\em Current Biology}, 25(21):2774--2784, 2015.

\bibitem{aw2016transient}
Wen~Yih Aw, Bryan~W Heck, Bradley Joyce, and Danelle Devenport.
\newblock Transient tissue-scale deformation coordinates alignment of planar
  cell polarity junctions in the mammalian skin.
\newblock {\em Current Biology}, 26(16):2090--2100, 2016.

\bibitem{bertet2004myosin}
Claire Bertet, Lawrence Sulak, and Thomas Lecuit.
\newblock Myosin-dependent junction remodelling controls planar cell
  intercalation and axis elongation.
\newblock {\em Nature}, 429(6992):667--671, 2004.

\bibitem{shimada2006polarized}
Yuko Shimada, Shigenobu Yonemura, Hiroyuki Ohkura, David Strutt, and Tadashi
  Uemura.
\newblock Polarized transport of frizzled along the planar microtubule arrays
  in drosophila wing epithelium.
\newblock {\em Developmental cell}, 10(2):209--222, 2006.

\bibitem{harumoto2010atypical}
Toshiyuki Harumoto, Masayoshi Ito, Yuko Shimada, Tetsuya~J Kobayashi, Hiroki~R
  Ueda, Bingwei Lu, and Tadashi Uemura.
\newblock Atypical cadherins dachsous and fat control dynamics of
  noncentrosomal microtubules in planar cell polarity.
\newblock {\em Developmental cell}, 19(3):389--401, 2010.

\bibitem{matis2014microtubules}
Maja Matis, David~A Russler-Germain, Qie Hu, Claire~J Tomlin, and Jeffrey~D
  Axelrod.
\newblock Microtubules provide directional information for core pcp function.
\newblock {\em Elife}, 3:e02893, 2014.

\bibitem{burak2009order}
Yoram Burak and Boris~I Shraiman.
\newblock Order and stochastic dynamics in drosophila planar cell polarity.
\newblock {\em PLoS computational biology}, 5(12):e1000628, 2009.

\bibitem{schamberg2010modelling}
Sabine Schamberg, Paul Houston, Nick~AM Monk, and Markus~R Owen.
\newblock Modelling and analysis of planar cell polarity.
\newblock {\em Bulletin of mathematical biology}, 72(3):645--680, 2010.

\bibitem{abley2013intracellular}
Katie Abley, Pierre~Barbier De~Reuille, David Strutt, Andrew Bangham,
  Przemyslaw Prusinkiewicz, Athanasius~FM Mar{\'e}e, Ver{\^o}nica~A Grieneisen,
  and Enrico Coen.
\newblock An intracellular partitioning-based framework for tissue cell
  polarity in plants and animals.
\newblock {\em Development}, 140(10):2061--2074, 2013.

\bibitem{le2006establishment}
Jean-Fran{\c{c}}ois Le~Garrec, Philippe Lopez, and Michel Kerszberg.
\newblock Establishment and maintenance of planar epithelial cell polarity by
  asymmetric cadherin bridges: a computer model.
\newblock {\em Developmental dynamics: an official publication of the American
  Association of Anatomists}, 235(1):235--246, 2006.

\bibitem{mani2013collective}
Madhav Mani, Sidhartha Goyal, Kenneth~D Irvine, and Boris~I Shraiman.
\newblock Collective polarization model for gradient sensing via dachsous-fat
  intercellular signaling.
\newblock {\em Proceedings of the National Academy of Sciences},
  110(51):20420--20425, 2013.

\bibitem{meinhardt2007computational}
Hans Meinhardt.
\newblock Computational modelling of epithelial patterning.
\newblock {\em Current opinion in genetics \& development}, 17(4):272--280,
  2007.

\bibitem{maree2006polarization}
Athanasius~FM Mar{\'e}e, Alexandra Jilkine, Adriana Dawes, Ver{\^o}nica~A
  Grieneisen, and Leah Edelstein-Keshet.
\newblock Polarization and movement of keratocytes: a multiscale modelling
  approach.
\newblock {\em Bulletin of mathematical biology}, 68(5):1169--1211, 2006.

\bibitem{wu2008frizzled}
Jun Wu and Marek Mlodzik.
\newblock The frizzled extracellular domain is a ligand for van gogh/stbm
  during nonautonomous planar cell polarity signaling.
\newblock {\em Developmental cell}, 15(3):462--469, 2008.

\bibitem{struhl2012dissecting}
Gary Struhl, Jos{\'e} Casal, and Peter~A Lawrence.
\newblock Dissecting the molecular bridges that mediate the function of
  frizzled in planar cell polarity.
\newblock {\em Development}, 139(19):3665--3674, 2012.

\bibitem{fisher2017integrating}
Katherine~H Fisher, David Strutt, and Alexander~G Fletcher.
\newblock Integrating planar polarity and tissue mechanics in computational
  models of epithelial morphogenesis.
\newblock {\em Current Opinion in Systems Biology}, 5:41--49, 2017.

\bibitem{cetera2017planar}
Maureen Cetera, Liliya Leybova, Frank~W Woo, Michael Deans, and Danelle
  Devenport.
\newblock Planar cell polarity-dependent and independent functions in the
  emergence of tissue-scale hair follicle patterns.
\newblock {\em Developmental biology}, 428(1):188--203, 2017.

\bibitem{chen2008asymmetric}
Wei-Shen Chen, Dragana Antic, Maja Matis, Catriona~Y Logan, Michael Povelones,
  Graham~A Anderson, Roel Nusse, and Jeffrey~D Axelrod.
\newblock Asymmetric homotypic interactions of the atypical cadherin flamingo
  mediate intercellular polarity signaling.
\newblock {\em Cell}, 133(6):1093--1105, 2008.

\bibitem{ma2003fidelity}
Dali Ma, Chung-hui Yang, Helen McNeill, Michael~A Simon, and Jeffrey~D Axelrod.
\newblock Fidelity in planar cell polarity signalling.
\newblock {\em Nature}, 421(6922):543--547, 2003.

\bibitem{adler1997tissue}
Paul~N Adler, Randi~E Krasnow, and Jingchun Liu.
\newblock Tissue polarity points from cells that have higher frizzled levels
  towards cells that have lower frizzled levels.
\newblock {\em Current Biology}, 7(12):940--949, 1997.

\bibitem{adler2000domineering}
Paul~N Adler, Job Taylor, and Jeannette Charlton.
\newblock The domineering non-autonomy of frizzled and van gogh clones in the
  drosophila wing is a consequence of a disruption in local signaling.
\newblock {\em Mechanisms of development}, 96(2):197--207, 2000.

\bibitem{chang2013responses}
Hao Chang and Jeremy Nathans.
\newblock Responses of hair follicle--associated structures to loss of planar
  cell polarity signaling.
\newblock {\em Proceedings of the National Academy of Sciences},
  110(10):E908--E917, 2013.

\bibitem{zhu2013damped}
Hao Zhu and Markus~R Owen.
\newblock Damped propagation of cell polarization explains distinct pcp
  phenotypes of epithelial patterning.
\newblock {\em Scientific reports}, 3, 2013.

\bibitem{shimizu2002light}
Sae Shimizu-Sato, Enamul Huq, James~M Tepperman, and Peter~H Quail.
\newblock A light-switchable gene promoter system.
\newblock {\em Nature biotechnology}, 20(10):1041, 2002.

\bibitem{cho2015clustering}
Bomsoo Cho, Gandhy Pierre-Louis, Andreas Sagner, Suzanne Eaton, and Jeffrey~D
  Axelrod.
\newblock Clustering and negative feedback by endocytosis in planar cell
  polarity signaling is modulated by ubiquitinylation of prickle.
\newblock {\em PLoS genetics}, 11(5):e1005259, 2015.

\end{thebibliography}
\bibliographystyle{unsrt}
\end{tiny}

%\nextpage
\newpage
\clearpage
%\documentclass[onecolumn]{article}
%\onecolumn
\numberwithin{equation}{section}
\numberwithin{figure}{section}
\appendix
\begin{center}
{\LARGE{Supporting Information}}
\end{center}

\section{The Model and its Ingredients}{\label{model}}
We begin by writing the full reaction-diffusion equations for the binding/unbinding dynamics of the local density of the complexes complexes [Fz:Fmi-Fmi:Vang], namely $u_{ij}(\rr)$. The corresponding equation for [Vang:Fmi-Fmi:Fz] or $v_{ij}(\rr)$, is obtained by simultaneous replacements: Fz:Fmi $\leftrightarrow$ Fmi:Vang and $u\leftrightarrow v$. The primary complexes Fz:Fmi and Fmi:Vang, are denoted by F and G, respectively. Their corresponding concentrations are then $f$ and $g$.
%\begin{widetext}
\begin{align}{\label{mainRD-SI}}
\frac{du_{ij}(\rr)}{dt} &=\kappa f_i^{\text{ubd}}g_j^{\text{ubd}}\bigg(1+\alpha\sum_{\{k\}_i}\int_{i\cap k} d\rr'\mathcal K_{uu}(\rr-\rr')u_{ik}(\rr')\bigg)\nonumber\\
&- \gamma u_{ij}(\rr)\bigg(1+\beta\sum_{\{k\}_i}\int_{i\cap k} d\rr' \mathcal K_{uv}(\rr-\rr')v_{ik}(\rr')\bigg)\nonumber\\
&+ \eta(\rr, t),
\end{align}
%\end{widetext}
where $\gamma^{-1}$ is the timescale associated with complex dissociation. As mentioned in the Main Text, the kernels $\mathcal K(\rr)$ are assumed to be of the form of $\exp(-|\rr|/\lambda)$, which was motivated by the diffusive nature of the cytoplasmic proteins carrying the interactions. Although the kernels couple the concentrations of the complexes on the boundaries of the cells, the coordinate $\rr$ can in principle represent any point within the cytoplasm as well as the junctions. Suppose that a cytoplasmic protein C, obeys a diffusion equation with degradation time $\tau_{\text{c}}$:
\begin{equation}
\frac{\partial c(\rr,t)}{\partial t} = D_{\text{c}}\nabla^2 c(\rr,t) - \tau_{\text{c}}^{-1} c(\rr,t).
\end{equation}
Assuming the degradation and hence dynamics of C takes place on a much faster timescale than that of PCP, i.e. $\gamma\tau_{\text{c}}\ll 1$, it suffices to only consider the steady-state solutions of protein C. With $D_{\text{c}}$, the diffusion constant of the cytoplasmic proteins C, we get the following equations for C:
\begin{subequations}
\begin{equation}
D_{\text{c}}\nabla^2 c(\rr,t) - \tau_{\text{c}}^{-1}c(\rr,t) = 0, %\Longrightarrow \qquad 
\end{equation}
\begin{equation}
c(\rr , t) = \int_{\varhexagon} d\rr' c_0(\rr') \exp(|\rr-\rr'|/\lambda_{\text{c}}).
\end{equation}
\end{subequations}
Here, $\rr=0$ corresponds to the specific point at which the concentration of C is measured by superimposing the concentration of proteins diffused from $\rr'$. Next, $\lambda_{\text{c}} = \sqrt{D_{\text{c}}\tau_{\text{c}}}$ is the diffusion length of protein C. The diffusing proteins enhance the formation of like complexes and suppress that of unlike complexes. This nonlinear effect in turn, depends on the respective interactions and their binding affinities with the target complexes. Altogether, the coefficients are lumped into the phenomenological constants $\alpha$ and $\beta$.\\

For notational convenience, in the following paragraphs Greek letters label edges, e.g. $\mu \equiv i\cap j$. Using Eq. (\ref{mainRD-SI}) and the definitions of $f^{\text{ubd}}_i$ and $g^{\text{ubd}}_j$, we can solve for the dynamics and the steady states of $u_{\mu}(\rr)$. Evidently, solving the above integro-differential equations is a cumbersome task. One possible simplification is junctional averaging: $ u_{\mu} = \int_{\mu}d\rr' u_{\mu}(\rr')/\ell_{\mu}$, in terms of which we have: 
%\begin{widetext}
\begin{align}{\label{Kmunu}}
\frac{du_\mu}{dt} =&\kk f^{\text{ubd}}_ig^{\text{ubd}}_j\bigg(1+\alpha_{\mu}\sum_{\nu \in i}\mathcal K_{\mu\nu}u_{\nu}\bigg)\nonumber\\
&-\gamma u_{\mu}\bigg(1+\beta_{\mu}\sum_{\nu\in i}\mathcal K_{\mu\nu}v_{\nu}\bigg) + \eta_\mu(t).
\end{align}
%\end{widetext}
In the above equation, $\alpha_{\mu}= \alpha/\ell_\mu$, $\beta_{\mu}= \beta/\ell_\mu$, and for the kernels $\mathcal K_{\mu\nu} = \mathcal K_{\nu\mu} =\iint_{\mu,\nu}d\rr d\rr'\mathcal K(\rr-\rr')$. The diagonal elements equal
\begin{equation}
\mathcal K_{\mu\mu} = 2\ell^2_{\mu}x_{\mu}^{-2}\left(e^{-x_{\mu}} + x_{\mu} - 1\right), 
\end{equation}
in which $x_{\mu} = \ell_{\mu}/\lambda$. The effective stochastic noise on a junction $\mu$ reads: $\langle \eta_{\mu}(t)\eta_\mu(t')\rangle = \eta_0^2\ell_\mu\delta(t-t')$. It is noteworthy that, as we see in Eq. (\ref{Kmunu}), the junctional averaging is a very simplifying approximation that reduces the integrations into matrix products. However, in reality the core proteins for example Flamingo and Frizzled in the prepupal and pupal wing of {\textit{Drosophila}}, are observed to be persistently localized at subdomains of plasma membranes, called ``puncta'' \cite{strutt2011dynamics}.

\section{Definitions of Junctional and Cellular Polarity}{\label{pdef}}

Planar cell polarity can be defined at either junctional or cellular level. For individual junctions, polarity is defined as the difference between the concentrations of $ u_{ij}=[$F$_i:\;$G$_j]$ and the opposite dimer, $u_{ji}$, thus $p_{ij} = u_{ij} - u_{ji} = u_{ij} - v{ij}$. 

Cellular polarization, on the other hand, is referred to as the asymmetric distribution of PCP proteins in the cells. Cell polarity, depending on the symmetry of binding complexes, might be either a vector quantity, called vectorial polarity, or a nematic, which is then called axial polarity. Vectorial and axial polarities are used when PCP proteins form heterodimers and homodimers, respectively. The former is a vector identified by a magnitude and angle $\in[0,2\pi)$, whereas the latter is a traceless nematic tensor with a magnitude and an angle $\in[0,\pi)$. In our case of study, the polarization involves two distinct complexes, Fz and Vang, hence vectorial polarity. However, we shall introduce both vectorial and axial cell polarities. \\

\noindent {\textbf{(i) Vectorial Polarization.}} This definition is used to calculate the ``dipole moments'' of the distribution of bound Fz (or Vang), around each cell. The polarization vector associated with cell $i$ is defined as:
\begin{align}{\label{pol}}
\pp_i = \frac{1}{2}\;\int_{\varhexagon_i} d\rr'\;\frac{\rr'-\mathbf R_i}{|\rr'-\mathbf R_i|}\;\big(u_{i}(\rr') - v_i(\rr')\big) = p_i^x \hat x + p_i^y \hat y.
\end{align}
Here $\mathbf R_i$ is a reference point within cell $i$, with respect to which the polarization is defined. We take this to be the geometrical center of mass of each cell. The polarization vector is identified by a magnitude $|\pp_i|$ and an angle $\theta_{i} \in [0, 2\pi)$ measured from the $x$-axis:
\begin{equation}
|\pp_i| = \sqrt{{(p_i^x)}^2 + {(p_i^y)}^2}, \qquad \text{and} \qquad\phi^p_i = \tan^{-1}(p_i^y/p_i^x).
\end{equation}
The polarization of the tissue with $N_c$ cells, and its global order are characterized by the following quantities: (1) average polarization (2) average magnitude of polarization, and (3) the ratio of (1) and (2):
\begin{align}
&(1)\;:\;\overline \pp = \frac{1}{N_c}\sum_{i=1}^{N_c} \pp_i ,\nonumber\\
&(2)\;:\;\overline Q = \frac{1}{N_c}\sum_{i=1}^{N_c}|\pp_i|, \nonumber\\
&(3)\;:\;\mathcal O = |\overline \pp|/\overline Q.
\end{align}

\noindent {\textbf{(ii) Axial Polarization.}} The second definition is a measure for cellular polarity that is used especially when dealing with axial nematic PCP (like Fmi (Celsr) homodimers) which merely determines the axis of polarization by measuring the traceless nematic tensor of the polarity:
\begin{align}{\label{Pmatrix}}
\widehat{\mathtt P}_i &= 
\begin{pmatrix} \mathcal P_{i,1} & \mathcal P_{i,2}\\ \mathcal P_{i,2} & -\mathcal P_{i,1} \end{pmatrix},\nonumber\\
\mathcal P_{i,1} &= \int_0^{2\pi}d\phi \;u_i(\rr)\,\cos(2\phi), \nonumber\\
\mathcal P_{i,2} &= \int_0^{2\pi}d\phi \;u_i(\rr)\,\sin(2\phi). 
\end{align}
In the above equation, $\phi$ is the polar angle of point $\rr$ on the perimeter of cell $i$ and is measured, with respect to a reference axis. The magnitude of polarization equals $\mathcal P_i = (\mathcal P_{i,1}^2 + \mathcal P_{i,2}^2)^{1/2} = |\det {\mathtt P}_i|^{1/2}$. Its orientation is determined by angle $\phi^p_i$ satisfying $\cos(2\phi^p_i) = \mathcal P_{i,1}/\mathcal P_i$ and $\sin(2\phi^p_i) = \mathcal P_{i,2}/\mathcal P_i$. In Sec. (\ref{eloncells}), we use the same general formalism for the cell shape nematic tensor. 

\section{Measure of Quenched Disorder and Defects in 2D Systems}
\begin{figure*}[ht]
\center
\includegraphics[width=0.8\linewidth]{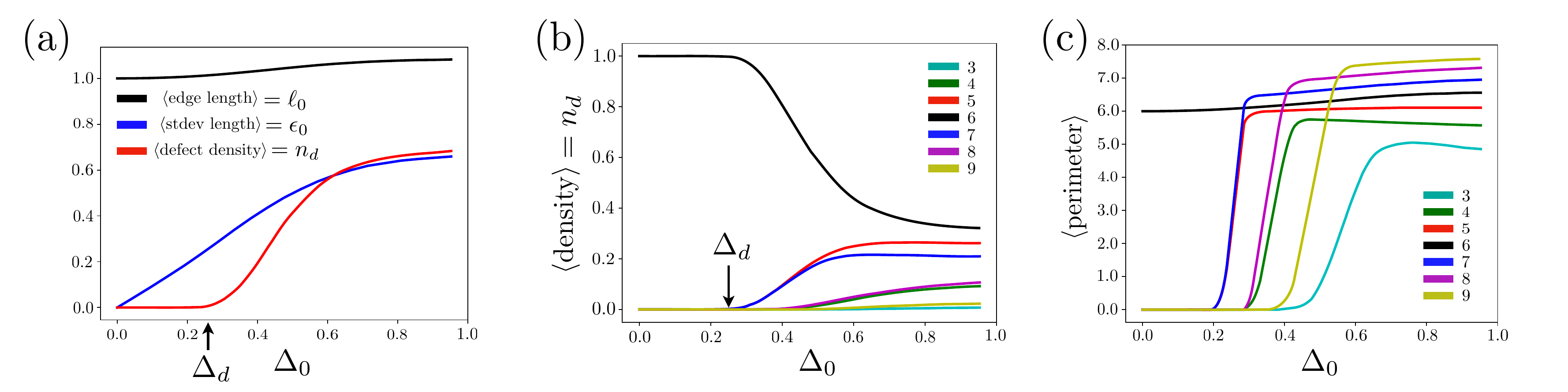}\\
\caption{(a) Black curve represents the ensemble averaged lengths of edges versus the disorder in the position of cell centroids. This curve, as expected, is approximately one sixth of average perimeter of hexagons, depicted in (c), except for very large magnitudes where the effect of defects sets in. Blue curve, shows $\epsilon_0$, the standard deviation of the edge length in the actual lattice. For small values of disorder $\Delta_0\lesssim 0.5$, the edge length disorder grows linearly, with the slope equal to one, and decreases for larger values of disorder. Red curve corresponds to the ensemble average of total density of defects as the disorder is enhanced. The disorders start at around $\Delta \simeq 0.25$. (b), (c) Ensemble averaged densities and perimeters of polygons with different number of sides versus the disorder in the position of cell centroids. The ensemble average is carried out over $10000$ realizations of $50\times 50$ lattices. }{\label{SI-disorderstat}}
\end{figure*}

{\label{app-disorder}}
In the case of 1D systems, disorder is only in the lengths of junctions, and there is no room for the change in the topology of the network. The edge lengths are $\ell_i = \ell_0 (1 + \epsilon_i)$, where $\epsilon_i\in[-\epsilon_0, +\epsilon_0]$ with {\textit{uniform}} distribution, and $\langle \epsilon_i\epsilon_j\rangle = \epsilon_0^2\delta_{ij}/6$. In 2D the quenched disorder refers to (a) unequal edge lengths, and (b) defects defined as the non-hexagonal polygons, tiling the plane. The level of quenched disorder is controlled by randomizing the sites of a triangular lattice, based on which the polygonal lattice is generated using Voronoi tessellation. The edge lengths of the Voronoi lattice $\epsilon_i$, and density of defects $n_d$ are then obtained by ensemble averaging over all the realizations of the disordered triangular lattice. Perturbing the sites of a triangular lattice, we get: $\rr_i = \rr_i^0 (1 + \bm \Delta_i)$, with $\{\rr_i^0\}$. The spatial disorder term $\Delta_i$ is uniformly distributed in range $[-\Delta_0, +\Delta_0]$, with local correlations: $\langle \bm\Delta_{i}\cdot\bm\Delta_{j}\rangle =\Delta_0^2\delta_{ij}/3$, where $\delta_{ij}$ is the Kronecker delta function. For $\Delta_0 = 0$, the corresponding Voronoid lattice would be an ordered hexagonal lattice. By displacing randomly the sites of triangular lattice, we can distort the resultant Voronoi lattice. In order to obtain the disorder statistics of the Voronoi lattice, i.e. variations in the edge lengths $\epsilon_i$, as well as the density of defects $n_d$, we average over ensemble of disordered triangular lattices. Defects with finite (i.e. nonzero) density of defects in the limit of infinitely large systems (i.e. thermodynamic limit), appear above a certain threshold of disorder $\Delta_d\simeq0.25$, in the triangular lattice (see Fig. (\ref{SI-disorderstat})). The edge-length disorder in the Voronoi lattice increases linearly with $\Delta_0$ for $\Delta_0\lesssim 0.5$.

\section{Mean-field and Numerical Solutions}{\label{MF-SI}}
We define the mean-field approximation in this system as constant $f_i^{\text{ubd}}g_j^{\text{ubd}}$ for both sides of all junctions. As mentioned in the Main Text, the validity of this assumption follows from the diffusion-like dynamics of $p, s$. Bound proteins redistribute across the tissue until a relatively smooth state is reached, therefore the free proteins $f_i^{\text{ubd}}g_j^{\text{ubd}}$ too, distribute uniformly. We also check this assumption numerically; see Fig. (\ref{SI-afbf}b1) and (\ref{SI-afbf}b2). Here we first elaborate on the MF solutions in 1D, then discuss the 2D case where the MF solutions are divided into two different classes: trivial and nontrivial. 

\subsection{\small{Mean-Field Solutions in One Dimension}}
\begin{figure}[h]
\center\includegraphics[width=1\linewidth]{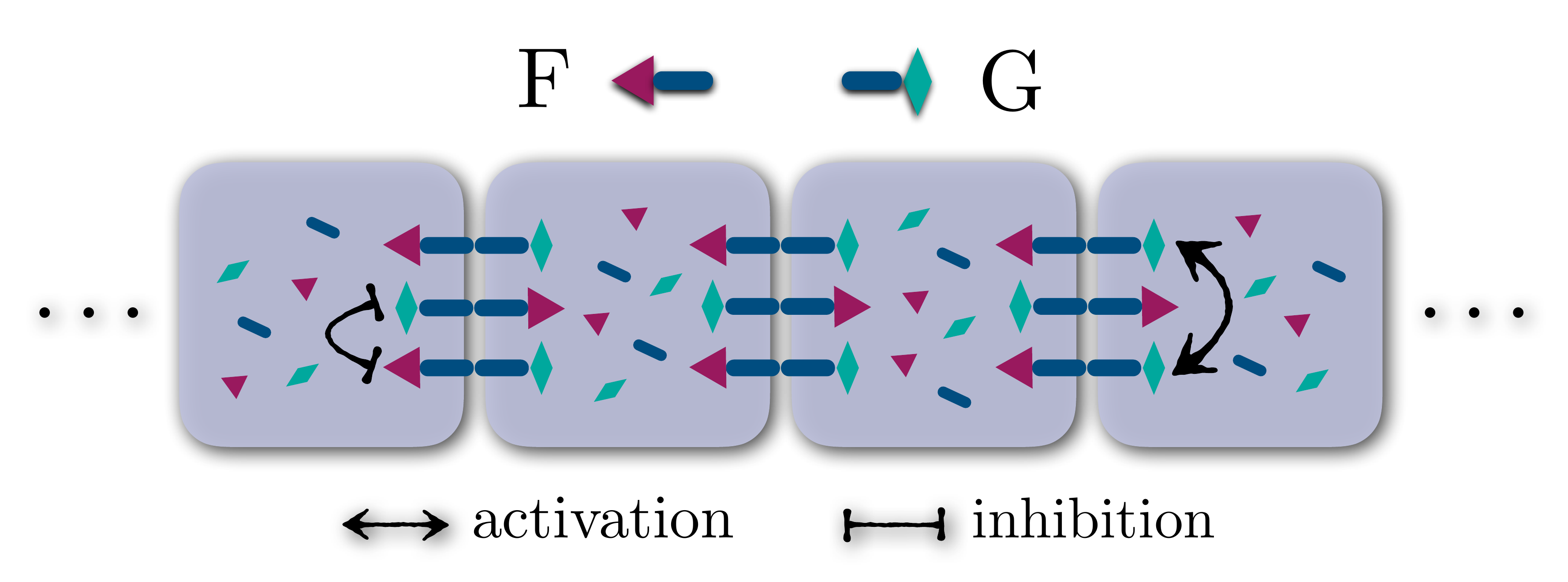}\\
\caption{\small{A cartoon of a one-dimensional array of cells. Free PCP proteins with concentrations $f^{\text{ubd}}, g^{\text{ubd}}$, are available to bind the transmembrane proteins T, and form cross-junctional dimers. The sum of concentrations of [F : G] and [G : F] dimers at each junction is the total concentration $s_{i,i+1} = u_{i,i+1} + u_{i+1,i}$, and the difference is defined as the polarization $p_{i,i+1} = u_{i,i+1} - u_{i+1,i}$. The like/unlike dimers activate/inhibit one another on each junction (shown in the far right/left cells).}}
{\label{SI-1Dscheme}}
\end{figure}

In one dimension the cells are juxtaposed in an array and are separated by junctions. The proteins localize on both sides of these junctions and form dimers. A general scheme of one-dimensional arrays can be seen in Fig. (\ref{SI-1Dscheme}). 
The reaction-diffusion (RD) equations governing $u$ and $v$ complexes on a junction shared by cells $i$ and $i+1$ read:
\begin{align}{\label{SLIRD}}
\frac{du_{i,i+1}}{dt} =&\,\kk f^{\text{ubd}}_ig^{\text{ubd}}_{i+1}(1+\alpha u_{i,i+1}) -\gamma u_{i,i+1}(1+\beta u_{i+1,i}).
\end{align}
A similar equation exists for the opposite complex $v_{i , i+1} = u_{i+1,i}$. The above equations were originally proposed by Mani, et.al. in Ref. \cite{mani2013collective}, namely our general RD equations introduced in the Main Text, reduce to the above equations in 1D with strictly interactions. Therefore we refer the reader to Ref. \cite{mani2013collective}, where the properties and predictions of Eq. (\ref{SLIRD}) in 1D systems are investigated elaborately. Here we briefly reproduce the main results and show that in the steady state, the mean field (MF) solutions of ordered systems, exhibit a bifurcation at a critical value of the control parameter $g_0/f_0$, i.e. the ratio of total amounts of proteins F and G. We start by deriving equations for $p_{i,i+1}$ and $s_{i,i+1}$. This can be done by adding and subtracting Eq. (\ref{SLIRD}) and its counterpart for the opposite complex. We get:
\begin{widetext}
\begin{subequations}{\label{SI-1DFullRD}}
\begin{equation}
\frac{dp_{i,i+1}}{dt} =\,\kk (f^{\text{ubd}}_ig^{\text{ubd}}_{i+1} - f^{\text{ubd}}_{i+1}g^{\text{ubd}}_{i}) + \kk\alpha (f^{\text{ubd}}_ig^{\text{ubd}}_{i+1}u_{i,i+1} - f^{\text{ubd}}_{i+1}g^{\text{ubd}}_{i}u_{i+1,i}) -\gamma (u_{i,i+1} - u_{i+1,i}),
\end{equation}
\begin{equation}
\frac{ds_{i,i+1}}{dt} =\,\kk (f^{\text{ubd}}_ig^{\text{ubd}}_{i+1} + f^{\text{ubd}}_{i+1}g^{\text{ubd}}_{i}) + \kk\alpha (f^{\text{ubd}}_ig^{\text{ubd}}_{i+1}u_{i,i+1} + f^{\text{ubd}}_{i+1}g^{\text{ubd}}_{i}u_{i+1,i}) -\gamma (u_{i,i+1} + u_{i+1,i}) -2\gamma \beta -u_{i,i+1}u_{i+1,i} .\\
\end{equation}
\text{In the above equations we have:}
\begin{equation}
f^{\text{ubd}}_i = \frac{F_0 - (u_{i,i+1}\ell_{i,i+1} + u_{i,i-1}\ell_{i-1,i})}{\ell_{i,i+1} + \ell_{i-1,i}} =  f_{0,i} - \frac{u_{i,i+1}\ell_{i,i+1} + u_{i,i-1}\ell_{i-1,i}}{\ell_{i,i+1} + \ell_{i-1,i}},
\end{equation}
\end{subequations}
\end{widetext}
in which $F_0$ is the total amount of protein  in a cell, assumed to be uniform across the tissue, and $f_{0,i}$ is the concentration of free F available to the two junctions of cell $i$. Similar relation can be written for $g^{\text{ubd}}_i$ and $g_{0,i}$. In ordered systems and within the MF approximation, all of the above variables become independent of index $i$. A few lines of simple algebra shows that the critical value reads:
\begin{align}{\label{bcrit}}
p^2 = &s^2 - \frac{4}{\alpha\beta}\qquad\Rightarrow \qquad s^* = \frac{2}{\sqrt{\alpha\beta}} \nonumber\\
& g_0^* = \frac{\gamma/\kappa\alpha}{1-\sqrt{1/\alpha\beta}}+\sqrt{\frac{1}{\alpha\beta}}.
\end{align}
For the values of the parameters used in the simulations, i.e. $\alpha = \beta = 5$, and $\gamma = 1$ , $\kappa = 10$, the value of $g_0^*$ is found to be $g_0^* = 0.225$, which can also be seen in Fig. (2) of Main Text. Note that in the above equations, for $\alpha\beta \to 0$, the bifurcation point diverges, implying that cooperative interactions are essential to the emergence of collective polarization. 
A closer look at Eqs. (\ref{SI-1DFullRD}), makes it clear that the relevant parameters in the {\textit{local}} bistability of the polarization are $f^{\text{ubd}}_i g^{\text{ubd}}_{i+1}$, and $f^{\text{ubd}}_{i+1} g^{\text{ubd}}_{i}$. When normalized by the total concentration $F_0/\ell_0$ as the unit of concentration, one can rewrite the above parameters in terms of $f_{0,i}g_{0,i+1}$ and $f_{0,i+1}g_{0,i}$.
Therefore, as mentioned in the Main Text, for disordered systems where $f_{0,i}$ and $g_{0,i}$ are no longer uniform, the local bifurcation point is randomized and the singular transition between unpolarized and polarized states becomes a smooth cross-over.
Below we discuss the MF solutions in ordered two dimensional systems.

\subsection{\small{Trivial Mean-Field Solutions}}
Defining the trivial mean-field approximation as the translational invariance of polarization along each of the three axes separately, the RD equations take the same form as in 1D, except the total amount of proteins Fz and Vang are shared by six junctions instead of two. Therefore, the solutions are identical to those of 1D case. This can also be understood intuitively, by noting that in a polarized trivial MF state, each cell is partitioned into a positive and a negative side. Each partition can be thought of as an ``effective edge'' in the 1D case. Although the lengths of these ``effective edges" are different from those in 1D case, the concentrations are equal. Now, in the hexagonal lattice, each edge carries the same $|p|,s$, equal to those found in 1D case. The net polarity of each cell thus reads, $p_c=p_e (1 - 2 \cos\theta)$, with $p_e$ the magnitude of polarization of one edge calculated above, and $\theta$ is the angle between the two adjacent edges in ordered hexagons, which equals $p_c = 2\sqrt{s^2-4/\alpha\beta}$, for $\theta = 2\pi/3$. The trivial MF solutions take two distinct configurations illustrated in Figs. (\ref{SI-afbf}a1) and (\ref{SI-afbf}a2), respectively. While the former represents a state with nonzero net polarization the magnitude of which is calculated above, the latter has zero net polarization. The solutions of the second type, are however destabilized when NLCI is included in the model. 

Furthermore, we consider a different situation in which edges have unequal $\alpha,\beta$'s. 
At this point, different values of $\alpha$, and $\beta$ can have various origins, that are beyond the scope of our discussion. However in the Sec. (\ref{eloncells}), we argue that unequal parameters can be a consequence of nonlocal interactions in elongated cells. Assume the three pairs of parallel edges acquire coefficients $\alpha_{1,2,3}$ and $\beta_{1,2,3}$. Without loss of generality we consider two scenarios: (i) $\alpha_1>\alpha_2 = \alpha_3$, and (ii) $\alpha_1 = \alpha_2 >\alpha_3$. From the results we found in the case of sixfold symmetric lattices, the onset of bifurcation is inversely proportion to $s^*\sim1/\sqrt{\alpha}$. Therefor in case (i), axis 1 is the first axis that shows instability upon increasing $g_0$ above $g_0^*$. Therefore, we have $f^{\text{ubd}}g^{\text{ubd}} = {\gamma}/{(k\alpha_1)}$, where,
\begin{equation}
f^{\text{ubd}} = f_0 - \frac{s_1\ell_1+s_2\ell_2 + s_3\ell_3}{2(\ell_1+\ell_2+\ell_3)},
\end{equation}
and similarly for $g^{\text{ubd}}$. Again, as derived above, the axes where the polarization is zero, i.e. 2, 3, we have:
\begin{equation}
s_{2,3}^2+\frac{2}{\gamma\beta_{2,3}}\left(\gamma - \kk f^{\text{ubd}}g^{\text{ubd}}\alpha_{2,3}\right)s_{2,3}-2\kk f^{\text{ubd}}g^{\text{ubd}} = 0.
\end{equation}
Using the fact that $kf^{\text{ubd}}g^{\text{ubd}}=\gamma/\alpha_1$, we get:
\begin{equation}
s_{2,3} = \frac{1}{\beta_{2,3}}\left(1-\frac{\alpha_{2,3}}{\alpha_1}\right) + \sqrt{\frac{1}{\beta^2_{2,3}}\left(1-\frac{\alpha_{2,3}}{\alpha_1}\right)^2+\frac{2\gamma}{\alpha_1}}.
\end{equation}
Defining $f_0^{\text{eff}}$ and $g_0^{\text{eff}}$,
\begin{align}
f_0^{\text{eff}} &= f_0 - \frac{s_2\ell_2 +s_3\ell_3}{2(\ell_1+\ell_2+\ell_3)}, \nonumber\\
f^{\text{ubd}} &= f_0^{\text{eff}} - \frac{s_1\ell_1}{2(\ell_1+\ell_2+\ell_3)},
\end{align}
and similar expressions for $g_0^{\text{eff}}$ and $g^{\text{ubd}}$, we get for $p_1,s_1$:
\begin{align}
p_1^2 =& s_1^2 - \frac{4}{\alpha_1\beta_1},\nonumber\\
 s_1 = \left(f_0^{\text{eff}} +g_0^{\text{eff}} \right) -& \sqrt{\left(f_0^{\text{eff}} -g_0^{\text{eff}} \right)^2+\frac{4\gamma}{\kk \alpha_1}}.
\end{align}
From the above analysis, we learn that in systems with unequal $\alpha$'s and $\beta$'s, the system essentially reduces to a one dimensional problem, with effective values for the concentrations of the proteins Fz and Vang. There remains one more interesting case in which $\alpha_1 = \alpha_2>\alpha_3$. In this case, the third axis remains unpolarized as the other two are effectively more absorbent, and share the total bound proteins, thus are equally polarized with fourfold symmetry. As mentioned above, one situation of interest in which the coefficients $\alpha,\beta$ acquire unequal values for the edges is the elongated tissues. In such systems, the above two cases correspond to Figs. (\ref{SI-elon}a1), (\ref{SI-elon}a2), and (\ref{SI-elon}a3), respectively; see Sec. (\ref{eloncells}).

\subsection{Nontrivial Mean-Field Solutions}
\begin{figure*}[ht]
\center
\includegraphics[width=1\linewidth]{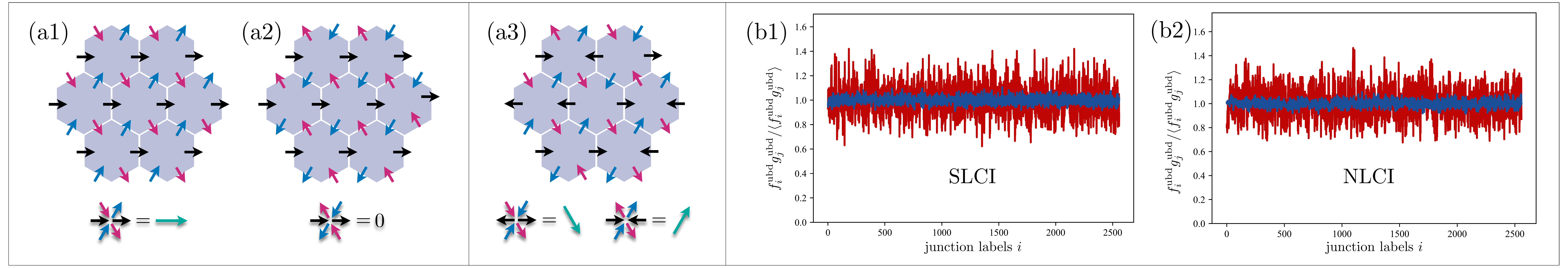}
\caption{(a1) and (a2) show the trivial MF solutions with nonzero and zero net polarizations, respectively. While (a1) is a stable solution, (a2) is destabilized by NLCI. (a3) is an illustration of a nontrivial MF solution where cellular polarities point in random directions, that is only stable in SLCI limit, but unstable in NLCI. For SLCI and NLCI, respectively, (b1) and (b2) show the initial (red) and final (dark blue) distributions of $f^{\text{ubd}}_ig^{\text{ubd}}_j$ for randomized edges labeled ($ij$), i.e. shared between cells $i$ and $j$. The variance of the distributions decreases from initial to final state, validating the MF approximation.}{\label{SI-afbf}}
\end{figure*}

The nontrivial solutions in 2D do not possess the translational invariance of polarity \textit{vectors}, but only satisfy a weaker constraint. The full analytic treatment of this problem is cumbersome. We only briefly touch upon this subject to provide some intuition into how large the basin of attraction is, for a system in SLCI limit. The only assumption in nontrivial MF solutions is that for each junction shared by adjacent cells $i$ and $j$, the products $f_i^{\text{ubd}}g_j^{\text{ubd}}$ and $f_j^{\text{ubd}}g_i^{\text{ubd}}$ are uniform across the tissue. This assumption is backed up by the following numerical analysis. We compute the r.m.s. of the fluctuations of $f^{\text{ubd}}_i g^{\text{ubd}}_j$, for all the junctions with fully randomized labels, and compare the initial and final values. We see in Figs. (\ref{SI-afbf}b1) and (\ref{SI-afbf}b2), that for both SLCI and NLCI mechanisms, the final distribution of $f^{\text{ubd}}_ig^{\text{ubd}}_j$ is noticeably narrower than the initial distribution. Intuitively this approximation can be justified by the fact that the linearized reaction-diffusion equations governing $u$ and $v$, obey diffusion-like equations in the continuum limit \cite{mani2013collective}. Therefore, given that the total concentrations $f_0$ and $g_0$ are uniform across the tissue, the free proteins too, spread diffusively into a rather uniform state. In SLCI limit and/or for ordered lattices, by virtue of sixfold symmetry, all junctions equally absorb the proteins. Therefore above the bifurcation point, net polarization $p$ and the total amount of proteins $s$, are identical for all junctions. The only constraint is that three junctions have net positive and the other three have negative polarizations; i.e. three outgoing and three incoming dipoles; see Fig. (\ref{SI-afbf}). 

Besides trivial MF solutions depicted in Figs. (\ref{SI-afbf}a1) and (\ref{SI-afbf}a2), nontrivial solutions like that in (\ref{SI-afbf}a3) constitute another class of MF solutions. In order to make it easier to picture such configurations of dipoles, imagine we start from a trivial solution of (a1), where three adjacent junctions carry outgoing dipoles and the other three carry incoming dipoles. Flipping one of the outgoing dipoles violates the MF assumption of $f_i^{\text{ubd}}g_j^{\text{ubd}}=$ constant, unless one of the incoming dipoles flips as well. Since each dipole is shared between two cells, in order to satisfy the constraint in any finite-size system, this flipping process must continue until it forms a loop ending at the first flipped dipole. It is easy to see that the paths can be broken down into self-avoiding loops. Furthermore, all such loops preserve the net polarization of the tissue, set by the value of the control parameter $g_0/f_0$. However, they disrupts the uniform polarization of the trivial solutions, namely create excitations above the uniform configuration. Incidentally this is what we observe in simulations; the net polarization at the steady state is to a high accuracy independent of initial configuration. The small deviations is however understandable: the constraint of constant $f^{\text{ubd}}g^{\text{ubd}}$ is not guaranteed to be accurately satisfied in real systems with arbitrary initial conditions and stochastic noise. Moreover, in an ensemble of the systems starting from all possible initial conditions, all nontrivial configurations preserving the constraint as well as the net polarization, are equally accessible to the system. The above simple analysis provides insight into why the system in SLCI limit is not guaranteed to reach a state with large-scale polarization, and instead NLCI mechanism stabilizes the states with nearly uniform polarization, by promoting the adjacent junctions to carry the same kind of dipoles, hence segregation. In order for the cells to meet the two criteria simultaneously, the directions of the cell polarizations must also be parallel, whereas in SLCI case, different cells can have different $|\pp_i|$, hence different orientations. \\

\subsection{Equal vs. Unequal Length Scales of Cytoplasmic Interactions}
In the Main Text we discussed the effects of unequal nonlocal cytoplasmic interactions that appear in intra- and interspecies kernels, i.e. $\lambda_{\text{uu}}$ and  $\lambda_{\text{uv}}$. Here we elaborate on that discussion and consider various cases to investigate the role of these length scales, and the behavior of angular correlation of polarity. In each case the geometry of the tissue is held fixed. The geometrical disorder is denoted by $\epsilon_0$.\\ 

\noindent (1) For $\epsilon_0 = 0.5$, find the angular correlation for different values of $\lambda_{\text{uu}} = \lambda_{\text{uv}} =$ 0.01 -- 0.8. 

\noindent (2),(3) For $\epsilon_0 =$ 0.45 and 0.6, and $\lambda_{\text{uv}} =$ 0.5, compare the angular correlation for $\lambda_{\text{uu}}=$ 0.01 -- 0.8.

\noindent (4) For $\epsilon_0 = 0.5$, and $\lambda_{\text{uu}} =$ 0.5, compare the angular correlation for $\lambda_{\text{uv}}=$ 0.01 -- 0.8.\\

\begin{figure*}[t]
\center
\includegraphics[width=0.8\linewidth]{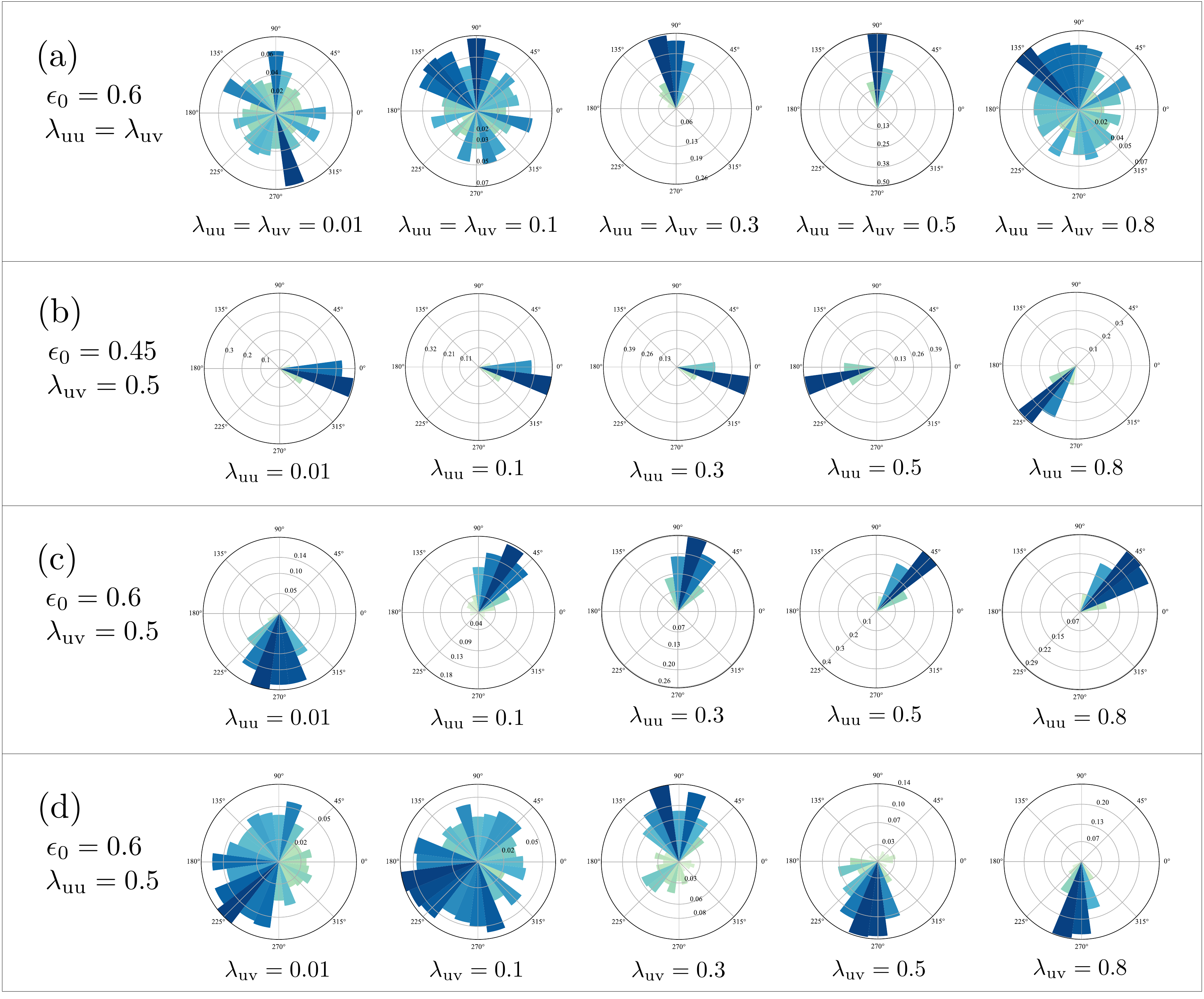}\\
\caption{Rose plots demonstrate the angular distribution of polarization field in systems with equal and unequal $\lambda_{\text{uu}}$ and $\lambda_{\text{uv}}$, and for different values of geometrical disorder, $\epsilon_0$. (a) Equal $\lambda$'s, with $\epsilon_0 = 0.6$. (b),(c) For fixed $\lambda_{\text{uv}} = 0.5$ and variable $\lambda_{\text{uu}}=$ 0.01 -- 0.8, angular distributions are compared for $\epsilon_0 = 0.45$ and $\epsilon_0 =0.6$. (d) Illustration of the angular distributions for fixed $\lambda_{\text{uu}}=0.5$ and variable $\lambda_{\text{uv}}=$ 0.01 -- 0.8, for $\epsilon_0 = 0.6$. }{\label{SI-LAGI}}
\end{figure*}

\noindent The findings are as follows:\\

\noindent (1) For equal length scales, we observe that the angular correlation increases with $\lambda = \lambda_{\text{uu}} = \lambda_{\text{uv}}$. The standard deviation of polarity is found to be smaller than 40 (degrees), for $\lambda \simeq$ 0.2 -- 0.7, and is maximized at around $\lambda \simeq$ 0.4 -- 0.5. For $\lambda \gtrsim$ 0.7 the stochastic noise destroys the long-range polarization drastically. The reason is stabilizing and destabilizing forces compete at all points on the perimeter of each cell, namely all points are strongly coupled to one another. Thus the segregation becomes progressively more unstable as $\lambda$ is increased; see Fig. (\ref{SI-LAGI}a). \\

\noindent (2),(3) For geometrical disorder $\epsilon_0 \lesssim0.45$, and for $\lambda_{\text{uv}}\simeq$ 0.2 -- 0.7 there is no detectable difference between the angular correlation of systems with different values of $\lambda_{\text{uu}} \lesssim$ 0.8. Beyond this value, the correlation declines slightly, and is eventually destroyed for larger values. In order to further examine the importance $\lambda_{\text{uu}}$ in disordered tissues, we repeat the same simulations, but for larger geometrical disorder $\epsilon_0=0.6$. Interestingly we realize that larger disorder impedes the long-range correlation of polarity for $\lambda_{\text{uu}}\lesssim 0.4$. The correlated polarity is retained for $0.4\lesssim\lambda_{\text{uu}}\lesssim 0.7$ and is declined again for larger values; see Fig. (\ref{SI-LAGI}b) and (\ref{SI-LAGI}c). \\

\noindent (4) Systems with nonlocal activation and local inhibition are incapable of achieving long-range polarization. We fixed $\epsilon_0 = 0.6$ and $\lambda_{\text{uu}}=0.5$ and vary the inhibition length scale between $\lambda_{\text{uv}}=$ 0.01 -- 0.8. The correlation grows as $\lambda_{\text{uv}}$ increases; see Fig. (\ref{SI-LAGI}d). An intuitive argument for why small $\lambda_{\text{uv}}$ fails to work is as follows: for $\lambda_{\text{uv}}\gtrsim 0.8$, the unlike complexes on the opposite sides of the cell inhibit each other, e.g. Fz on the right side, inhibits Vang on the left side. Thus, the Vang proteins sitting on the left side of the same cell are destabilized by Fz from across the cell. However, as far as polarity is concerned, the Vang protein on the opposite side of the cell is indeed contributing positively to the same direction of polarity as Fz. Therefore, for $\lambda_{\text{uv}}\gtrsim 0.8$, the polarization, even if initially correlated over long distances, becomes highly unstable against the stochastic noise. The same logic holds true for mutual stabilizing feedback of the like complexes if $\lambda_{\text{uu}}\gtrsim 0.8$, which also contributes negatively to the alignment of the polarization field. \\

\noindent Comparing the charts in Fig. (\ref{SI-LAGI}), we conclude that: (1) In general geometrical disorder distorts the alignment of polarization on large length scales. (2) Equal length scales of activation and inhibition is more efficient in retaining the long-range correlation, especially in highly disordered tissues. (3) Unlike the case of equal $\lambda_{\text{uu}} = \lambda_{\text{uv}}$, fixing $\lambda_{\text{uu}}$ at an intermediate value, e.g. 0.5, allows for the other length scale to exceed $\lambda_{\text{uv}}\simeq0.7$, without losing the angular correlation. Though the correlation eventually decays for larger values, $\lambda_{\text{uv}}\gtrsim 0.9$.

\begin{figure*}
\begin{center}
    \includegraphics[width=1\textwidth]{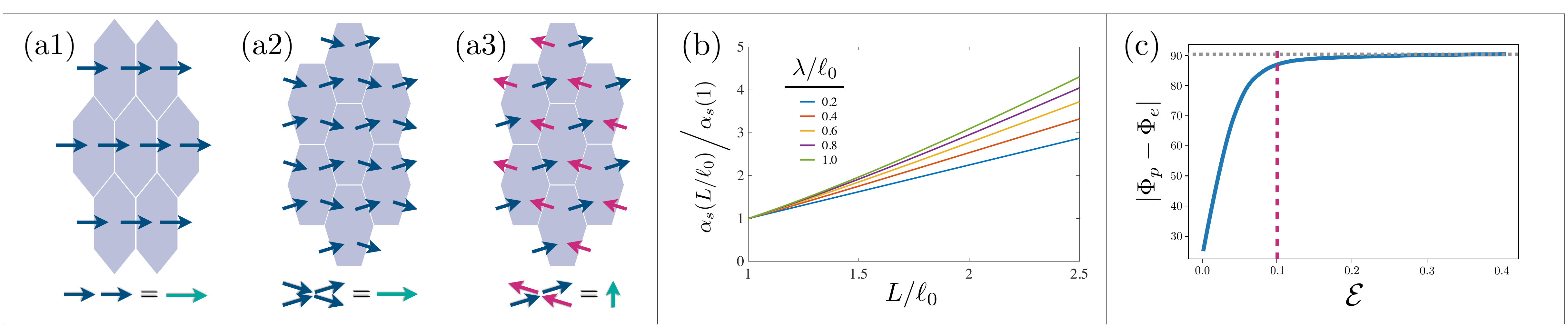}
\end{center}
\caption{ (a1) Shows the elongated system, with elongation axis passing thorough a vertex. Since the cooperative interactions increase with length, the long junctions will be polarized before the shorted edges can absorb complexes. This case is twofold symmetric, like a 1D array of cells, extended perpendicular to the elongation axis. (a2) and (a3) show the alternative elongation axis perpendicular to one of the edges. In these cases there are four long edges competing to absorb complexes and get polarized. The possible configurations are now fourfold, two of which (a2) are polarized perpendicular to the axis of elongation, and the other two are parallel (a3). The latter is destabilized by the NLCI. (b) The cooperative self-interaction $\alpha_s$ as a function of $L/\ell_0$, normalized by those calculated for $L=\ell_0$, is plotted for different values of $\lambda/\ell_0$. (c) The angle between the axis of net polarization with $y$-axis, i.e. the axis of elongation, as a function of elongation index $\mathcal E$, for a system with the same initial condition and primary lattice (see the explanation below, on the precise definition of $\Phi_p$). At $\mathcal E\simeq 0.1$, the axes of polarization and elongation are almost orthogonal, with $|\Phi_p-\Phi_e|\simeq 87$ (degrees).}{\label{SI-elon}}
\end{figure*}

\section{Elongated Cell Geometry}{\label{eloncells}}
In anisotropic cells, the effective values of cooperative interactions are larger for the long junctions. Here, without deriving explicit expressions for the effective parameters in elongated system, we only argue that, should we solve the full NLCI equations in elongated cells, the longer junctions will acquire larger coefficients $\alpha,\beta$. As in the case of one dimension, searching for mean-field (MF) solutions, we assume that in steady state, the concentrations of bound proteins are translationally invariant along the three main axes of the lattice. In a self-consistent approach, the effective $\alpha,\beta$ are dependent on the concentrations of dimers on other edges. Therefore, assuming the system has reached its steady state, we can write the cooperative interactions as functions of the concentrations of dimers on all the edges, weighted by the geometrical factors originating from nonlocal interactions. For small ranges of NLCI, $\lambda\ll 0.7$, the effect of other edges are negligible and only the self-interaction of each edge is to be taken into account. Intuitively and also from the expression given in Sec. (1) of the Main Text for nonlocal interactions, it can be understood that the self-interaction is a monotonically increasing function of the edge length. Therefore longer edges with NLCI, have larger values of $\alpha,\beta$. Upon increasing $\lambda$, the interactions between all pairs of edges increase. However the qualitative behavior of effective $\alpha,\beta$'s for different edges does not change. With this in mind, and using the symmetry arguments between junctions with equal lengths, one can see that the cartoons in Figs. (\ref{SI-elon}a1), and (\ref{SI-elon}a2, \ref{SI-elon}a3) correspond to, i.e. $\alpha_1>\alpha_2,\alpha_3$ and $\alpha_1=\alpha_2>\alpha_3$, respectively. Using the analysis sketched in Sec. (\ref{MF-SI}.2), it is easy to see that in such systems the junctions with larger $\alpha,\beta$, are more unstable towards polarization transition (remember the critical value decreases as $1/\sqrt{\alpha\beta}$, as $\alpha,\beta$ increase). In Fig. (\ref{SI-elon}a1) the polarization points toward the middle of the junction parallel to the elongation axis, whereas (a2) polarization vector passes through a vertex. Both cases of Fig. (\ref{SI-elon}a1) and (\ref{SI-elon}a2) acquire polarizations perpendicular to the elongation axis. In disordered systems (and in real systems), the cell geometry can be a combination of both types. There also exists an unstable configuration Fig. (\ref{SI-elon}a3) which is unpolarized. The instability is again due to nonlocal interactions which prohibits adjacent cells carrying opposite dimers; like the twofold symmetric trivial MF solutions of equilateral cells. Following the above discussion on different coefficients cooperative self-interactions $\alpha_s$, and the equations derived in Sec. (\ref{MF-SI}.2), we calculate these coefficients as functions of $L/\ell_0$, for different $\lambda$'s. Fig. (\ref{SI-elon}b) shows the numerical results for $0.2\leq\lambda/\ell_0\leq1$. In order to discern the effect of the elongation from that of nonlocal interactions, we normalize all the curves by the coefficients calculated for $L = \ell_0$, such that $\alpha_s(1) = 1$ for all $\lambda$'s. The coefficient of self-interaction is a monotonically increasing function of the length of junction; hence polarity is perpendicular to the longer junctions, as depicted in Figs. (\ref{SI-elon}a1) and (\ref{SI-elon}a2).\\

\begin{figure*}[t]
\center
\includegraphics[width=0.8\linewidth]{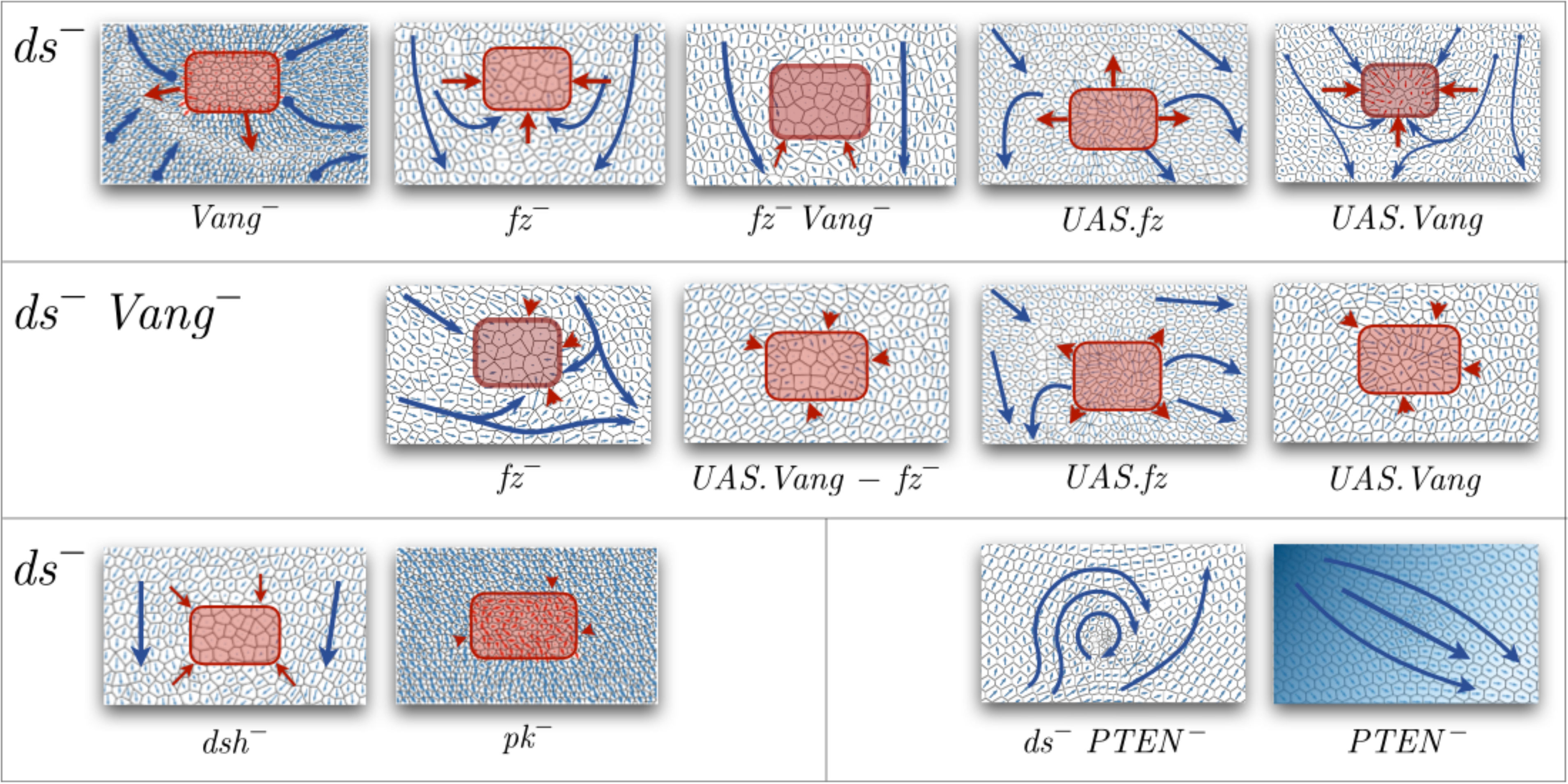}\\
\caption{Illustrations of type I, II, and III mutants. The layout of the table is the same as that in the Main Text. The read arrows are in the direction of distortion with respect to the wild-type polarity.}{\label{SI-MUT}}
\end{figure*}

\noindent \textbf{Measure of elongation: cell nematic tensor.}
Elongation is commonly parametrized using a traceless matrix $\widehat{\mathcal E}$. This matrix, the magnitude and the angle of elongation with respect to $x$-axis read:
\begin{align}{\label{ShapeTensor}}
\widehat{\mathcal E}_i &= 
\begin{pmatrix}  \varepsilon_{i,1} & \varepsilon_{i,2}\\ \varepsilon_{i,2} & -\varepsilon_{i,1} \end{pmatrix},\nonumber\\
 \mathcal E_i &= \sqrt{\varepsilon_{i,1}^2 + \varepsilon_{i,2}^2}\;,\nonumber\\
\phi^e_i &= \frac{1}{2}\cos^{-1}(\varepsilon_{i,1}/\mathcal E_i).
\end{align}
For a tissue consisting of $N_c$ cells, and with $\varepsilon_1 = N_c^{-1}\sum_{i=1}^{N_c}\varepsilon_{i,1}$, we get for the average elongation:
\begin{equation}
\mathcal E = N_c^{-1}\sum_{i=1}^{N_c}\mathcal E_i\;,\qquad
\Phi_e = \frac{1}{2}\cos^{-1}(\varepsilon_1/\mathcal E).
\end{equation}

In terms of the elongation index $\mathcal E$, we plot in Fig. (\ref{SI-elon}b), the angle between $y$-axis and the steady-state polarization vector $|\Phi_p-\Phi_e|$ (in degrees), versus different values of $\mathcal E$, for a system with the same initial condition and same \textit{initial} lattice (i.e. before stretching), but elongated along $y$-axis. Here, $\Phi_p$ is defined as the angle of the axis between polarization and the $x$-axis, and over the domain of $\Phi_p\in[0,\pi)$. Since $\Phi_e = \pi/2$ per definition, we have $|\Phi_p-\Phi_e|\in[0,\pi/2]$. Of course for $\mathcal E = 0$, the axis of elongation is not defined, yet we measure it with respect to $y$-axis. It is important to note that this figure is only an example where the polarization for $\mathcal E = 0$, happens to make a small angle with $y$-axis (25 degrees), hence the pronounced effect of elongation. It is clear that depending on the geometry of the lattice and initial condition, the polarization can be almost perpendicular to $y$-axis, even for $\mathcal E = 0$. Therefore among different simulations, we chose one with a relatively large effect of elongation.

\section{Mutants and The Corresponding Phenotypes}
In the Main Text we tabulated the schematics of phenotypes of three types of mutants. The schematics in the Main Text Fig. (9) show the non-autonomous effects of each clone. Here we show the actual polarization field obtained from the simulations, in which the autonomous effects within the clones are also illustrated. \\

\noindent\textbf{Type-I Mutants: Lack of Membrane Proteins.}
Type I lack one or both of the membrane proteins, namely $f_0, g_0$. The non-autonomy is evident in all cases. In the first row of Fig. (\ref{SI-MUT}), i.e. $\textit{ds}^-$ background, non-autonomy is extended to multiple layers of cells from the clone boundaries. In second row where the background lacks Vang as well, the non-autonomy is mostly limited to a single layer of surrounding cells. \\

\noindent\textbf{Type-II Mutants: Lack of Cytoplasmic Interactions.}
The role of cytoplasmic proteins is deduced by comparing the \textit{in silico} and \textit{in vivo} phenotypes. We concluded that Dsh is mostly responsible for the local cooperative interactions with some contributions to the nonlocal part. Pk on the other hand is mainly involved in nonlocal interactions, through cytoplasmic diffusion. It can be seen in bottom left panel of Fig. (\ref{SI-MUT}), that in agreement with experiments, e.g. \cite{amonlirdviman2005mathematical}, the phenotypes show almost no non-autonomy, thought distinct autonomous patterns. Lacking Dsh removes the polarity altogether, whereas absence of Pk has only minor effects on the polarity of the clone. \\

\noindent\textbf{Type-III Mutants: Disordered Geometry.}
We tested the predictions of our model in the case of geometrical irregularity. In Ref. \cite{ma2008cell}, Ma, et.al. suggested that clones with enhanced geometrical disorder, induced by under-expression of PTEN, are obstacles to the propagation of polarization field. This effect appears in tissues where global cues are absent. Polarity was observed to be retained when the global cue was added. As can be seen in the two right panels of the last row in Fig. (\ref{SI-MUT}), the system with lack of global cue shows swirly pattern of polarization. The WT polarity is retained upon adding the gradient cue.

%
%\footnotesize
%\bibliography{pcpref}{}
%\bibliographystyle{unsrt}

\end{document}